%% file: main.tex
\documentclass[preprint]{iacrtrans}
\usepackage[utf8]{inputenc}

\usepackage{amsmath,amssymb,amsfonts}
\usepackage{algorithm}
\usepackage{algorithmic}
\usepackage{stfloats}
\usepackage{multicol}
\usepackage{multirow}
\usepackage{tablefootnote}
\usepackage[flushleft]{threeparttable}
\usepackage{color, colortbl}

\usepackage{hyperref}

\usepackage{graphicx}
\graphicspath{ {Figures/} }

\usepackage{lipsum}

\newcommand{\Fq}{\mathbb{F}_q}

\newcommand{\Z}{\mathbb{Z}}

\newcommand{\Zq}{\mathbb{Z}_q}
\newcommand{\Zqn}{\mathbb{Z}_q^n}

\definecolor{Gray}{gray}{0.9}
\definecolor{Comment}{gray}{0.65}
\definecolor{IFDEF}{gray}{0.35}

\author{Utsav Banerjee \and Tenzin S. Ukyab \and Anantha P. Chandrakasan}
\institute{Dept. of EECS, Massachusetts Institute of Technology, Cambridge, MA, USA}
\title[Sapphire: A Configurable Lattice Crypto-Processor]{Sapphire: A Configurable Crypto-Processor for Post-Quantum Lattice-based Protocols}

\begin{document}

\maketitle


\begin{abstract}
Public key cryptography protocols, such as RSA and elliptic curve cryptography, will be rendered insecure by Shor’s algorithm when large-scale quantum computers are built. Cryptographers are working on quantum-resistant algorithms, and lattice-based cryptography has emerged as a prime candidate. However, high computational complexity of these algorithms makes it challenging to implement lattice-based protocols on low-power embedded devices. To address this challenge, we present Sapphire -- a lattice cryptography processor with configurable parameters. Efficient sampling, with a SHA-3-based PRNG, provides two orders of magnitude energy savings; a single-port RAM-based number theoretic transform memory architecture is proposed, which provides 124k-gate area savings; while a low-power modular arithmetic unit accelerates polynomial computations. Our test chip was fabricated in TSMC 40nm low-power CMOS process, with the Sapphire cryptographic core occupying 0.28 mm$^2$ area consisting of 106k logic gates and 40.25 KB SRAM. Sapphire can be programmed with custom instructions for polynomial arithmetic and sampling, and it is coupled with a low-power RISC-V micro-processor to demonstrate NIST Round 2 lattice-based CCA-secure key encapsulation and signature protocols Frodo, NewHope, qTESLA, CRYSTALS-Kyber and CRYSTALS-Dilithium, achieving up to an order of magnitude improvement in performance and energy-efficiency compared to state-of-the-art hardware implementations. All key building blocks of Sapphire are constant-time and secure against timing and simple power analysis side-channel attacks. We also discuss how masking-based DPA countermeasures can be implemented on the Sapphire core without any changes to the hardware.
\end{abstract}


\input{body/01_intro.tex}
\input{body/02_background.tex}
\input{body/03_ntt.tex}
\input{body/04_sampling.tex}

\input{body/05_arch.tex}
\input{body/06_measurements.tex}

\input{body/07_conclusion.tex}

\section*{Acknowledgements}

The authors would like to thank Texas Instruments for funding this work, the TSMC University Shuttle Program for chip fabrication support, and Bluespec, Xilinx, Cadence, Synopsys and Mentor Graphics for providing CAD tools. The authors also thank the anonymous reviewers for their valuable comments and suggestions.

\bibliographystyle{ieeetr}
\bibliography{references.bib}

\clearpage

\appendix
\renewcommand{\thesection}{Appendix \Alph{section}}
\renewcommand{\thealgorithm}{}

\input{body/appendix_a.tex}

\input{body/appendix_b.tex}

\end{document}

%% file: body/01_intro.tex
\section{Introduction}
\label{sec:intro}

Modern public key cryptography relies on hard mathematical problems such as integer factorization, discrete logarithms over finite fields and discrete logarithms over elliptic curve groups. However, these problems can be solved by a large-scale quantum computer in polynomial time using Shor’s algorithm \cite{shor_quantum_1997}, thus making today's public key protocols like RSA and ECC vulnerable to quantum attacks. Given the rapid advancement in quantum computing technology over the past few years, cryptographers are developing quantum-secure public key algorithms to protect today’s data from tomorrow’s threats. Lattice-based cryptography is being considered one of the most promising candidates for post-quantum cryptographic protocols because of its extensive security analysis as well as small public key and signature sizes.

The National Institute of Standards and Technology (NIST) formally initiated the process of standardizing post-quantum cryptography in 2016 \cite{nist_pq1_2016}. The first round of candidates were announced in late 2017, with lattice-based cryptography accounting for 48\% of the public-key encryption and key encapsulation (PKE/KEM) schemes and 25\% of the signature schemes. In early 2019, the candidates moving on to the second round were announced \cite{nist_pq2_2019}, and lattice-based cryptography accounts for 53\% (9 out of 17) and 33\% (3 out of 9) of the candidates for PKE/KEM and signature schemes respectively. The theoretical foundation of several of these lattice-based protocols lies in the \textit{learning with errors} (LWE) problem \cite{regev_lwe_2005} and its variants such as Ring-LWE \cite{vadim_ringlwe_2013} and Module-LWE \cite{langlois_module_2015}, and the hardness of LWE has been well-studied in the presence of both classical and quantum adversaries \cite{brakerski_hardness_2013, regev_quantum_2004}. This has been accompanied by several software and hardware implementations \cite{ingrid_ringlwe_2014, ingrid_software_2015, alkim_newhopearm_2016, kuo_newhopefpga_2017, guneysu_newhopefpga_2017, cammarota_ringlwe_2018, bos_frodom4_2018, guneysu_frodo_2018, zhang_leia_2018, albrecht_rsa_2018, liu_rlwe_2019, basu_pqchw_2019} of LWE and Ring-LWE-based public key encryption and key encapsulation protocols, each supporting specific lattice parameters chosen for increased performance and efficiency. Existing lattice-based cryptography implementations, both in software and hardware, have been thoroughly surveyed in \cite{cammarota_survey_2019}. Most of the hardware implementations focus on FPGA demonstration in order to support reconfigurability of lattice parameters, which is especially important for a fast evolving field like lattice-based cryptography, while existing ASIC implementations either lack configurability or have power and area overheads. Some of the key challenges of implementing lattice-based cryptography in ASICs have been discussed in \cite{guneysu_survey_2016}, and this work presents a solution using a combination of architectural and algorithmic techniques.

\textbf{Our contributions:} In this work, we present Sapphire -- a configurable lattice cryptography processor -- which combines low-power modular arithmetic, area-efficient memory architecture and fast sampling techniques to achieve high energy-efficiency and low cycle count, ideal for securing low-power embedded systems. The key technical aspects of our work are as follows:
\begin{enumerate}
\itemsep0em
\item A low-power modular arithmetic core, with configurable prime modulus, is used to accelerate polynomial arithmetic operations; a pseudo-configurable modular multiplier is also implemented, which provides up to $3 \times$ improvement in energy-efficiency.
\item A single-port SRAM-based number theoretic transform (NTT) memory architecture provides 124k-gate area savings without any loss in performance or energy-efficiency.
\item An efficient Keccak core is combined with fast sampling techniques to speed up polynomial sampling, while supporting a wide variety of discrete distribution parameters suitable for lattice-based schemes.
\item These efficient hardware building blocks are integrated together with an instruction memory and decoder to build our crypto-processor, which can be programmed with custom instructions for polynomial sampling and arithmetic.
\item The Sapphire crypto-processor is coupled with an efficient RISC-V micro-processor to demonstrate several NIST Round 2 lattice-based key encapsulation and signature protocols such as Frodo \cite{alkim_frodo_2019}, NewHope \cite{alkim_newhope_2019}, qTESLA \cite{bindel_qtesla_2019}, CRYSTALS-Kyber \cite{bos_kyber_2019} and CRYSTALS-Dilithium \cite{vadim_dilithium_2019}, achieving more than an order of magnitude improvement in performance and energy-efficiency compared to state-of-the-art assembly-optimized software and hardware implementations.
\item All the key building blocks, such as NTT, polynomial arithmetic and binomial sampling, are constant-time and secure against timing and simple power analysis attacks. While our baseline protocol implementations are not secure against differential power analysis attacks, we discuss how the programmability of our crypto-processor can be utilized to implement masking-based countermeasures.
\item Our ASIC implementation was fabricated in the TSMC 40nm low-power CMOS process, and all protocol-level demonstrations and side-channel measurements have been conducted on our test chip.
\end{enumerate}

The rest of the paper is organized as follows: Section \ref{sec:background} provides a brief mathematical background on LWE and associated computations; in Section \ref{sec:ntt}, we present our implementation of energy-efficient modular arithmetic along with an area-efficient NTT memory architecture; in Section \ref{sec:sampling}, we describe our discrete distribution sampler accelerated by a low-power SHA-3 core; Section \ref{sec:arch} describes the overall chip architecture; Section \ref{sec:meas} presents detailed measurement results obtained from evaluating lattice-based protocols on our test chip, comparison with state-of-the-art software and hardware implementations as well as side-channel analysis; a summary of our key conclusions along with future research directions are discussed in Section \ref{sec:conclusion}. This version is same as our CHES 2019 paper except for fixed typos and the addition of Frodo-1344 implementation results.

%% file: body/02_background.tex
\section{Background}
\label{sec:background}

In this section, we provide a brief introduction to LWE, Ring-LWE and Module-LWE along with the associated computations. We use bold lower-case symbols to denote vectors and bold upper-case symbols to denote matrices. The symbol $\text{lg}$ is used to denote all logarithms with base 2. The set of all integers is denoted as $\Z$ and the quotient ring of integers modulo $q$ is denoted as $\Zq$. For two $n$-dimensional vectors $\boldsymbol{a}$ and $\boldsymbol{b}$, their inner product is written as $\langle \boldsymbol{a}, \boldsymbol{b} \rangle = \sum_{i=0}^{n-1} a_i \cdot b_i$. The concatenation of two vectors $\boldsymbol{a}$ and $\boldsymbol{b}$ is written as $\boldsymbol{a}$ $||$ $\boldsymbol{b}$.

\subsection{LWE and Related Lattice Problems}

The Learning with Errors (LWE) problem \cite{regev_lwe_2005} acts as the foundation for several modern lattice-based cryptography schemes. The LWE problem states that given a polynomial number of samples of the form $(\boldsymbol{a}, \langle \boldsymbol{a}, \boldsymbol{s} \rangle + e)$, it is difficult to determine secret vector $\boldsymbol{s} \in \Zqn$, where vector $\boldsymbol{a} \in \Zqn$ is sampled uniformly at random and error $e$ is sampled from the appropriate error distribution $\chi$. Examples of secure LWE parameters are $(n, q) = (640, 2^{15})$, $(n, q) = (976, 2^{16})$ and $(n, q) = (1344, 2^{16})$ for Frodo \cite{alkim_frodo_2019}.

LWE-based cryptosystems involve large matrix operations which are computationally expensive and also result in large key sizes. To solve this problem, the Ring-LWE problem \cite{vadim_ringlwe_2013} was proposed, which uses ideal lattices. Let $R_q = \Zq[x]/(x^n+1)$ be the ring of polynomials where $n$ is power of 2. The Ring-LWE problem states that given samples of the form $(a, a \cdot s + e)$, it is difficult to determine the secret polynomial $s \in R_q$, where the polynomial $a \in R_q$ is sampled uniformly at random and the coefficients of the error polynomial $e$ are small samples from the error distribution $\chi$. Examples of secure Ring-LWE parameters are $(n, q) = (512, 12289)$ and $(n, q) = (1024, 12289)$ for NewHope \cite{alkim_newhope_2019}.

Module-LWE \cite{langlois_module_2015} provides a middle ground between LWE and Ring-LWE. By using module lattices, it reduces the algebraic structure present in Ring-LWE and increases security while not compromising too much on the computational efficiency. The Module-LWE problem states that given samples of the form $(\boldsymbol{a}, \boldsymbol{a}^T \boldsymbol{s} + e)$, it is difficult to determine the secret vector $\boldsymbol{s} \in R_q^k$, where the vector $\boldsymbol{a} \in R_q^k$ is sampled uniformly at random and the coefficients of the error polynomial $e$ are small samples from the error distribution $\chi$. Examples of secure Module-LWE parameters are $(n, k, q) = (256, 2, 7681)$, $(n, k, q) = (256, 3, 7681)$ and $(n, k, q) = (256, 4, 7681)$ for CRYSTALS-Kyber \cite{bos_kyber_2019}.


\subsection{Number Theoretic Transform}

While the protocols based on standard lattices (LWE) involve matrix-vector operations modulo $q$, all the arithmetic is performed in the ring of polynomials $R_q = \Zq[x]/(x^n+1)$ when working with ideal and module lattices. There are several efficient algorithms for polynomial multiplication \cite{bernstein_mult_2008}, and the Number Theoretic Transform (NTT) is one such technique widely used in lattice-based cryptography.

\begin{algorithm}[!b]
\caption{Iterative In-Place NTT \cite{cormen_algo_2009}}
\label{algo:iter_ntt}
\begin{algorithmic}[1]
\REQUIRE Polynomial $a(x) \in R_q$ and $n$-th primitive root of unity $\omega_n \in \Zq$
\ENSURE Polynomial $\hat{a}(x) \in R_q$ such that $\hat{a}(x) = \text{NTT}(a(x))$
\STATE $\hat{a} \leftarrow \text{PolyBitRev}(a)$ 
\FOR{$(s = 1; s \le \text{lg} \, n; s = s + 1)$}
\STATE $m \leftarrow 2^s$
\STATE $\omega_m \leftarrow \omega_n^{n/m}$
\FOR{$(k = 0; k < n; k = k + m)$}
\STATE $\omega \leftarrow 1$
\FOR{$(j = 0; j <  m/2; j = j + 1)$}
\STATE $t \leftarrow \omega \cdot \hat{a}[k+j+m/2] \; \text{mod} \; q$
\STATE $u \leftarrow \hat{a}[k+j]$
\STATE $\hat{a}[k+j] \leftarrow u + t \; \text{mod} \; q$
\STATE $\hat{a}[k+j+m/2] \leftarrow u - t \; \text{mod} \; q$
\STATE $\omega \leftarrow \omega \cdot \omega_m \; \text{mod} \; q$
\ENDFOR
\ENDFOR
\ENDFOR
\RETURN $\hat{a}$
\end{algorithmic}
\end{algorithm}

The NTT is a generalization of the well-known Fast Fourier Transform (FFT) where all the arithmetic is performed in a finite field instead of complex numbers. Instead of working with powers of the $n$-th complex root of unity exp$(-2\pi j/n)$, NTT uses the $n$-th primitive root of unity $\omega_n$ in the ring $\Zq$, that is, $\omega_n$ is an element in $\Zq$ such that $\omega_n^n = 1 \, \text{mod} \, q$ and $\omega_n^i \ne 1 \, \text{mod} \, q$ for $i \ne n$. In order to have elements of order $n$, the modulus $q$ is chosen to be a prime such that $q \equiv 1 \, \text{mod} \, n$. A polynomial $a(x) \in R_q$ with coefficients $a(x) = (a_0, a_1, \cdots, a_{n-1})$ has the NTT representation $\hat{a}(x) = (\hat{a_0}, \hat{a_1}, \cdots, \hat{a_{n-1}})$, where
\[
\hat{a_i} = \sum_{j=0}^{n-1} a_{j} \omega_n^{ij} \; \text{mod} \; q \; \forall \; i \in [0, n-1]
\]
The inverse NTT (INTT) operation converts $\hat{a}(x) = (\hat{a_0}, \hat{a_1}, \cdots, \hat{a_{n-1}})$ back to $a(x)$ as
\[
a_i = \frac{1}{n} \sum_{j=0}^{n-1} \hat{a_{j}} \omega_n^{-ij} \; \text{mod} \; q \; \forall \; i \in [0, n-1]
\]
Note that the INTT operation is similar to NTT, except that $\omega_n$ is replaced by $\omega_n^{-1} \, \text{mod} \, q$ and the final results is divided by $n$. An iterative in-place version of the NTT algorithm is provided in Algorithm \ref{algo:iter_ntt} \cite{cormen_algo_2009, bai_polymul_2016}. The PolyBitRev function performs a permutation on the input polynomial $a$ such that $\hat{a}[i] = \text{PolyBitRev}(a)[i] = a[\text{BitRev}(i)]$, where BitRev is formally defined as $\text{BitRev}(i) = \sum_{j=0}^{\text{lg} \, n - 1} (((i \gg j) \, \& \, 1) \ll (\text{lg} \, n - 1 - i))$ (for positive integer $i$ and power-of-two $n$), that is, bit-wise reversal of the binary representation of the index $i$. Since there are $\text{lg} \, n$ stages in the NTT outer loop, with $O(n)$ operations in each stage, its time complexity is $O(n \, \text{lg} \, n)$. The factors $\omega$ are called the \textit{twiddle factors}, similar to FFT.

The NTT provides a fast multiplication algorithm in $R_q$ with time complexity $O(n \, \text{lg} \, n)$ instead of $O(n^2)$ for schoolbook multiplication. Given two polynomials $a, b \in R_q$, their product $c = a \cdot b \in R_q$ can be computed as
\[
c = \text{INTT} \; ( \; \text{NTT}(a) \; \odot \; \text{NTT}(b) \; )
\]
where $\odot$ denotes coefficient-wise multiplication of the polynomials. Since the product of $a$ and $b$, before reduction modulo $f(x) = x^n + 1$, has $2n$ coefficients, using the above equation directly to compute $a \cdot b$ will require padding both $a$ and $b$ with $n$ zeros. To eliminate this overhead, the \textit{negative-wrapped convolution} \cite{howell_algo_2012} is used, with the additional requirement $q \equiv 1 \, \text{mod} \, 2n$ so that both the $n$-th and $2n$-th primitive roots of unity modulo $q$ exist, respectively denoted as $\omega_n$ and $\psi = \sqrt{\omega_n} \, \text{mod} \, q$. By multiplying $a$ and $b$ coefficient-wise by powers of $\psi$ before the NTT computation, and by multiplying $\text{INTT} ( \, \text{NTT}(a) \odot \text{NTT}(b) \, )$ coefficient-wise by powers of $\psi^{-1} \, \text{mod} \, q$, no zero padding is required and the $n$-point NTT can be used directly.

Similar to FFT, the NTT inner loop involves butterfly computations. There are two types of butterfly operations -- Cooley-Tukey (CT) and Gentleman-Sande (GS) \cite{naehrig_ntt_2016}. The CT butterfly-based NTT requires inputs in normal order and generates outputs in bit-reversed order, similar to the \textit{decimation-in-time} FFT. The GS butterfly-based NTT requires inputs to be in bit-reversed order while the outputs are generated in normal order, similar to the \textit{decimation-in-frequency} FFT. Using the same butterfly for both NTT and INTT requires a bit-reversal permutation. However, the bit-reversal can be avoided by using CT for NTT and GS for INTT, as proposed in \cite{naehrig_ntt_2016}.


\subsection{Sampling}

In lattice-based protocols, the public vectors $\boldsymbol{a}$ are generated from the uniform distribution over $\Zq$ through rejection sampling. The secret vectors $\boldsymbol{s}$ and error terms $e$ are sampled from the distribution $\chi$ typically with zero mean and appropriate standard deviation $\sigma$. Accurate sampling of $\boldsymbol{s}$ and $e$ is critical to the security of these protocols, and the sampling must be constant-time to prevent side-channel leakage of the secret information. Although the original LWE proof used discrete Gaussian distributions for sampling the error terms, several lattice-based schemes use binomial, uniform and ternary distributions for efficiency. A detailed survey of different sampling techniques is available in \cite{cammarota_survey_2019}.

%% file: body/03_ntt.tex
\section{Modular Arithmetic and NTT}
\label{sec:ntt}

The core arithmetic and logic unit (ALU) of Sapphire consists of a 24-bit data-path, with modular operations in $\Fq$ for configurable $q$. In this section, we describe the details of our energy-efficient modular arithmetic implementation, the ALU design and our area-efficient NTT memory architecture.

\subsection{Modular Arithmetic Implementation}

\begin{figure}[!t]
\centering
\includegraphics[width=5.5in]{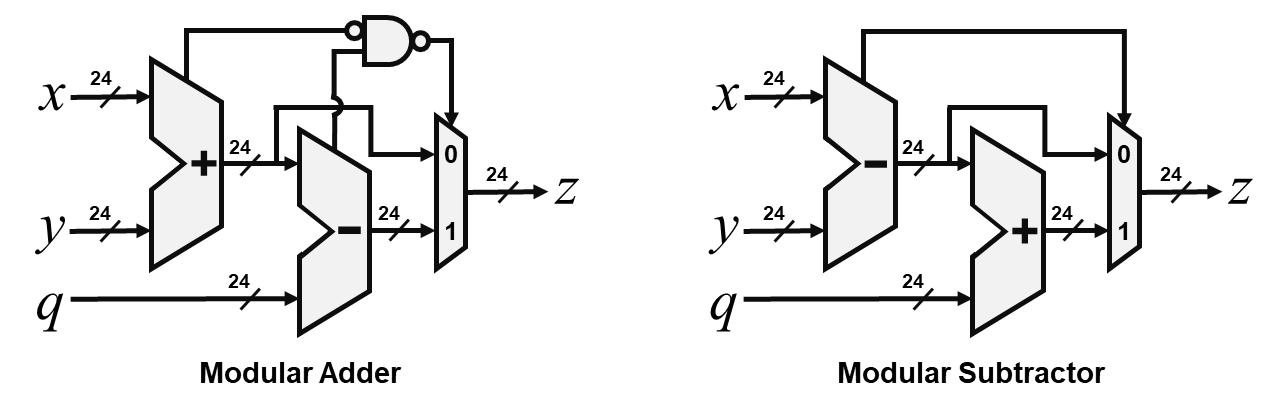}
\caption{Design of our modular adder and subtractor with configurable modulus $q$.}
\label{mod_add_sub_arch}
\end{figure}

The modular arithmetic core consists of a 24-bit adder, a 24-bit subtractor and a 24-bit multiplier along with associated modular reduction logic. Our modular adder and subtractor designs are shown in Fig. \ref{mod_add_sub_arch}, and the corresponding pseudo-codes are shown in Algorithms \ref{algo:mod_add} and \ref{algo:mod_sub}. Both designs use a pair of adder and subtractor, with the sum, carry bit, difference and borrow bit denoted as $s$, $c$, $d$ and $b$ respectively. Modular reduction is performed using conditional subtraction and addition, which are computed in the same cycle to avoid timing side-channels. The synthesized areas of the adder and the subtractor are around 550 GE (gate equivalent) each in area.

\begin{figure*}[!t]
\begin{minipage}[t]{3.0in}
\begin{algorithm}[H]
\caption{Modular Addition}
\label{algo:mod_add}
\begin{algorithmic}[1]
\REQUIRE $x, y \in \Zq$
\ENSURE $z = x + y \; \text{mod} \; q$
\STATE $(c, s) \leftarrow x + y$
\STATE $(b, d) \leftarrow s - q$
\IF{$c = 1$ \OR $b = 0$}
\STATE $z \leftarrow d$
\ELSE
\STATE $z \leftarrow s$
\ENDIF
\RETURN $z$
\end{algorithmic}
\end{algorithm}
\end{minipage}
\hfill
\begin{minipage}[t]{3.0in}
\begin{algorithm}[H]
\caption{Modular Subtraction}
\label{algo:mod_sub}
\begin{algorithmic}[1]
\REQUIRE $x, y \in \Zq$
\ENSURE $z = x - y \; \text{mod} \; q$
\STATE $(b, d) \leftarrow x - y$
\STATE $(c, s) \leftarrow d + q$
\IF{$b = 1$}
\STATE $z \leftarrow s$
\ELSE
\STATE $z \leftarrow d$
\ENDIF
\RETURN $z$
\end{algorithmic}
\end{algorithm}
\end{minipage}
\end{figure*}

For modular multiplication, we use a 24-bit multiplier followed by Barrett reduction \cite{barrett_red_1986} modulo a prime $q$ of size up to 24 bits. Barrett reduction does not exploit any special property of the modulus $q$, thus making it ideal for supporting configurable moduli. Let $z$ be the 48-bit product to be reduced to $\Zq$, then Barrett reduction computes $z \, \text{mod} \, q$ by estimating the quotient $\lfloor z/q \rfloor$ without performing any division, as shown in Algorithm \ref{algo:mod_mul}. Barrett reduction involves two multiplications, one subtraction, one bit-shift and one conditional subtraction. The value of $1/q$ is approximated as $m/2^k$, with the error of approximation being $e = 1/q - m/2^k$, therefore the reduction is valid as long as $ze < 1$. Since $z < q^2$, $k$ is set to be the smallest number such that $e = 1/q - (\lfloor 2^k / q \rfloor / 2^k) < 1/q^2$. Typically, $k$ is very close to $2 \, \lceil \text{lg} \, q \rceil$, that is, the bit-size of $q^2$.

\begin{algorithm}[b]
\caption{Modular Multiplication with Barrett Reduction \cite{barrett_red_1986}}
\label{algo:mod_mul}
\begin{algorithmic}[1]
\REQUIRE $x, y \in \Zq$, $m$ and $k$ such that $m = \lfloor 2^k / q \rfloor$
\ENSURE $z = x \cdot y \; \text{mod} \; q$
\STATE $z \leftarrow x \cdot y$
\STATE $t \leftarrow (z \cdot m) \gg k$
\STATE $z \leftarrow z - (t \cdot q)$
\IF{$z \ge q$}
\STATE $z \leftarrow z - q$
\ENDIF
\RETURN $z$
\end{algorithmic}
\end{algorithm}

\begin{figure}[!t]
\centering
\includegraphics[width=6.0in]{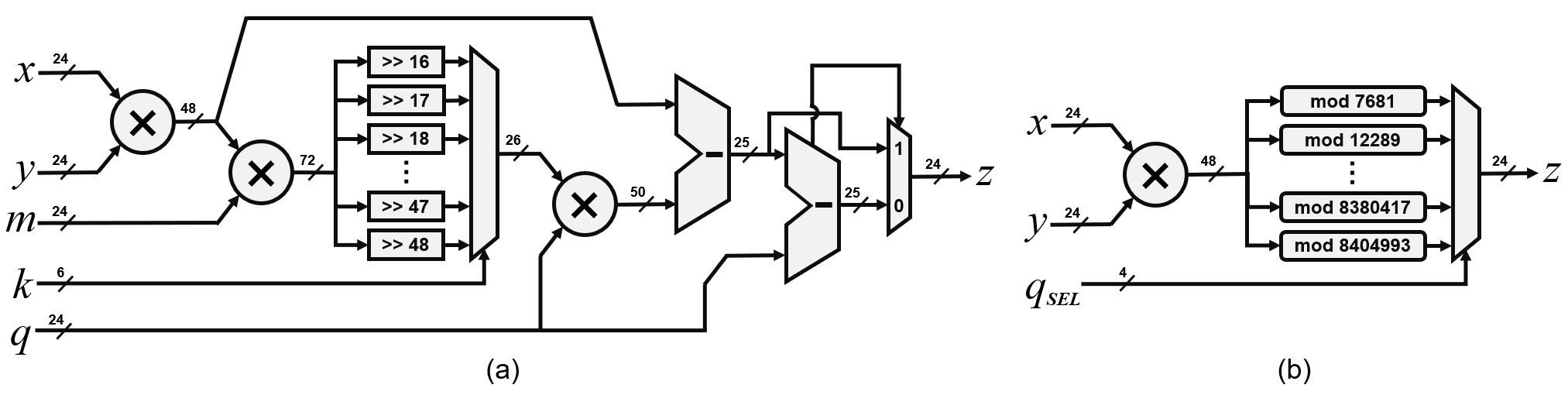}
\caption{Two different single-cycle modular multiplier architectures with (a) fully configurable and (b) pseudo-configurable modulus for Barrett reduction.}
\label{mod_mult_arch}
\end{figure}

In order to understand the trade-offs between flexibility and efficiency in modular multiplication, we have implemented two different architectures of Barrett reduction logic: (1) with fully configurable modulus ($q$ can be an arbitrary prime) and (2) with pseudo-configurable modulus ($q$ belongs to a specific set of primes), as shown in Fig. \ref{mod_mult_arch}.

Apart from the prime $q$ (which can be up to 24 bits), the fully configurable version requires two additional inputs $m$ and $k$ such that $m = \lfloor 2^k / q \rfloor$ ($m$ and $k$ are allowed to be up to 24 bits and 6 bits respectively). It consists of total 3 multipliers, as shown in Fig. \ref{mod_mult_arch}a, the first two being used to compute $z = x \cdot y$ and $z \cdot m$ respectively. For obtaining $t = (z \cdot m) \gg k$, the bit-wise shift is implemented purely using combinational logic (multiplexers) because shifting bits sequentially in registers can be extremely inefficient in terms of power consumption. We assume that $16 \le k \le 48$ since $q$ is not larger than 24 bits, $q$ is typically not smaller than 8 bits and we know that $k \approx 2 \, \lceil \text{lg} \, q \rceil$. The third multiplier is used to compute $t \cdot q$, and a pair of subtractors is used to calculate $z - (t \cdot q)$ and perform the final reduction step. All the steps are computed in a single cycle to avoid any potential timing side-channels. The design was synthesized at 100 MHz (with near-zero slack) and occupies around 11k GE area, which includes the area (around 4k GE) of the 24-bit multiplier used to compute $z = x \cdot y$.

\begin{figure}[!t]
\centering
\includegraphics[width=5.5in]{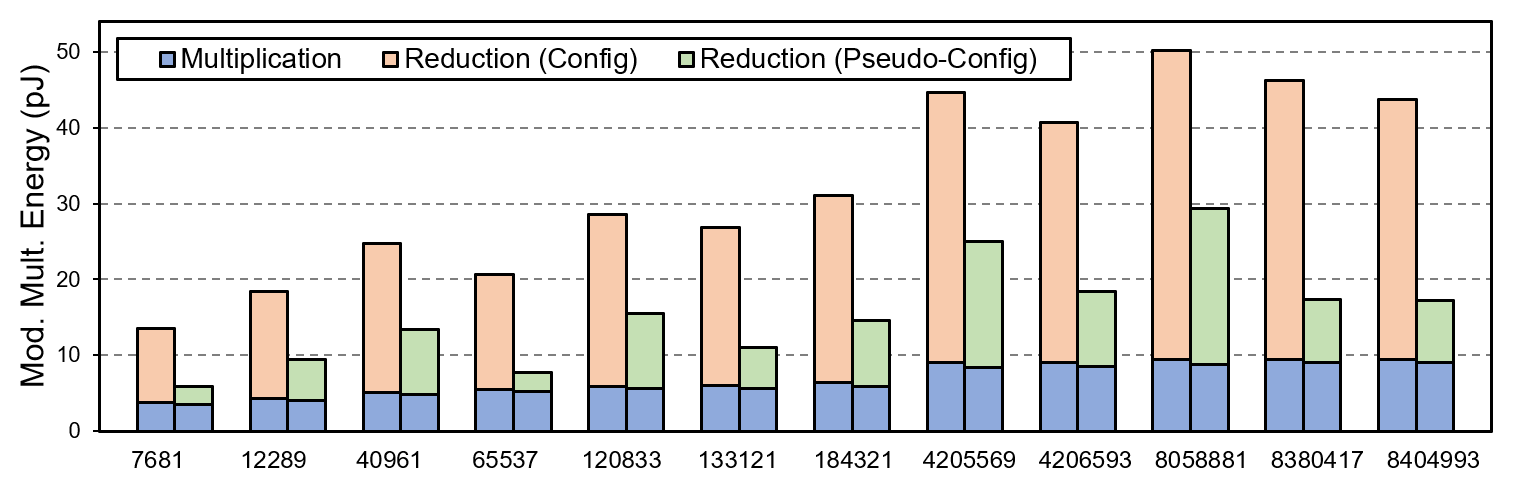}
\caption{Comparison of modular multiplication energy for the two reduction architectures.}
\label{mod_mult_analysis}
\end{figure}

\begin{figure*}[!b]
\begin{minipage}[b]{3.0in}
\begin{algorithm}[H]
\caption{Reduction $\text{mod} \; 7681$}
\label{algo:mod_7681}
\begin{algorithmic}[1]
\REQUIRE $q = 7681, x \in [0, q^2)$
\ENSURE $z = x \; \text{mod} \; q$
\STATE $t \leftarrow (x \ll 8) + (x \ll 4) + x$
\STATE $t \leftarrow t \gg 21$
\STATE $t \leftarrow (t \ll 13) - (t \ll 9) + t$
\STATE $z \leftarrow x - t$
\IF{$z \ge q$}
\STATE $z \leftarrow z - q$
\ENDIF
\RETURN $z$
\end{algorithmic}
\end{algorithm}
\end{minipage}
\hfill
\begin{minipage}[b]{3.0in}
\begin{algorithm}[H]
\caption{Reduction $\text{mod} \; 8380417$}
\label{algo:mod_8380417}
\begin{algorithmic}[1]
\REQUIRE $q = 8380417, x \in [0, q^2)$
\ENSURE $z = x \; \text{mod} \; q$
\STATE $t \leftarrow (x \ll 23) + (x \ll 13) + (x \ll 3) - x$
\STATE $t \leftarrow t \gg 46$
\STATE $t \leftarrow (t \ll 23) - (t \ll 13) + t$
\STATE $z \leftarrow x - t$
\IF{$z \ge q$}
\STATE $z \leftarrow z - q$
\ENDIF
\RETURN $z$
\end{algorithmic}
\end{algorithm}
\end{minipage}
\end{figure*}

The pseudo-configurable modular multiplier implements Barrett reduction logic for the following primes used by NIST Round 1 lattice-based candidates: 7681 (CRYSTALS-Kyber) \cite{bos_kyber_2019}, 12289 (NewHope) \cite{alkim_newhope_2019}, 40961 (R.EMBLEM) \cite{seo_emblem_2017}, 65537 (pqNTRUSign) \cite{zhang_pqntrusign_2017}, 120833 (Ding Key Exchange) \cite{ding_kex_2017}, 133121 / 184321 (LIMA) \cite{smart_lima_2017}, 8380417 (CRYSTALS-Dilithium) \cite{vadim_dilithium_2019}, 8058881 (qTESLA v1.0) and 4205569 / 4206593 / 8404993 (qTESLA v2.0) \cite{bindel_qtesla_2019}. As shown in Fig. \ref{mod_mult_arch}b, there is dedicated reduction block for each of these primes, and the $q_{SEL}$ input is used to select the output of the appropriate block while the inputs to the other blocks are data-gated to save power. Since the reduction blocks have the parameters $m$, $k$ and $q$ coded in digital logic and do not require explicit multipliers, they involve lesser computation than the fully configurable reduction circuit from Fig. \ref{mod_mult_arch}a, albeit at the cost of some additional area and decrease in flexibility. The reduction becomes particularly efficient when at least one of $m$ and $q$ or both can be written in the form $2^{l_1} \pm 2^{l_2} \pm \cdots \pm 1$, where $l_1, l_2, \cdots$ are not more than four positive integers. For example, we consider the CRYSTALS primes: for $q = 7681 = 2^{13} - 2^{9} + 1$ we have $k = 21$ and $m = 273 = 2^{8} + 2^{4} + 1$, and for $q = 8380417 = 2^{23} - 2^{13} + 1$ we have $k = 46$ and $m = 8396807 = 2^{23} + 2^{13} + 2^{3} - 1$. Therefore, the multiplications by $q$ and $m$ can be converted to significantly cheaper bit-shifts and additions / subtractions, as shown in Algorithms \ref{algo:mod_7681} and \ref{algo:mod_8380417}. Implementation details and reduction parameters for each customized modular reduction block are provided in \ref{sec:appendix_a}. This design also performs modular multiplication in a single cycle. It was synthesized at 100 MHz (with near-zero slack) and occupies around 19k GE area, including the area of the 24-bit multiplier.

In Fig. \ref{mod_mult_analysis}, we compare the simulated energy consumption of the fully configurable and pseudo-configurable modular multiplier architectures for all the primes mentioned earlier. As expected, the multiplication itself consumes the same energy in both cases, but the modular reduction energy is up to $6 \times$ lower for the pseudo-configurable design. The overall decrease in modular multiplication energy, considering both multiplication and reduction together, is up to $3 \times$, clearly highlighting the benefit of the dedicated modular reduction data-paths when working with prime moduli. For reduction modulo $2^m$ ($m < 24$), e.g., in the case of Frodo, the output of the 24-bit multiplier is simply bit-wise AND-ed with $2^m - 1$ implying that the modular reduction energy is negligible.

\subsection{Butterfly Unit and ALU}

\begin{figure}[!t]
\centering
\includegraphics[width=5.2in]{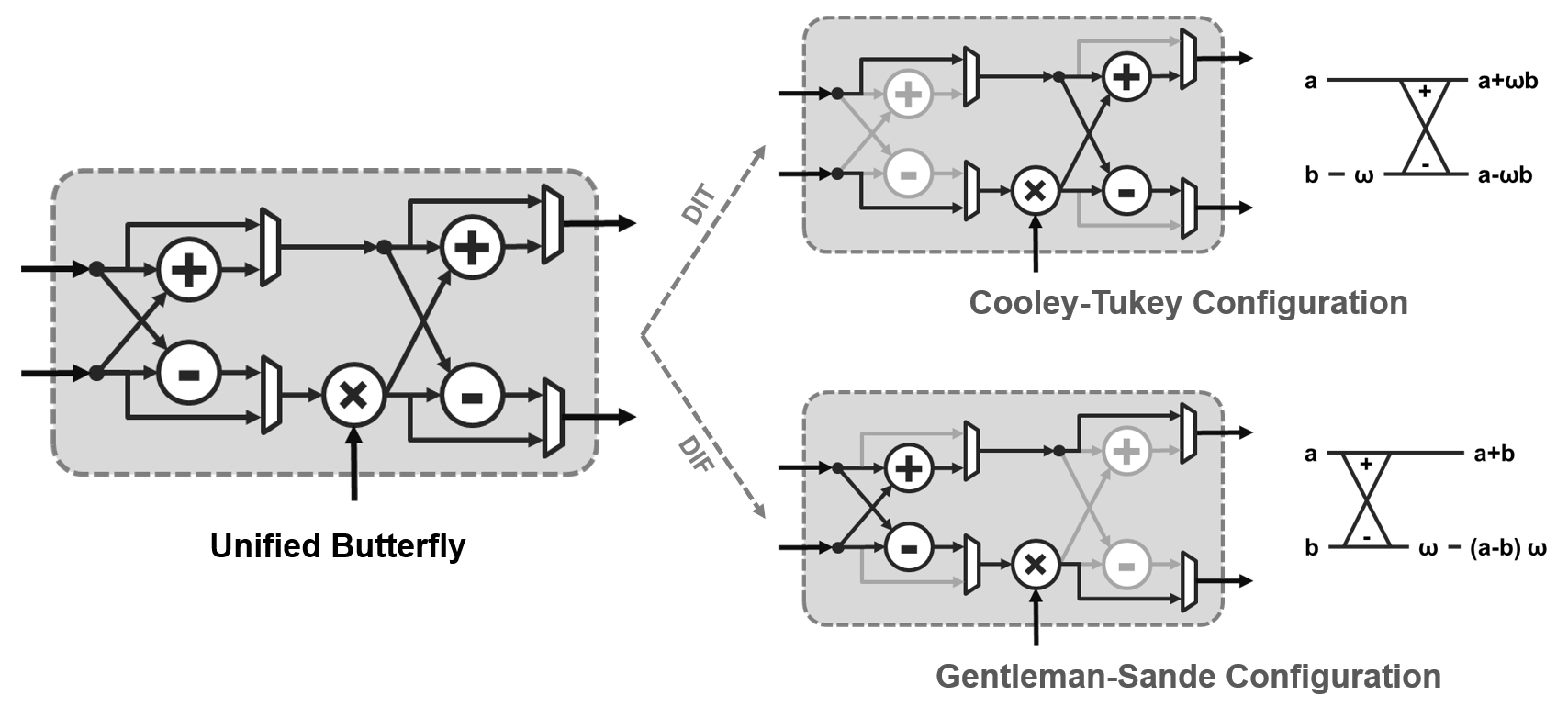}
\caption{Unified butterfly in Cooley-Tukey and Gentleman-Sande configurations.}
\label{butterfly_arch}
\end{figure}

Next, we elaborate how the modular arithmetic units described earlier are integrated together to build the butterfly module. As discussed in Section \ref{sec:background}, NTT computations involve butterfly operations similar to the Fast Fourier Transform, with the only difference being that all arithmetic is performed modulo $q$ instead of complex numbers. There are two butterfly configurations -- Cooley-Tukey (or DIT) and Gentleman-Sande (or DIF). In terms of arithmetic, the DIT butterfly computes $(a + \omega b \; \text{mod} \; q, a - \omega b \; \text{mod} \; q)$ and the DIF butterfly computes $(a + b \; \text{mod} \; q, (a - b) \omega \; \text{mod} \; q)$, where $a$ and $b$ are the inputs to the butterfly and $\omega$ is the twiddle factor. The DIT butterfly requires inputs to be in bit-reversed order and the DIF butterfly generates outputs in bit-reversed order, thus making DIF and DIT suitable for NTT and INTT respectively. While software implementations have the flexibility to program both configurations, hardware designs typically implement either DIT or DIF, thus requiring bit-reversals. To solve this problem, we have implemented a unified butterfly architecture \cite{banerjee_isscc_2019} which can be configured as both DIT and DIF, as shown in Fig. \ref{butterfly_arch}. It consists of two sets of modular adders and subtractors along with some multiplexing circuitry to select whether the multiplication with $\omega$ is performed before or after the addition and subtraction. Since the critical path of the design is inside the modular multiplier, there is no impact on system performance. The associated area overhead is also negligible.

The modular arithmetic blocks inside the butterfly are re-used for coefficient-wise polynomial arithmetic operations as well as for multiplying polynomials with the appropriate powers of $\psi$ and $\psi ^ {-1}$ during negative-wrapped convolution. Apart from butterfly and arithmetic modulo $q$, the Sapphire ALU also supports the following bit-wise operations -- AND, OR, XOR, left shift and right shift.

\subsection{NTT Memory Architecture}

Hardware architectures for polynomial multiplication using NTT consist of memory banks for storing the polynomials along with the ALU which performs butterfly computations. Since each butterfly needs to read two inputs and write two outputs all in the same cycle, these memory banks are typically implemented using dual-port RAMs \cite{ingrid_ringlwe_2014, ingrid_polymul_2015, bai_polymul_2016, liu_rlwe_2019} or four-port RAMs \cite{zhang_leia_2018}. Although true dual-port memory is easily available in state-of-the-art commercial FPGAs in the form of block RAMs (BRAMs), use of dual-port SRAMs in ASIC can pose large area overheads in resource-constrained devices. Compared to a simple single-port SRAM, a dual-port SRAM has double the number of row and column decoders, write drivers and read sense amplifiers. Also, the bit-cells in a low-power dual-port SRAM consist of ten transistors (10T) compared to the usual six transistor (6T) bit-cells in a single-port SRAM \cite{noguchi_dpsram_2008}. Therefore, the area of a dual-port SRAM can be as much as double the area of a single-port SRAM with the same number of bits and column muxing. To reduce this area overhead, we implement an area-efficient NTT memory architecture \cite{banerjee_isscc_2019} which uses the constant-geometry FFT data-flow \cite{pease_fft_1968} and consists of single-port SRAMs only.

\begin{algorithm}[!t]
\caption{Constant Geometry Out-of-Place NTT \cite{pollard_fft_1971}}
\label{algo:cg_ntt}
\begin{algorithmic}[1]
\REQUIRE Polynomial $a(x) \in R_q$ and $n$-th primitive root of unity $\omega_n \in \Zq$
\ENSURE Polynomial $\hat{a}(x) \in R_q$ such that $\hat{a}(x) = \text{NTT}(a(x))$
\STATE $a \leftarrow \text{PolyBitRev}(a)$ 
\FOR{$(s = 1; s \le \text{lg} \, n; s = s + 1)$}
\FOR{$(j = 0; j <  n/2; j = j + 1)$}
\STATE $k \leftarrow \lfloor j / 2^{\text{lg} \, (n-s)} \rfloor \cdot 2^{\text{lg} \, (n-s)}$
\STATE $\hat{a}[j] \leftarrow a[2j] + a[2j+1] \cdot \omega_n^{k} \; \text{mod} \; q$
\STATE $\hat{a}[j+n/2] \leftarrow a[2j] - a[2j+1] \cdot \omega_n^{k} \; \text{mod} \; q$
\ENDFOR
\IF{$s \ne \text{lg} \, n$}
\STATE $a \leftarrow \hat{a}$
\ENDIF
\ENDFOR
\RETURN $\hat{a}$
\end{algorithmic}
\end{algorithm}

The constant geometry NTT is described in Algorithm \ref{algo:cg_ntt} \cite{pollard_fft_1971, ingrid_polymul_2015}. Clearly, the coefficients of the polynomial are accessed in the same order for each stage, thus simplifying the read/write control circuitry. For constant geometry DIT NTT, the butterfly inputs are $a[2j]$ and $a[2j+1]$ and the outputs are $\hat{a}[j]$ and $\hat{a}[j+n/2]$, while the inputs are $a[j]$ and $a[j+n/2]$ and the outputs are $\hat{a}[2j]$ and $\hat{a}[2j+1]$ for DIF NTT. However, the constant geometry NTT is inherently out-of-place, therefore requiring storage for both polynomials $a$ and $\hat{a}$. For our hardware implementation, we create two memory banks -- \textit{left} and \textit{right} -- to store these two polynomials while allowing the butterfly inputs and outputs to \textit{ping-pong} between them during each stage of the transform. Although out-of-place NTT requires storage for both the input and output polynomials, this does not affect the total memory requirements of the crypto-processor because the total number of polynomials required to be stored during the protocol execution is greater than two, e.g., four polynomials are involved in any computation of the form $b = a \cdot s + e$.

\begin{figure}[!b]
\centering
\includegraphics[width=5.0in]{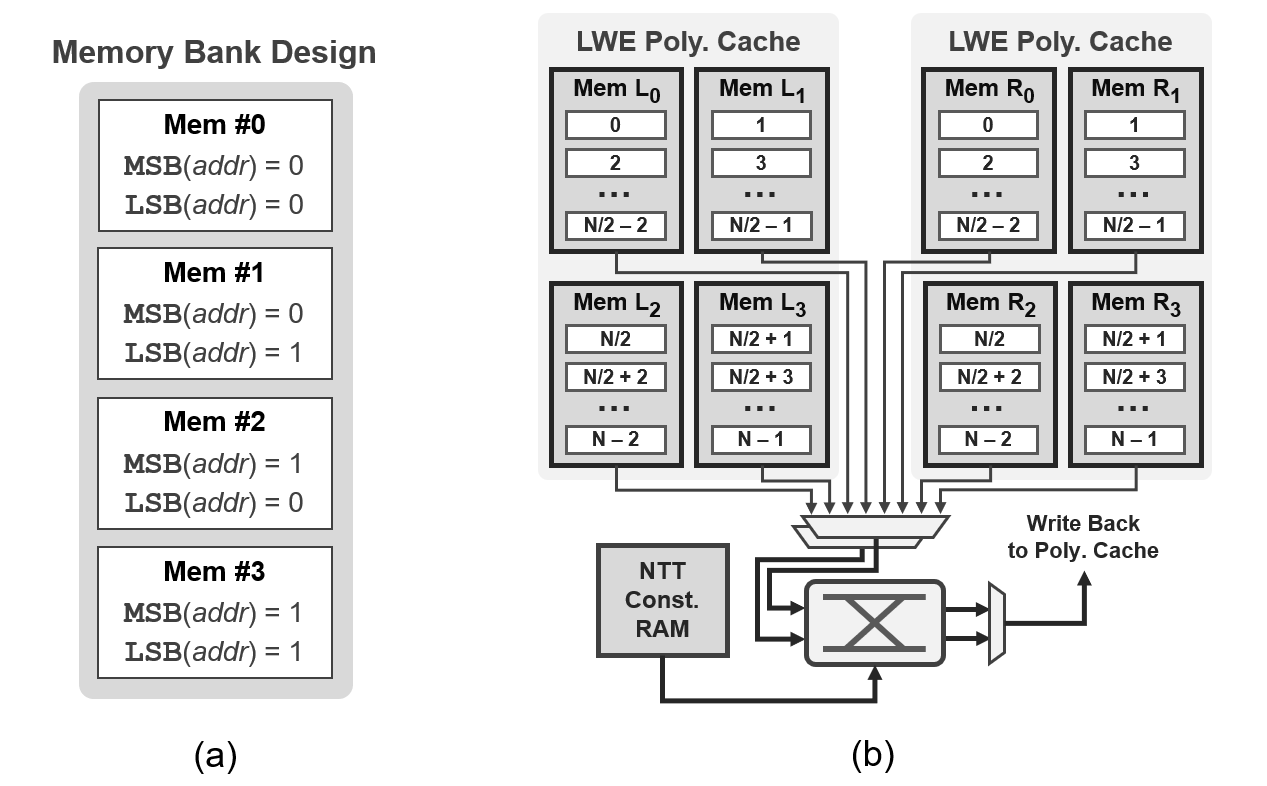}
\caption{(a) Memory bank construction using single-port SRAMs and (b) proposed area-efficient NTT architecture using two such memory banks.}
\label{ntt_arch}
\end{figure}

\begin{figure}[!t]
\centering
\includegraphics[width=5.5in]{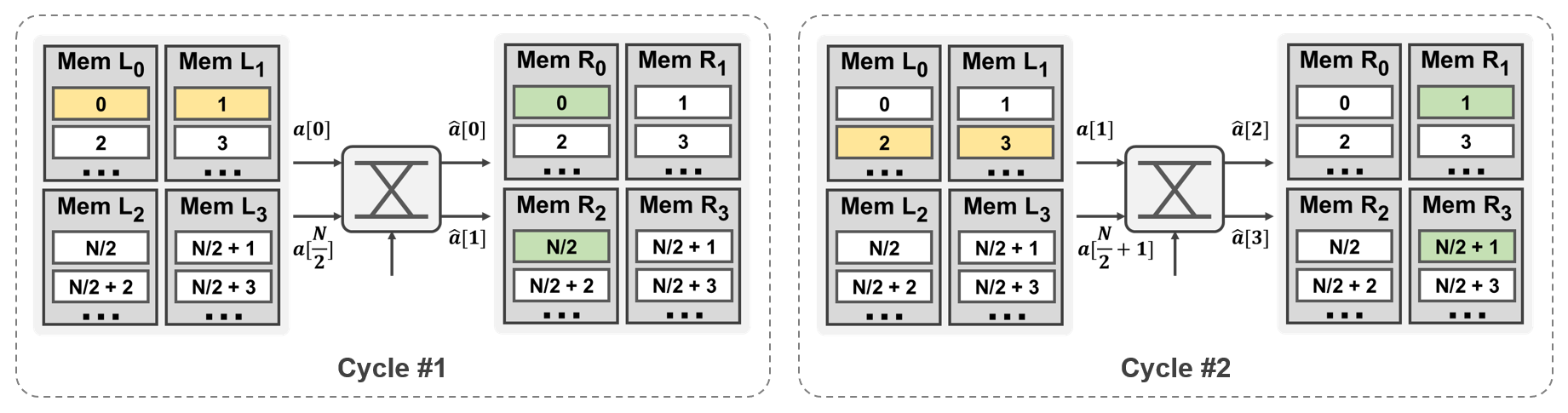}
\caption{Data-flow of our NTT memory architecture in the first two cycles (butterfly inputs are in yellow and outputs are in green).}
\label{ntt_dataflow}
\end{figure}

\begin{figure}[!t]
\centering
\includegraphics[width=5.5in]{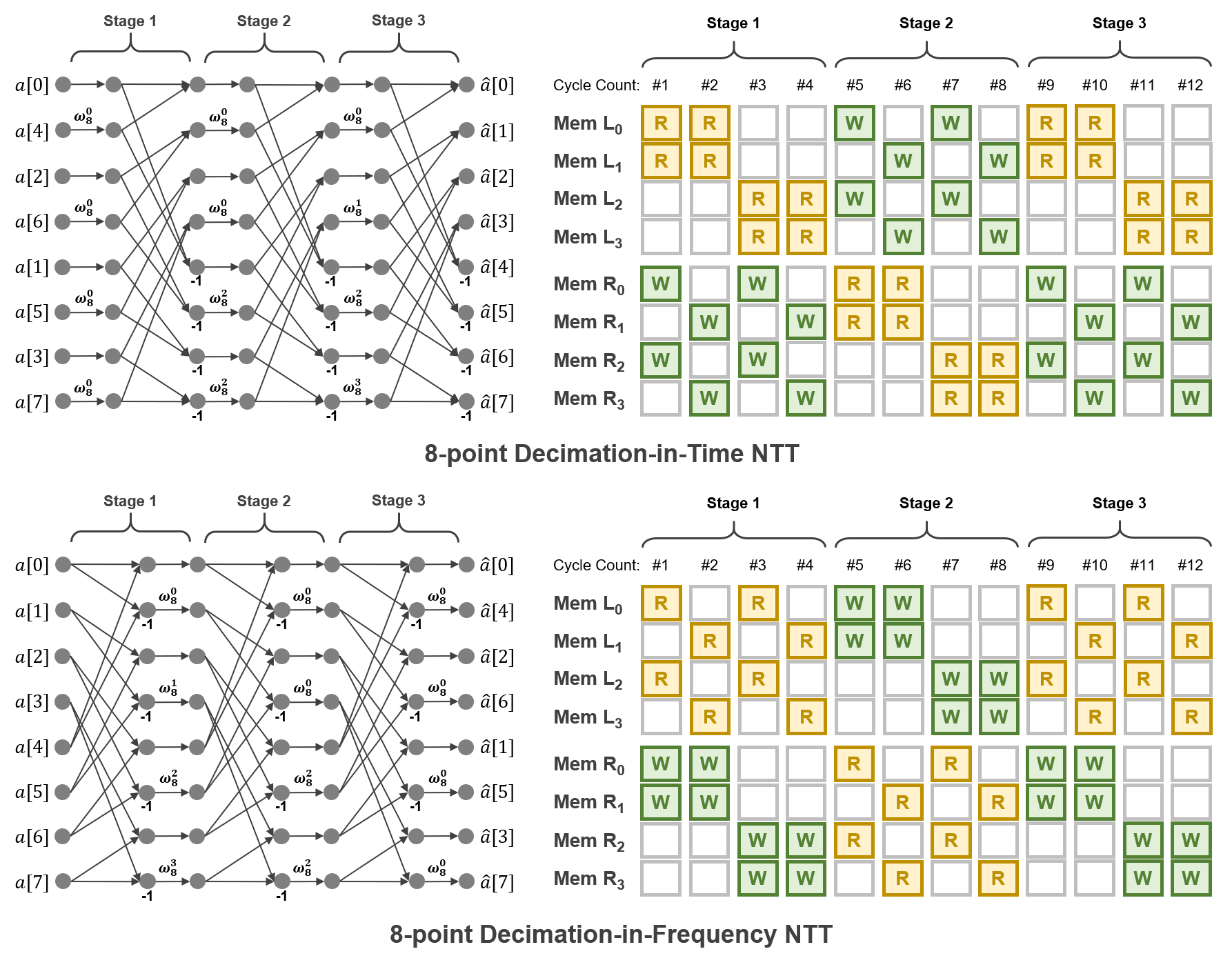}
\caption{Memory access patterns for 8-point DIT and DIF NTT using our single-port SRAM-based memory architecture (R and W denote read and write respectively).}
\label{ntt_mem_access}
\end{figure}

Next, we describe how these memory banks are constructed using single-port SRAMs so that each butterfly can be computed in a single cycle without causing read/write hazards. As shown in Fig. \ref{ntt_arch}a, each polynomial is split among four single port SRAMs \textit{Mem 0-3} on the basis of the least and most significant bits (LSB and MSB) of the coefficient index (or address $addr$). This allows simultaneously accessing coefficient index pairs of the form $(2j, 2j+1)$ and $(j, j+n/2)$. Our NTT memory architecture is shown in Fig. \ref{ntt_arch}b, which consists of two such memory banks labelled as \textit{LWE Poly Cache}. In every cycle, the butterfly inputs are read from two different single-port SRAMs (out of four SRAMs in the input memory bank) and the outputs are also written to two different single-port SRAMs (out of four SRAMs in the output memory bank), thus avoiding hazards. The data flow in the first two cycles of NTT is shown in Fig. \ref{ntt_dataflow}, where the input polynomial $a$ is stored in the left bank and the output polynomial $\hat{a}$ is stored in the right bank. As the input and output polynomials exchange their memory banks from one stage to the next, our NTT control circuitry ensures that the same data-flow is maintained. To illustrate this, the memory access patterns for all three stages of an 8-point NTT are shown in Fig. \ref{ntt_mem_access} for both decimation-in-time and decimation-in-frequency.

The two memory banks consist of four $1024 \times 24$-bit single-port SRAMs each (24 KB total). Together they store 8192 entries, which can be split into four 2048-dimension polynomials or eight 1024-dimension polynomials or sixteen 512-dimension polynomials or thirty-two 256-dimension polynomials or sixty-four 128-dimension polynomials or one-hundred-twenty-eight 64-dimension polynomials. By constructing this memory using single-port SRAMs (and some additional read-data multiplexing circuitry), we have achieved area savings equivalent to 124k GE compared to a dual-port SRAM-based implementation. This is particularly important since SRAMs account for a large portion of the total hardware area in ASIC implementations of lattice-based cryptography \cite{zhang_leia_2018, sepulveda_ntt_2019}.

In order to allow configurable parameters, our NTT hardware also requires additional storage (labelled as \textit{NTT Constants RAM} in Fig. \ref{ntt_arch}) for the pre-computed twiddle factors: $\omega_{2^i}^{j}$, $\omega_{2^i}^{-j} \, \text{mod} \, q$ for $i \in [1, \text{lg} \, n]$ and $j \in [0, 2^{i-1})$ and $\psi^{i}$, $n^{-1} \psi^{-i} \, \text{mod} \, q$ for $i \in [0, n)$. Since $n \le 2048$ and $q < 2^{24}$, this would require another 24 KB of memory. To reduce this overhead, we exploit the following properties of $\omega$ and $\psi$: $\omega_{n/2} = \omega_{n}^{2}$, $\omega_{n}^{-j} = \omega_{n}^{n-j}$ and $\omega = \psi^{2}$ \cite{bai_polymul_2016}. Then, it's sufficient to store only $\omega_{n}^{j}$ for $j \in [0, n/2)$ and $\psi^{i}$, $n^{-1} \psi^{-i} \, \text{mod} \, q$ for $i \in [0, n)$, thus reducing the twiddle factor memory size by 37.5\% down to 15 KB.

Finally, we compare the energy-efficiency and performance of our NTT with state-of-the-art software and ASIC hardware implementations in Table \ref{table:ntt_comparison}. For the software implementation, we have used assembly-optimized code for ARM Cortex-M4 from the PQM4 crypto library \cite{pqm4}, and measurements were performed using the NUCLEO-F411RE development board \cite{nucleo_f411re}. Total cycle count of our NTT is $(\frac{n}{2} + 1) \, \text{lg} \, n + (n + 1)$, including the multiplication of polynomial coefficients with powers of $\psi$. All measurements for our NTT implementation were performed on our test chip operating at clock frequency 72 MHz and nominal supply voltage 1.1 V. Our hardware-accelerated NTT is up to $11 \times$ more energy-efficient than the software implementation, after accounting for voltage scaling. It is $2.5 \times$ more energy-efficient compared to the fast NTT design from \cite{cammarota_ringlwe_2018} with similar cycle count, and $1.5 \times$ more energy-efficient compared to the slow NTT design from \cite{cammarota_ringlwe_2018} with $4 \times$ cycle count. Our NTT is almost twice as fast as \cite{sepulveda_ntt_2019}, since our memory architecture allows computing one butterfly per cycle even with single-port SRAMs, while having similar energy consumption. The energy-efficiency of our NTT implementation is largely due to the careful design of low-power modular arithmetic, as discussed earlier, which decreases overall modular reduction complexity and simplifies the logic circuitry. However, our NTT is still about $4 \times$ less energy-efficient compared to \cite{zhang_leia_2018}, primarily due to the fact that \cite{zhang_leia_2018} uses 16 parallel butterfly units along with dedicated four-port scratch-pad buffers to achieve higher parallelism and lower energy consumption at the cost of significantly larger chip area (2.05 mm$^2$) compared to our design (0.28 mm$^2$). As will be discussed in Section \ref{sec:meas}, sampling accounts for majority of the computational cost in Ring-LWE and Module-LWE schemes, therefore justifying our choice of area-efficient NTT architecture at the cost of some energy overhead.


\begin{table}[!t]
\small
\renewcommand{\arraystretch}{1.2}
\caption{Comparison of our NTT performance with state-of-the-art}
\label{table:ntt_comparison}
\centering
\begin{tabular}{|l|c|c|c|c|c|c|c|}
\hline
\rowcolor{Gray}
\textbf{Design} & \textbf{Platform} & \textbf{Tech} & \textbf{VDD} & \textbf{Freq} & \textbf{Parameters} & \textbf{NTT} & \textbf{NTT} \\
\rowcolor{Gray}
& & \textbf{(nm)} & \textbf{(V)} & \textbf{(MHz)} & & \textbf{Cycles} & \textbf{Energy} \\
\hline
\multirow{3}{1.7cm}{\textbf{This work}} & \multirow{3}{1.7cm}{\centering ASIC} & \multirow{3}{*}{40} & \multirow{3}{*}{1.1} & \multirow{3}{*}{72} & $(n = 256, q = 7681)$ & 1,289 & 165.98 nJ \\
 & & & & & $(n = 512, q = 12289)$ & 2,826 & 410.52 nJ \\
 & & & & & $(n = 1024, q = 12289)$ & 6,155 & 894.28 nJ \\
\hline
\multirow{3}{1.7cm}{Software \cite{pqm4}} & \multirow{3}{1.7cm}{\centering ARM Cortex-M4} & \multirow{3}{*}{-} & \multirow{3}{*}{3.0} & \multirow{3}{*}{100} & $(n = 256, q = 7681)$ & 22,031 & 13.55 $\mu$J \\
 & & & & & $(n = 512, q = 12289)$ & 34,262 & 21.07 $\mu$J \\
 & & & & & $(n = 1024, q = 12289)$ & 75,006 & 46.13 $\mu$J \\
\hline
\multirow{2}{1.7cm}{Song et al. \cite{zhang_leia_2018}} & \multirow{2}{1.7cm}{\centering ASIC} & \multirow{2}{*}{40} & \multirow{2}{*}{0.9} & \multirow{2}{*}{300} & $(n = 256, q = 7681)$ & 160 & 31 nJ \\
 & & & & & $(n = 512, q = 12289)$ & 492 & 96 nJ \\
\hline
\multirow{2}{1.7cm}{Nejatollahi et al. \cite{cammarota_ringlwe_2018}} & \multirow{2}{1.7cm}{\centering ASIC} & \multirow{2}{*}{45} & \multirow{2}{*}{1.0} & \multirow{2}{*}{100} & \multirow{2}{*}{$(n = 512, q = 12289)$} & 2,854 & 1016.02 nJ \\
 & & & & & & 11,053 & 596.86 nJ \\
\hline
\multirow{3}{1.7cm}{Fritzmann et al. \cite{sepulveda_ntt_2019}} & \multirow{3}{1.7cm}{\centering ASIC} & \multirow{3}{*}{65} & \multirow{3}{*}{1.2} & \multirow{3}{*}{25} & $(n = 256, q = 7681)$ & 2,056 & 254.52 nJ \\
 & & & & & $(n = 512, q = 12289)$ & 4,616 & 549.98 nJ \\
 & & & & & $(n = 1024, q = 12289)$ & 10,248 & 1205.03 nJ \\
\hline
\multirow{2}{1.7cm}{Roy et al. \cite{ingrid_ringlwe_2014}} & \multirow{2}{1.7cm}{\centering FPGA} & \multirow{2}{*}{-} & \multirow{2}{*}{-} & 313 & $(n = 256, q = 7681)$ & 1,691 & - \\
 & & & & 278 & $(n = 512, q = 12289)$ & 3,443 & - \\
\hline
\multirow{2}{1.7cm}{Du et al. \cite{bai_polymul_2016}} & \multirow{2}{1.7cm}{\centering FPGA} & \multirow{2}{*}{-} & \multirow{2}{*}{-} & \multirow{2}{*}{233} & $(n = 256, q = 7681)$ & 4,066 & - \\
 & & & & & $(n = 512, q = 12289)$ & 8,806 & - \\
\hline
\end{tabular}
\end{table}

%% file: body/04_sampling.tex
\section{Discrete Distribution Sampler}
\label{sec:sampling}

Hardness of the LWE problem is directly related to statistical properties of the error samples. Therefore, an accurate and efficient sampler is a critical component of any lattice cryptography implementation. Sampling accounts for a major portion of the computational overhead in software implementations of ideal and module lattice-based protocols \cite{guneysu_masked_2018}. A cryptographically secure pseudo-random number generator (CS-PRNG) is used to generate uniformly random numbers, which are then post-processed to convert them into samples from different discrete probability distributions. In this section, we describe our design of energy-efficient CS-PRNG along with fast sampling techniques for configurable distribution parameters.

\subsection{Energy-Efficient CS-PRNG}


\begin{table}[!t]
\renewcommand{\arraystretch}{1.25}
\caption{Comparison of CS-PRNG designs}
\label{table:prng_comparison}
\centering
\begin{tabular}{|l|c|c|c|c|}
\hline
\rowcolor{Gray}
\textbf{PRNG} & \textbf{Area (kGE)} \textsuperscript{a} & \textbf{Cycles /} & \textbf{No. of} & \textbf{Energy} \\
\rowcolor{Gray}
& & \textbf{Round} & \textbf{PRNG Bits} & \textbf{(pJ/bit)} \textsuperscript{b} \\
\hline
SHAKE-128 & \multirow{2}{*}{34.5 (23.5)} & \multirow{2}{*}{24} & 1344 & 1.67 \\
\cline{1-1} \cline{4-5}
SHAKE-256 & & & 1088 & 2.07 \\
\hline
ChaCha20 & 21.1 (17.5) & 20 & 512 & 3.53 \\
\hline
AES-128-CTR & \multirow{2}{*}{15.0 (11.1)} & 11 & 128 & 5.10 \\
\cline{1-1} \cline{3-5}
AES-256-CTR & & 15 & 128 & 7.56 \\
\hline
\multicolumn{5}{l}{\small \textsuperscript{a} Area of placed-and-routed design (post-synthesis area in brackets)} \\
\multicolumn{5}{l}{\small \textsuperscript{b} Energy measured from test chip operating at 1.1 V}
\end{tabular}
\end{table}

\begin{figure}[t]
\centering
\includegraphics[width=6.0in]{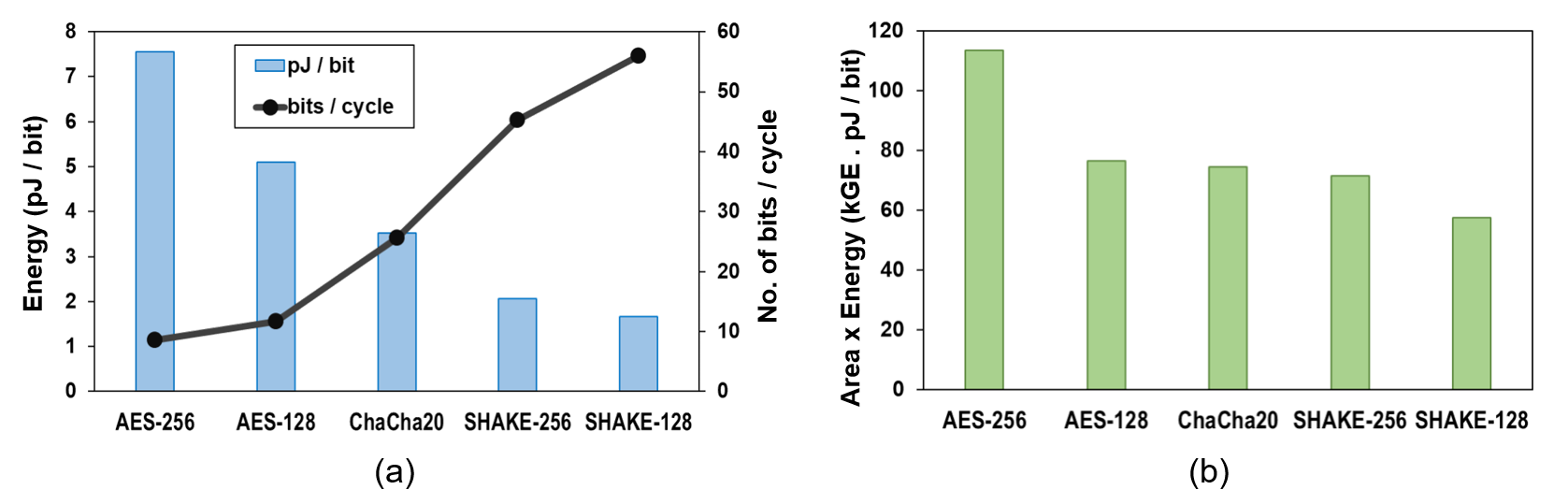}
\caption{Analysis of SHAKE-128, SHAKE-256, AES-128-CTR, AES-256-CTR and ChaCha20 in terms of energy per bit, bits per cycle and area-energy product.}
\label{prng_comparison}
\end{figure}

\begin{figure}[!b]
\centering
\includegraphics[width=5.0in]{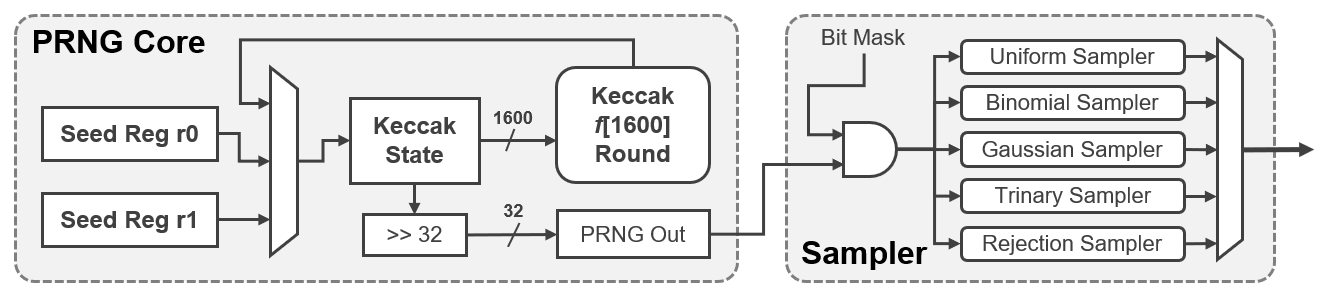}
\caption{Architecture of discrete distribution sampler with Keccak-based PRNG core.}
\label{sampler_arch}
\end{figure}

Some of the standard choices for CS-PRNG are SHA-3 in the SHAKE mode \cite{nist_sha3_2015}, AES in counter mode \cite{nist_aes_2001} and ChaCha20 \cite{bernstein_chacha_2008}. In order to identify the most efficient among these, we have compared them in terms of area, pseudo-random bit generation performance and energy consumption, as shown in Table \ref{table:prng_comparison}. Only place-and-route area and measured energy are considered for all analysis, and synthesis area is reported for reference. For fair comparison, all the three primitives -- SHA-3, AES and ChaCha20 -- were implemented as full data path architectures. From Fig. \ref{prng_comparison}, we observe that although all three primitives have comparable area-energy product, SHA-3 is $2 \times$ more energy-efficient than ChaCha20 and $3 \times$ more energy-efficient than AES; and this is largely due to the fact that SHA-3 generates the highest number of pseudo-random bits per round.

The basic building block of SHA-3 is the Keccak permutation function \cite{bertoni_keccak_2009}. Therefore, our PRNG consists of a 24-cycle Keccak-f[1600] core \cite{banerjee_isscc_2019} which can be configured in different SHA-3 modes and consumes 2.33 nJ per round at nominal voltage of 1.1 V (and 0.89 nJ per round at 0.68 V). Its 1600-bit state is processed in parallel, thus avoiding expensive register shifts and multiplexing required in serial architectures. Fig. \ref{sampler_arch} shows the overall architecture our discrete distribution sampler with the energy-efficient SHA-3 core. Pseudo-random bits generated by SHAKE-128 or SHAKE-256 are stored in the 1600-bit Keccak state register, and shifted out 32 bits at a time as required by the sampler. The sampler then feeds these bits, AND-ed with the appropriate bit mask to truncate them to desired size, to the post-processing logic to perform one of the following five types of operations -- rejection sampling in $[0, q)$, binomial sampling with standard deviation $\sigma$, discrete Gaussian sampling with standard deviation $\sigma$ and desired precision up to 32 bits, uniform sampling in $[-\eta,\eta]$ for $\eta < q$ and trinary sampling in $\{-1, 0, +1\}$ with specified weights for the $+1$ and $-1$ samples.

\subsection{Rejection Sampling}

The public polynomial $a$ in Ring-LWE and the public vector $\boldsymbol{a}$ in Module-LWE have their coefficients uniformly drawn from $\Zq$ through rejection sampling, where uniformly random numbers of desired bit size are obtained from the PRNG as candidate samples and only numbers smaller than $q$ are accepted. The probability that a random number is not accepted is known as the rejection probability.

\begin{table}[ht]
\renewcommand{\arraystretch}{1.2}
\caption{Rejection probabilities for different primes with and without fast sampling}
\label{table:rej_prob}
\centering
\begin{tabular}{|l|c|c|c|c|c|}
\hline
\rowcolor{Gray}
\textbf{Prime} & \textbf{Bit} & \textbf{Rej. Prob.} & \textbf{Scaling} & \textbf{Rej. Prob.} & \textbf{Decrease in} \\
\rowcolor{Gray}
& \textbf{Size} & \textbf{(w/o. scaling)} & \textbf{Factor} & \textbf{(w. scaling)} & \textbf{Rej. Prob.} \\
\hline
7681 & 13 & 0.06 & 1 & 0.06 & - \\
\hline
12289 & 14 & 0.25 & 5 & 0.06 & 0.19 \\
\hline
40961 & 16 & 0.37 & 3 & 0.06 & 0.31 \\
\hline
65537 & 17 & 0.50 & 7 & 0.12 & 0.38 \\
\hline
120833 & 17 & 0.08 & 1 & 0.08 & - \\
\hline
133121 & 18 & 0.49 & 7 & 0.11 & 0.38 \\
\hline
184321 & 18 & 0.30 & 11 & 0.03 & 0.27 \\
\hline
8380417 & 23 & $\approx$ 0 & 1 & $\approx$ 0 & - \\
\hline
8058881 & 23 & 0.04 & 1 & 0.04 & - \\
\hline
4205569 & 23 & 0.50 & 7 & 0.12 & 0.38 \\
\hline
4206593 & 23 & 0.50 & 7 & 0.12 & 0.38 \\
\hline
8404993 & 24 & 0.50 & 7 & 0.12 & 0.38 \\
\hline
\end{tabular}
\end{table}

For prime $q$, the rejection probability is calculated as $(1 - q / 2^{\lceil \text{lg} \, q \rceil})$. In Table \ref{table:rej_prob}, we list the rejection probabilities for primes mentioned earlier in Section \ref{sec:ntt}. Clearly, different primes have very different rejection probabilities, often as high as 50\%, which can be a bottleneck in lattice-based protocols. To solve this problem, we refer to \cite{gueron_ringlwe_2016} where pseudo-random numbers smaller than $5q$ are accepted for $q = 12289$, thus reducing the rejection probability from 25\% to 6\%. We extend this technique for any prime $q$ by scaling the rejection bound from $q$ to $kq$, for appropriate small integer $k$, so that the rejection probability is now $(1 - kq / 2^{\lceil \text{lg} \, kq \rceil})$. We list these scaling factors for the primes in Table \ref{table:rej_prob} along with the corresponding decrease in rejection probability.


\begin{table}[!b]
\small
\renewcommand{\arraystretch}{1.2}
\caption{Comparison of rejection sampling with software}
\label{table:rej_samp_comparison}
\centering
\begin{tabular}{|l|c|c|c|c|c|c|c|}
\hline
\rowcolor{Gray}
\textbf{Design} & \textbf{Platform} & \textbf{Tech} & \textbf{VDD} & \textbf{Freq} & \textbf{Parameters} & \textbf{Samp.} & \textbf{Samp.} \\
\rowcolor{Gray}
& & \textbf{(nm)} & \textbf{(V)} & \textbf{(MHz)} & & \textbf{Cycles} & \textbf{Energy} \\
\hline
\multirow{3}{1.7cm}{\textbf{This work}} & \multirow{3}{1.7cm}{\centering ASIC} & \multirow{3}{*}{40} & \multirow{3}{*}{1.1} & \multirow{3}{*}{72} & $(n = 256, q = 7681)$ & 461 & 50.90 nJ \\
 & & & & & $(n = 512, q = 12289)$ & 921 & 105.74 nJ \\
 & & & & & $(n = 1024, q = 12289)$ & 1,843 & 211.46 nJ \\
\hline
\multirow{3}{1.7cm}{Software \cite{pqm4}} & \multirow{3}{1.7cm}{\centering ARM Cortex-M4} & \multirow{3}{*}{-} & \multirow{3}{*}{3.0} & \multirow{3}{*}{100} & $(n = 256, q = 7681)$ & 60,433 & 37.17 $\mu$J \\
 & & & & & $(n = 512, q = 12289)$ & 139,153 & 85.58 $\mu$J \\
 & & & & & $(n = 1024, q = 12289)$ & 284,662 & 175.07 $\mu$J \\
\hline
\end{tabular}
\end{table}

Although this method reduces rejection rates, the output samples now lie in $[0, kq)$ instead of $[0, q)$. In \cite{gueron_ringlwe_2016}, for $q = 12289$ and $k = 5$, the accepted samples are reduced to $\Zq$ by subtracting $q$ from them up to four times. Since $k$ is not fixed for our rejection sampler, we employ Barrett reduction \cite{barrett_red_1986} for this purpose. Unlike modular multiplication, where the inputs lie in $[0, q^2)$, the inputs here are much smaller; so the Barrett reduction parameters are also quite small, therefore requiring little additional logic. In Table \ref{table:rej_samp_comparison}, we compare our rejection sampler performance (SHAKE-128 used as PRNG) with software implementation on ARM Cortex-M4 using assembly-optimized Keccak \cite{pqm4}.

\subsection{Binomial Sampling}


\begin{table}[!t]
\small
\renewcommand{\arraystretch}{1.2}
\caption{Comparison of binomial sampling with state-of-the-art}
\label{table:bin_samp_comparison}
\centering
\begin{tabular}{|l|c|c|c|c|c|c|c|}
\hline
\rowcolor{Gray}
\textbf{Design} & \textbf{Platform} & \textbf{Tech} & \textbf{VDD} & \textbf{Freq} & \textbf{Parameters} & \textbf{Samp.} & \textbf{Samp.} \\
\rowcolor{Gray}
& & \textbf{(nm)} & \textbf{(V)} & \textbf{(MHz)} & & \textbf{Cycles} & \textbf{Energy} \\
\hline
\multirow{3}{1.7cm}{\textbf{This work}} & \multirow{3}{1.7cm}{\centering ASIC} & \multirow{3}{*}{40} & \multirow{3}{*}{1.1} & \multirow{3}{*}{72} & $(n = 256, k = 4)$ & 505 & 58.20 nJ \\
 & & & & & $(n = 512, k = 8)$ & 1,009 & 116.26 nJ \\
 & & & & & $(n = 1024, k = 8)$ & 2,018 & 232.50 nJ \\
\hline
\multirow{3}{1.7cm}{Software \cite{pqm4}} & \multirow{3}{1.7cm}{\centering ARM Cortex-M4} & \multirow{3}{*}{-} & \multirow{3}{*}{3.0} & \multirow{3}{*}{100} & $(n = 256, k = 4)$ & 52,603 & 32.35 $\mu$J \\
 & & & & & $(n = 512, k = 8)$ & 155,872 & 95.86 $\mu$J \\
 & & & & & $(n = 1024, k = 8)$ & 319,636 & 196.58 $\mu$J \\
\hline
\multirow{2}{1.7cm}{Song et al. \cite{zhang_leia_2018}} & \multirow{2}{1.7cm}{\centering ASIC} & \multirow{2}{*}{40} & \multirow{2}{*}{0.9} & \multirow{2}{*}{300} & \multirow{2}{*}{$(n = 512, k = 16)$} & \multirow{2}{*}{3,704} & \multirow{2}{*}{1.25 $\mu$J} \\
 & & & & & & & \\
\hline
\multirow{2}{1.7cm}{Oder et al. \cite{guneysu_newhopefpga_2017}} & \multirow{2}{1.7cm}{\centering FPGA} & \multirow{2}{*}{-} & \multirow{2}{*}{-} & \multirow{2}{*}{125} & \multirow{2}{*}{$(n = 1024, k = 16)$} & \multirow{2}{*}{33,792} & \multirow{2}{*}{-} \\
 & & & & & & & \\
\hline
\end{tabular}
\end{table}

For binomial sampling, we take two $k$-bit chunks from the PRNG and computes the difference of their Hamming weights, as proposed in \cite{alkim_newhope_2019}. The resulting samples follow a binomial distribution with standard deviation $\sigma = \sqrt{k/2}$. We allow configuring $k$ to any value up to 32, thus providing the flexibility to support different standard deviations.

We compare our binomial sampling performance (SHAKE-256 used as PRNG) with state-of-the-art software and hardware implementations in Table \ref{table:bin_samp_comparison}. Our sampler is more than two orders of magnitude more energy-efficient compared to the software implementation on ARM Cortex-M4 which uses assembly-optimized Keccak \cite{pqm4}. It is also $14 \times$ more efficient than \cite{zhang_leia_2018} which uses Knuth-Yao sampling \cite{knuthyao_sample_1976} for binomial distributions with ChaCha20 as PRNG.

\subsection{Discrete Gaussian Sampling}

\begin{algorithm}[!b]
\caption{Discrete Gaussian Sampling using Inversion Method \cite{alkim_frodo_2019}}
\label{algo:cdt_sampling}
\begin{algorithmic}[1]
\REQUIRE Random inputs $r_0 \in \{0,1\}, r_1 \in [0, 2^{r})$ and table $T_{\chi} = (T_{\chi}[0], \cdots , T_{\chi}[s])$
\ENSURE Sample $e \in \Z$ from $\chi$
\STATE $e \leftarrow 0$ 
\FOR{$(z = 0; z < s; z = z + 1)$}
\IF{$r_1 > T_{\chi}[z]$}
\STATE $e \leftarrow e + 1$
\ENDIF
\ENDFOR
\STATE $e \leftarrow (-1)^{r_0} \cdot e$
\RETURN $e$
\end{algorithmic}
\end{algorithm}

Our discrete Gaussian sampler implements the inversion method of sampling \cite{follath_sampling_2014} from a discrete symmetric zero-mean distribution $\chi$ on $\Z$ with small support which approximates a rounded continuous Gaussian distribution, e.g., in Frodo \cite{alkim_frodo_2019} and R.EMBLEM \cite{seo_emblem_2017}. For a distribution with support $S_{\chi} = \{-s, \cdots, -1, 0, 1, \cdots, s\}$, where $s$ is a small positive integer, the probabilities $\text{Pr}(z)$ for $z \in S_{\chi}$, such that $\text{Pr}(z) = \text{Pr}(-z)$ can be derived from the cumulative distribution table (CDT) $T_{\chi} = (T_{\chi}[0], T_{\chi}[1], \cdots , T_{\chi}[s])$, where $2^{-r} \cdot T_{\chi}[0] = \text{Pr}(0)/2 - 1$ and $2^{-r} \cdot T_{\chi}[z] = \text{Pr}(0)/2 - 1+ \sum_{i=1}^{i=z} \text{Pr}(i)$ for $z \in [1, s]$ for precision $r$. Given random inputs $r_0 \in \{0,1\}, r_1 \in [0, 2^{r})$ and distribution table $T_{\chi}$, a sample $e \in \Z$ from $\chi$ can be obtained using Algorithm \ref{algo:cdt_sampling} \cite{alkim_frodo_2019}.

The sampling must be constant-time in order to eliminate timing side-channels, therefore the algorithm does a complete loop through the entire table $T_{\chi}$. The comparison $r_1 > T_{\chi}[z]$ must also be implemented in a constant-time manner. Our implementation adheres to these requirements and uses a $64 \times 32$ RAM to store the CDT, allowing the parameters $s \le 64$ and $r \le 32$ to be configured according to the choice of the distribution. In Table \ref{table:cdt_samp_comparison}, we have compared our Gaussian sampler performance (SHAKE-256 used as PRNG) with software implementation on ARM Cortex-M4 using assembly-optimized Keccak \cite{pqm4}, and we observe up to $40 \times$ improvement in energy-efficiency after accounting for voltage scaling. Hardware architectures for Knuth-Yao sampling have been proposed by \cite{ingrid_ringlwe_2014} and \cite{zhang_leia_2018}, but they are for discrete Gaussian distributions with larger standard deviation and higher precision, which we do not support.


\begin{table}[!t]
\small
\renewcommand{\arraystretch}{1.2}
\caption{Comparison of discrete Gaussian sampling with software}
\label{table:cdt_samp_comparison}
\centering
\begin{tabular}{|l|c|c|c|c|c|c|c|}
\hline
\rowcolor{Gray}
\textbf{Design} & \textbf{Platform} & \textbf{Tech} & \textbf{VDD} & \textbf{Freq} & \textbf{Parameters} & \textbf{Samp.} & \textbf{Samp.} \\
\rowcolor{Gray}
& & \textbf{(nm)} & \textbf{(V)} & \textbf{(MHz)} & & \textbf{Cycles} & \textbf{Energy} \\
\hline
\multirow{3}{1.7cm}{\textbf{This work}} & \multirow{3}{1.7cm}{\centering ASIC} & \multirow{3}{*}{40} & \multirow{3}{*}{1.1} & \multirow{3}{*}{72} & $(n = 512, \sigma = 25.0, s = 54)$ & 29,169 & 1232.71 nJ \\
 & & & & & $(n = 1024, \sigma = 2.75, s = 11)$ & 15,330 & 647.86 nJ \\
 & & & & & $(n = 1024, \sigma = 2.30, s = 10)$ & 14,306 & 604.58 nJ \\
\hline
\multirow{3}{1.7cm}{Software \cite{pqm4}} & \multirow{3}{1.7cm}{\centering ARM Cortex-M4} & \multirow{3}{*}{-} & \multirow{3}{*}{3.0} & \multirow{3}{*}{100} & $(n = 512, \sigma = 25.0, s = 54)$ & 397,921 & 244.72 $\mu$J \\
 & & & & & $(n = 1024, \sigma = 2.75, s = 11)$ & 325,735 & 200.33 $\mu$J \\
 & & & & & $(n = 1024, \sigma = 2.30, s = 10)$ & 317,541 & 195.29 $\mu$J \\
\hline
\end{tabular}
\end{table}

\subsection{Other Distributions}

Several lattice-based protocols, such as CRYSTALS-Dilithium \cite{vadim_dilithium_2019} and qTESLA \cite{bindel_qtesla_2019}, require polynomials to be sampled with coefficients uniformly distributed in the range $[-\eta, \eta]$ for a specified bound $\eta < q$. For this, we again use rejection sampling. Unlike rejection sampling from $\Zq$, we do not require any special techniques since $\eta$ is typically small or an integer close to a power of two.

\begin{figure*}[!b]
\begin{minipage}[b]{3.0in}
\begin{algorithm}[H]
\caption{Trinary Sampling with $m$ non-zero coefficients ($+1$'s and $-1$'s)}
\label{algo:tri_sampling_1}
\begin{algorithmic}[1]
\REQUIRE $m < n$ and a PRNG
\ENSURE $s = (s_0, s_1, \cdots, s_{n-1})$
\STATE $s \leftarrow (0, 0, \cdots, 0)$ ; $i \leftarrow 0$
\WHILE{$i < m$}
\STATE $pos \in_R [0,n)$
\STATE $sign \in_R \{0, 1\}$
\IF{$s_{pos} = 0$}
\IF{$sign = 0$}
\STATE $s_{pos} \leftarrow 1$
\ELSE
\STATE $s_{pos} \leftarrow -1$
\ENDIF
\STATE $i \leftarrow i + 1$
\ENDIF
\ENDWHILE
\RETURN $s$
\end{algorithmic}
\end{algorithm}
\end{minipage}
\hfill
\begin{minipage}[b]{3.0in}
\begin{algorithm}[H]
\caption{Trinary Sampling with $m_0$ $+1$'s and $m_1$ $-1$'s}
\label{algo:tri_sampling_2}
\begin{algorithmic}[1]
\REQUIRE $m_0 + m_1 < n$ and a PRNG
\ENSURE $s = (s_0, s_1, \cdots, s_{n-1})$
\STATE $s \leftarrow (0, 0, \cdots, 0)$ ; $i \leftarrow 0$
\WHILE{$i < m_0$}
\STATE $pos \in_R [0,n)$
\IF{$s_{pos} = 0$}
\STATE $s_{pos} \leftarrow +1$ ; $i \leftarrow i + 1$
\ENDIF
\ENDWHILE
\WHILE{$i < m_0 + m_1$}
\STATE $pos \in_R [0,n)$
\IF{$s_{pos} = 0$}
\STATE $s_{pos} \leftarrow -1$ ; $i \leftarrow i + 1$
\ENDIF
\ENDWHILE
\RETURN $s$
\end{algorithmic}
\end{algorithm}
\end{minipage}
\end{figure*}

Finally, we have also implemented a trinary sampler for polynomials with coefficients from $\{-1, 0, +1\}$. We classify these polynomials into three categories: (1) with $m$ non-zero coefficients, (2) with $m_0$ $+1$'s and $m_1$ $-1$'s, and (3) with coefficients distributed as $\text{Pr}(x=1) = \text{Pr}(x=-1) = \rho /2$ and $\text{Pr}(x=0) = 1 - \rho$ for $\rho \in \{ 1/2, 1/4, 1/8, \cdots, 1/128\}$. Their implementations are described in Algorithms \ref{algo:tri_sampling_1}, \ref{algo:tri_sampling_2} and \ref{algo:tri_sampling_3}. For the first two cases, we start with a zero-polynomial $s$ of size $n$. Then, uniformly random coefficient indices $\in [0, n)$ are generated, and the corresponding coefficients are replaced with $-1$ or $+1$ if they are zero \cite{bindel_qtesla_2019, zhang_pqntrusign_2017}. For the third case, sampling of the coefficients is based on the observation \cite{lee_lizard_2017} that for a uniformly random number $x \in [0, 2^k)$ we have $\text{Pr}(x=0) = 1/2^k$, $\text{Pr}(x=1) = 1/2^k$ and $\text{Pr}(x \in [2, 2^k)) = 1 - 1/2^k$. Therefore, for the appropriate value of $k \in [1,7]$, we can generate samples from the desired trinary distribution with $\rho = 1/2^k$. For all three algorithms, the symbol $\in_R$ denotes pseudo-random number generation using the PRNG.

\begin{algorithm}[!t]
\caption{Trinary Sampling with coefficients from $\{-1, 0, +1\}$ distributed according to $\text{Pr}(x=1) = \text{Pr}(x=-1) = \rho /2$ and $\text{Pr}(x=0) = 1 - \rho$}
\label{algo:tri_sampling_3}
\begin{algorithmic}[1]
\REQUIRE $k \in [1,7]$, $\rho = 1/2^k$ and a PRNG
\ENSURE $s = (s_0, s_1, \cdots, s_{n-1})$
\FOR{$(i = 0; i < n; i = i + 1)$}
\STATE $x \in_R [0, 2^k)$
\IF{$x = 0$}
\STATE $s_i \leftarrow 1$
\ELSIF{$x = 1$}
\STATE $s_i \leftarrow -1$
\ELSE
\STATE $s_i \leftarrow 0$
\ENDIF
\ENDFOR
\RETURN $s$
\end{algorithmic}
\end{algorithm}

%% file: body/05_arch.tex
\section{Chip Architecture}
\label{sec:arch}

\begin{figure}[!b]
\centering
\includegraphics[width=6.0in]{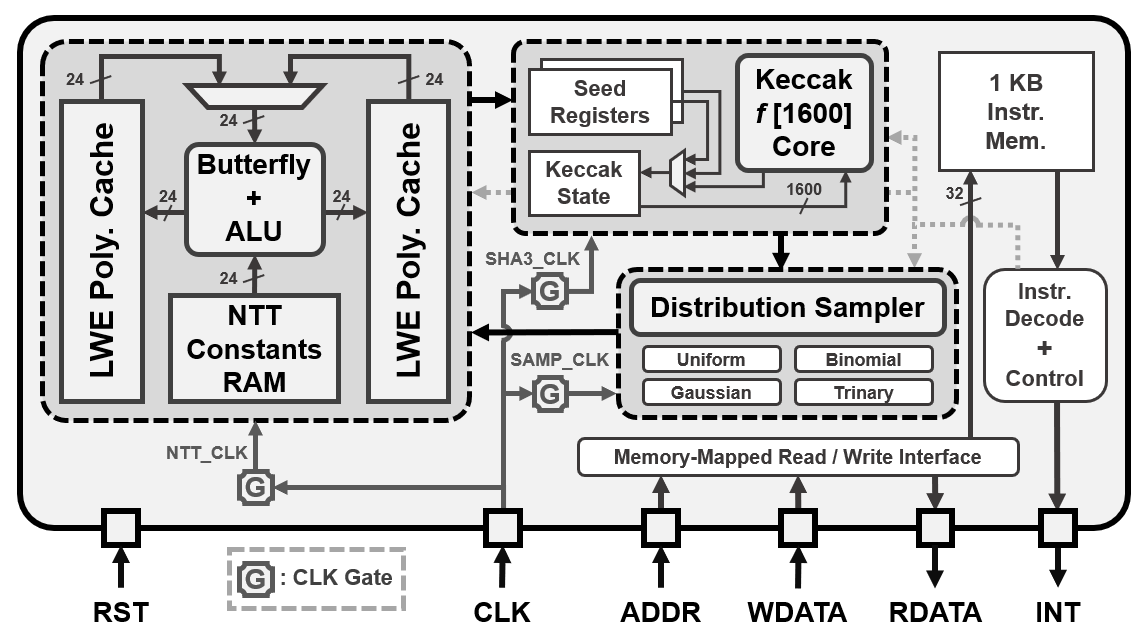}
\caption{Sapphire lattice crypto-processor top-level architecture.}
\label{skywalker_top_arch}
\end{figure}

The top-level architecture of Sapphire is shown in Fig. \ref{skywalker_top_arch}. The efficient building blocks described in Sections \ref{sec:ntt} and \ref{sec:sampling} are integrated with a 1 KB instruction memory and an instruction decoder to form the core of our crypto-processor. It can be programmed using 32-bit custom instructions to perform different polynomial arithmetic, transform and sampling operations, as well as simple branching. For example, the following instructions generate polynomials $a, s, e \in R_q$, and calculate $a \cdot s + e$, which is a typical computation in the Ring-LWE-based scheme NewHope-1024: \\ \\
\texttt{config (n = 1024, q = 12289)} \\
\textcolor{Comment}{\texttt{\# sample\_a}} \\
\texttt{rej\_sample (prng = SHAKE-128, seed = r0, c0 = 0, c1 = 0, poly = 0)} \\
\textcolor{Comment}{\texttt{\# sample\_s}} \\
\texttt{bin\_sample (prng = SHAKE-256, seed = r1, c0 = 0, c1 = 0, k = 8, poly = 1)} \\
\textcolor{Comment}{\texttt{\# sample\_e}} \\
\texttt{bin\_sample (prng = SHAKE-256, seed = r1, c0 = 0, c1 = 1, k = 8, poly = 2)} \\
\textcolor{Comment}{\texttt{\# ntt\_s}} \\
\texttt{mult\_psi (poly = 1)} \\
\texttt{transform (mode = DIF\_NTT, poly\_dst = 4, poly\_src = 1)} \\
\textcolor{Comment}{\texttt{\# a\_mul\_s}} \\
\texttt{poly\_op (op = MUL, poly\_dst = 0, poly\_src = 4)} \\
\textcolor{Comment}{\texttt{\# intt\_a\_mul\_s}}\\
\texttt{transform (mode = DIT\_INTT, poly\_dst = 5, poly\_src = 0)} \\
\texttt{mult\_psi\_inv (poly = 5)} \\
\textcolor{Comment}{\texttt{\# a\_mul\_s\_plus\_e}} \\
\texttt{poly\_op (op = ADD, poly\_dst = 1, poly\_src = 5)} \\ \\
The \texttt{config} instruction is first used to configure the protocol parameters $n$ and $q$ which, in this example, are the parameters from NewHope-1024. For $n = 1024$, the polynomial cache is divided into 8 polynomials, which are accessed using the \texttt{poly} argument in all instructions. For sampling, the seed can be chosen from a pair of 256-bit registers \texttt{r0} and \texttt{r1}, while two 16-bit registers \texttt{c0} and \texttt{c1} are used as counters for sampling multiple polynomials from the same seed. For coefficient-wise operations \texttt{poly\_op}, the \texttt{poly\_src} argument indicates the first source polynomial while the \texttt{poly\_dst} argument is used to denote the second source (and destination) polynomial. Similarly, the following set of instructions are used to generate matrix of polynomials $\boldsymbol{A} \in R_q^{2 \times 2}$ and vectors of polynomials $\boldsymbol{s}, \boldsymbol{e} \in R_q^{2}$, and calculate $\boldsymbol{A} \cdot \boldsymbol{s} + \boldsymbol{e}$, which is a typical computation in the Module-LWE-based scheme CRYSTALS-Kyber-512: \\ \\
\texttt{config (n = 256, q = 7681)} \\
\textcolor{Comment}{\texttt{\# sample\_s}} \\
\texttt{bin\_sample (prng = SHAKE-256, seed = r1, c0 = 0, c1 = 0, k = 3, poly = 4)} \\
\texttt{bin\_sample (prng = SHAKE-256, seed = r1, c0 = 0, c1 = 1, k = 3, poly = 5)} \\
\textcolor{Comment}{\texttt{\# sample\_e}} \\
\texttt{bin\_sample (prng = SHAKE-256, seed = r1, c0 = 0, c1 = 2, k = 3, poly = 24)} \\
\texttt{bin\_sample (prng = SHAKE-256, seed = r1, c0 = 0, c1 = 3, k = 3, poly = 25)} \\
\textcolor{Comment}{\texttt{\# ntt\_s}} \\
\texttt{mult\_psi (poly = 4)} \\
\texttt{transform (mode = DIF\_NTT, poly\_dst = 16, poly\_src = 4)} \\
\texttt{mult\_psi (poly = 5)} \\
\texttt{transform (mode = DIF\_NTT, poly\_dst = 17, poly\_src = 5)} \\
\textcolor{Comment}{\texttt{\# sample\_A0}} \\
\texttt{rej\_sample (prng = SHAKE-128, seed = r0, c0 = 0, c1 = 0, poly = 0)} \\
\texttt{rej\_sample (prng = SHAKE-128, seed = r0, c0 = 1, c1 = 0, poly = 1)} \\
\textcolor{Comment}{\texttt{\# A0\_mul\_s}} \\
\texttt{poly\_op (op = MUL, poly\_dst = 0, poly\_src = 16)} \\
\texttt{poly\_op (op = MUL, poly\_dst = 1, poly\_src = 17)} \\
\texttt{init (poly = 20)} \\
\texttt{poly\_op (op = ADD, poly\_dst = 20, poly\_src = 0)} \\
\texttt{poly\_op (op = ADD, poly\_dst = 20, poly\_src = 1)} \\
\textcolor{Comment}{\texttt{\# sample\_A1}} \\
\texttt{rej\_sample (prng = SHAKE-128, seed = r0, c0 = 0, c1 = 1, poly = 0)} \\
\texttt{rej\_sample (prng = SHAKE-128, seed = r0, c0 = 1, c1 = 1, poly = 1)} \\
\textcolor{Comment}{\texttt{\# A1\_mul\_s}} \\
\texttt{poly\_op (op = MUL, poly\_dst = 0, poly\_src = 16)} \\
\texttt{poly\_op (op = MUL, poly\_dst = 1, poly\_src = 17)} \\
\texttt{init (poly = 21)} \\
\texttt{poly\_op (op = ADD, poly\_dst = 21, poly\_src = 0)} \\
\texttt{poly\_op (op = ADD, poly\_dst = 21, poly\_src = 1)} \\
\textcolor{Comment}{\texttt{\# intt\_A\_mul\_s}} \\
\texttt{transform (mode = DIT\_INTT, poly\_dst = 8, poly\_src = 20)} \\
\texttt{mult\_psi\_inv (poly = 8)} \\
\texttt{transform (mode = DIT\_INTT, poly\_dst = 9, poly\_src = 21)} \\
\texttt{mult\_psi\_inv (poly = 9)} \\
\textcolor{Comment}{\texttt{\# A\_mul\_s\_plus\_e}} \\
\texttt{poly\_op (op = ADD, poly\_dst = 24, poly\_src = 8)} \\
\texttt{poly\_op (op = ADD, poly\_dst = 25, poly\_src = 9)} \\ \\
In this example, parameters from CRYSTALS-Kyber-512 have been used. For $n = 256$, the polynomial cache is divided into 32 polynomials, which are again accessed using the \texttt{poly} argument. The \texttt{init} instruction is used to initialize a specified polynomial with all zero coefficients. The matrix $\boldsymbol{A}$ is generated one row at a time, following a \textit{just-in-time} approach \cite{ingrid_saberm4_2018} instead of generating and storing all the rows together, to save memory, which becomes especially useful when dealing with larger matrices such as in CRYSTALS-Kyber-1024 and CRYSTALS-Dilithium-IV. We have written a Perl script to parse such plain-text programs and convert them into 32-bit binary instructions which can be decoded by the Sapphire crypto-processor. A complete list of supported instructions is provided in \ref{sec:appendix_b}.

\begin{figure}[!b]
\centering
\includegraphics[width=6.0in]{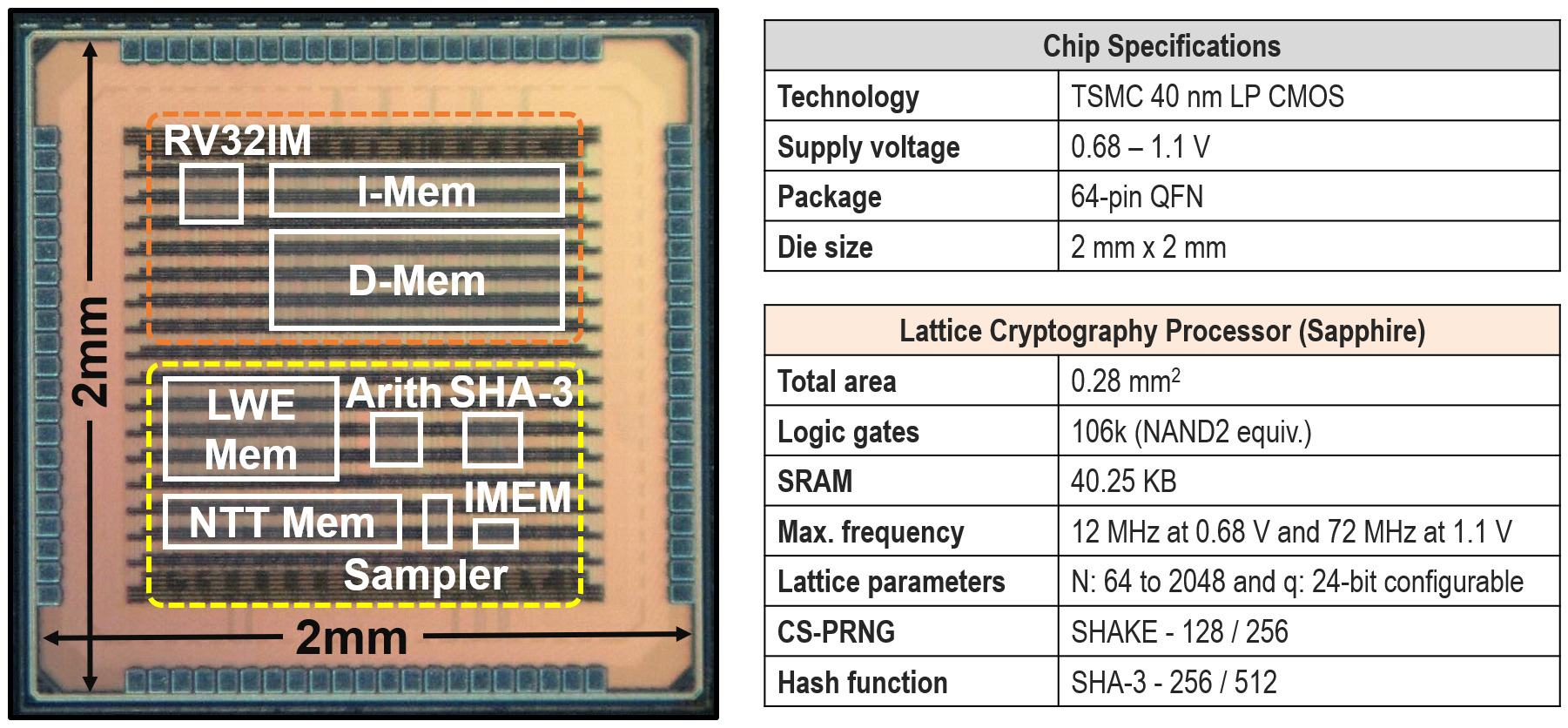}
\caption{Chip micrograph and test chip specifications.}
\label{chip_photo}
\end{figure}

We use dedicated clock gates for fine-grained power savings during program execution, and an interrupt pin is used to indicate completion of the program. Its memory and data registers can be accessed through a simple memory-mapped interface. Using the same interface, it is also coupled with a low-power RISC-V micro-processor \cite{banerjee_isscc_2018}, with 32 KB instruction memory and 64 KB data memory, which implements the RV32IM instruction set \cite{waterman_riscv_2014} and has Dhrystone performance similar to ARM Cortex-M0. When executing cryptographic workloads in the Sapphire core, the RISC-V core can be clock-gated using the \textit{wait-for-interrupt} (\texttt{wfi}) instruction. The processor is woken up by a dedicated interrupt from the Sapphire core, which is raised when the cryptographic operation is complete.  Using the memory-mapped interface ensures that the cryptographic core can be accessed through simple load and store instructions, without requiring any custom instructions or changes to the compilation toolchain. While the cryptographic core is used to accelerate all lattice cryptography computations, the RISC-V processor is used for scheduling the cryptographic workloads as well as for compression and decompression of public keys and ciphertexts. The Keccak-f[1600] core inside Sapphire can be accessed standalone through RISC-V software, and is used to accelerate SHA-3 hashing and extendable output functions according to the requirements of the protocol.

Our test chip was fabricated in the TSMC 40nm LP CMOS process, and the chip micrograph is shown in Fig. \ref{chip_photo} with the key design components highlighted. The final placed-and-routed design of our Sapphire core consists of 106k logic gates (76 kGE for synthesized design) and 40.25 KB SRAM, with a total area of 0.28 mm$^2$ (logic and memory combined). Our test chip supports supply voltage scaling from 0.68~V to 1.1~V. Although one of our key design objectives was to demonstrate a configurable lattice cryptography processor, our architecture can be easily scaled for more specific parameter sets. For example, in order to accelerate only NewHope-512 ($n = 512, q = 12289$), size of the polynomial cache can be reduced to 6.5 KB ($= 8 \times 512 \times 13$ bits) and the pre-computed NTT constants can be hard-coded in logic or stored in a 2.03 KB ROM ($= 2.5 \times 512 \times 13$ bits) instead of the 15 KB SRAM. Also, the modular arithmetic logic in the ALU can be simplified significantly to work with a single prime only.

We use the on-chip software-configurable clock gates (shown in Fig. \ref{skywalker_top_arch}) to accurately measure power consumption of different sub-modules inside the Sapphire core, e.g., sampling, NTT, arithmetic, etc. For example, the following instructions are executed to measure the average power consumption of NTT over 1000 executions: \\ \\
\texttt{clock\_config (keccak = GATE, ntt = UNGATE, sampler = GATE)} \\
\texttt{c0 = 0} \\
\texttt{loop: mult\_psi (poly = 0)} \\
\mbox{\texttt{~~~~~~    transform (mode = DIF\_NTT, poly\_dst = 4, poly\_src = 0)}} \\
\mbox{\texttt{~~~~~~    c0 = c0 + 1}} \\
\mbox{\texttt{~~~~~~    flag = compare (c0, 1000)}} \\
\mbox{\texttt{~~~~~~    if (flag == -1) goto loop}} \\ \\
The \texttt{clock\_config} instruction is used to control the clock gates, e.g., the PRNG and sampler clocks are gated when measuring NTT power (the RISC-V core is clock-gated using \texttt{wfi} as explained earlier). A simple loop is implemented using labels, comparison and conditional jump instructions, similar to assembly programs in general-purpose micro-controllers (please refer to \ref{sec:appendix_b} for details of our custom instructions). One of the chip GPIO pins is kept high during the execution of this program to indicate the measurement window, and the power consumption is measured using a source meter. This still includes leakage power from the rest of the chip, but it is only a small fraction of the total power compared to the dynamic power of the operation being measured. Similarly, power consumption of the RISC-V core is measured by clock-gating the Sapphire cryptographic core through software. Finally, leakage power of the chip is measured by externally gating the clock signal being supplied to the chip, so that all logic inside the chip is inactive.

The RISC-V processor consumes 45 $\mu$W/MHz at 1.1~V (18 $\mu$W/MHz at 0.68~V) when running the Dhrystone 2.1 benchmark. Power consumption of the cryptographic core is a strong function of the protocols being executed along with the associated parameters. Average power consumption of the lattice crypto-processor was measured to be around 8 mW at 1.1~V and 72~MHz (520 $\mu$W at 0.68~V and 12~MHz). Total leakage power of the chip was measured to be 391 $\mu$W at 1.1~V (70 $\mu$W at 0.68~V). Since our chip operates on a single power domain, it is not possible to measure leakage power of different components of the chip. We report the individual module-wise leakage and dynamic power consumption, as obtained from post-place-and-route simulations of our design operating at 1.1~V and 72~MHz, in the table below:

\begin{table}[!h]
\renewcommand{\arraystretch}{1.1}
\label{table:power_consumption}
\centering
\begin{tabular}{|l|c|c|c|}
\hline
\rowcolor{Gray}
\textbf{Module} & \textbf{P$_{\text{leak}}$ ($\mu$W)} & \textbf{P$_{\text{dyn}}$ ($\mu$W)} & \textbf{P$_{\text{tot}}$ ($\mu$W)} \\
\hline
Butterfly + ALU & 18.28 & 9210.04 & 9228.32 \\
LWE Polynomial Cache & 120.28 & 1660.18 & 1780.46 \\
NTT Constants RAM & 76.50 & 661.61 & 738.11 \\
Keccak Core + Sampler & 41.15 & 1053.58 & 1094.73 \\
RISC-V Processor + Memory & 320.15 & 2745.68 & 3065.83 \\
\hline
\end{tabular}
\end{table}

Before moving on to the protocol implementations and measurements, we summarize some key architectural design techniques we have used to achieve energy-efficiency:
\begin{itemize}
\itemsep0em
\item We have employed increased parallelism in the modular arithmetic and CS-PRNG modules in the form of single-cycle butterfly computation and 1600-bit 24-cycle Keccak data-path respectively. This reduces cycle count as well as data movement and control circuitry, thus decreasing overall energy consumption.
\item Based on overall computational complexity, we know that additions are much cheaper than multiplications. Therefore, we have exploited special properties of prime $q$ and parameter $m$, wherever possible, during Barrett reduction to convert expensive multiplications into cheaper bit-shifts and additions / subtractions.
\item Reading data from registers involves much smaller energy consumption compared to reading from SRAMs. We have used registers for storing PRNG seeds, temporary values and the Keccak state, and SRAMs are used to store only the polynomials. This significantly reduces overall energy consumption, especially for the Keccak core.
\item Software-controlled clock gates (explicitly inserted in RTL, apart from tool-inserted clock gates) for the sampler, PRNG and NTT allow fine-grained dynamic power savings by gating inactive modules as required during program execution.
\item The crypto-processor internal memory is efficiently utilized to store polynomials during protocol execution, thus avoiding access to the main processor's data memory as much as possible and reducing energy consumption.
\end{itemize}

%% file: body/06_measurements.tex
\section{Protocol Implementations and Measurement Results}
\label{sec:meas}

\begin{figure}[!t]
\centering
\includegraphics[width=5.2in]{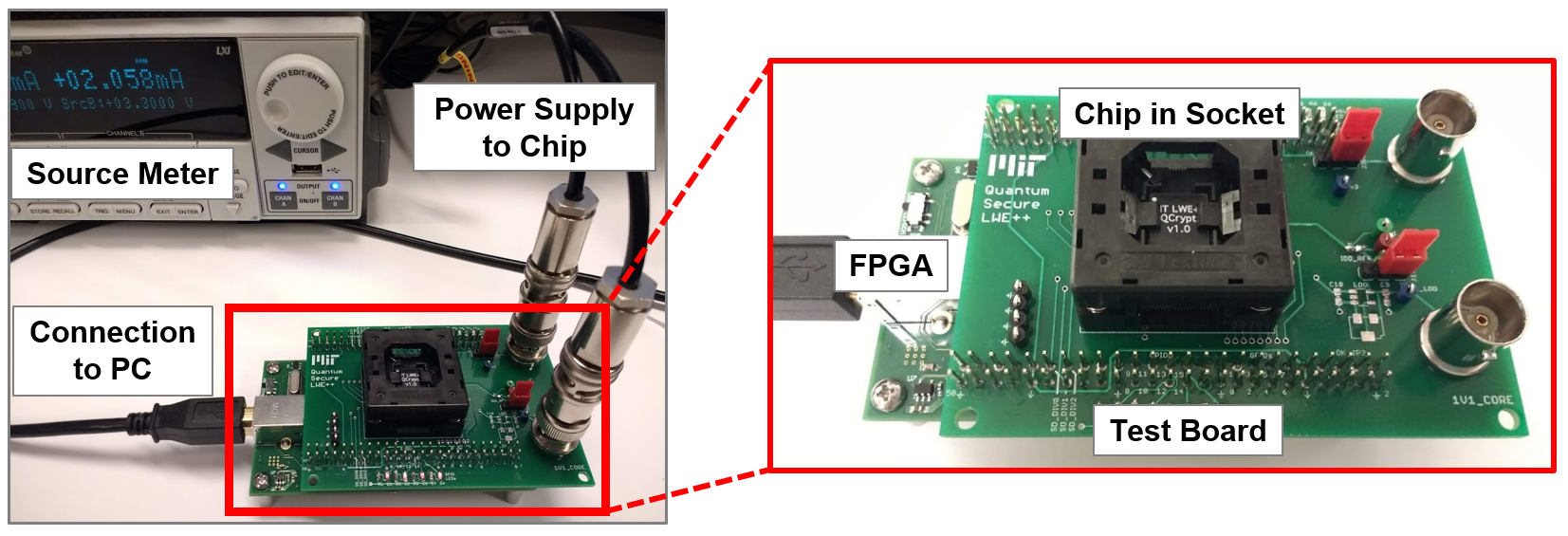}
\caption{Measurement setup with our test chip.}
\label{test_setup}
\end{figure}

To measure the efficiency of our design, we have implemented the following NIST Round 2 lattice-based cryptography protocols on our test chip:

\begin{table}[!h]
\renewcommand{\arraystretch}{1.05}
\label{table:protocols}
\centering
\begin{tabular}{|l|c|c|c|}
\hline
\rowcolor{Gray}
\textbf{Algorithm} & \textbf{Lattice Prob.} & \textbf{NIST Sec.} & \textbf{Parameter Set} \\
\hline
\multicolumn{4}{|c|}{\textbf{CCA-KEM Algorithms}} \\
\hline
\multirow{2}{*}{NewHope} & \multirow{2}{*}{Ring-LWE} & 1 & NewHope-512 \\
& & 5 & NewHope-1024 \\
\hline
\multirow{3}{*}{CRYSTALS-Kyber} & \multirow{3}{*}{Module-LWE} & 1 & Kyber-512 \\
& & 3 & Kyber-768 \\
& & 5 & Kyber-1024 \\
\hline
\multirow{3}{*}{Frodo} & \multirow{3}{*}{LWE} & 1 & Frodo-640 \\
& & 3 & Frodo-976 \\
& & 5 & Frodo-1344 \\
\hline
\multicolumn{4}{|c|}{\textbf{Signature Algorithms}} \\
\hline
\multirow{3}{*}{qTESLA} & \multirow{3}{*}{Ring-LWE} & 1 & qTESLA-I \\
& & 3 & qTESLA-III-size \\
& & 3 & qTESLA-III-speed \\
\hline
\multirow{3}{*}{CRYSTALS-Dilithium} & \multirow{3}{*}{Module-LWE} & 1 & Dilithium-II \\
& & 2 & Dilithium-III \\
& & 3 & Dilithium-IV \\
\hline
\end{tabular}
\end{table}

where NIST security levels 1-6 indicate brute-force security matching or exceeding that of AES-128, SHA3-256, AES-192, SHA3-384, AES-256 and SHA3-512 respectively. Fig. \ref{test_setup} shows our test board and measurement setup. The test chip is housed in a QFN64 socket soldered to the board, an Opal Kelly XEM7001 FPGA development board is used to interface with the chip, and a Keithley 2602A source meter supplies power to the chip. Both the FPGA and the source meter are controlled from a host computer through USB and GPIB interfaces respectively. The FPGA is used to transfer programs from the host computer to the instruction memory of our test chip. Also, a small ring-oscillator-based true random number generator \cite{dichtl_trng_2007} implemented on the FPGA is connected to our test chip through GPIO pins for providing fresh random inputs to the \texttt{randombytes} function which is part of the NIST API. All lattice cryptography programs are written using custom instructions and compiled with our script, while all RISC-V software is written in C and compiled using the \texttt{riscv-gcc} toolchain.

\subsection{Protocol Implementations and Evaluation Results}

Next, we describe some key aspects of our protocol implementations along with timing and energy profiling results. All polynomial arithmetic, transforms and sampling operations are accelerated using custom programs running in the Sapphire core, and all SHA-3 computations utilize the Keccak core inside Sapphire. The RISC-V processor is used only to read / write data and programs from / to the cryptographic core (both when executing polynomial computations and when utilizing the fast Keccak core for SHA-3 operations), generate initial randomness using the \texttt{randombytes} function, encode / decode messages and compress / decompress public keys and ciphertexts. For polynomials which need to be read from the polynomial cache and encoded (or decoded and written to the polynomial cache), we directly post-process the outputs (or pre-process the inputs) of the crypto-processor's internal memory, instead of first storing the data in intermediate temporary arrays and then processing them. This saves around 10-20\% cycles in overall protocol run-time. Also, the internal clock gates are strategically enabled and disabled during program execution using the \texttt{clock\_config} instruction (please refer to \ref{sec:appendix_b} for details of our custom instructions) to reduce overall energy consumption.

For the NewHope and CRYSTALS-Kyber key exchange schemes, each of the CPA-secure public key encryption functions -- \textsf{CPA-PKE.KeyGen}, \textsf{CPA-PKE.Encrypt} and \textsf{CPA-PKE.Decrypt} -- has been written entirely (excluding the encoding and decoding operations) using Sapphire custom instructions with each of the corresponding programs fitting completely in its 1 KB instruction memory. The CCA-secure key encapsulation functions -- \textsf{CCA-KEM.KeyGen}, \textsf{CCA-KEM.Encaps} and \textsf{CCA-KEM.Decaps} -- involve calls to SHA-3 and the CPA-PKE functions (according to the Fujisaki-Okamoto transform \cite{fujisaki_2013}), which are implemented in software. Since the signature schemes qTESLA and CRYSTALS-Dilithium both involve probabilistic rejection of intermediate values, the associated polynomial computations are split into multiple custom programs instead of one each for the \textsf{KeyGen}, \textsf{Sign} and \textsf{Verify} functions. These blocks of code are scheduled using RISC-V software, which also handles encoding and decoding operations. The only exception is the \textsf{KeyGen} step in qTESLA, where high-precision discrete Gaussian sampling using large CDT tables is implemented in software, with the SHA-3 functions accelerated in hardware.

Since Module-LWE algorithms involve working with vectors or matrices of polynomials, it is particularly important to ensure that these polynomials fit inside the crypto-processor memory as much as possible (because reads and writes to the internal memory through software are not cheap). When multiplying the public matrix \textbf{A} with the secret vector \textbf{s}, the matrix \textbf{A} is generated through rejection sampling, one row at a time, following the \textit{just-in-time} approach from \cite{ingrid_saberm4_2018}. This reduces memory footprint so that the entire computation can fit in the polynomial cache.

\begin{table}[!t]
\renewcommand{\arraystretch}{1.2}
\caption{Measured energy and performance of public key encryption schemes}
\label{table:cpapke_results}
\centering
\begin{tabular}{|l|c|c|c|c|c|}
\hline
\rowcolor{Gray}
\textbf{Protocol} & \multicolumn{2}{c|}{\textbf{Cortex-M4} \cite{pqm4}} & \multicolumn{3}{c|}{\textbf{This work} \textsuperscript{$\dagger$}} \\ \cline{2-6}
\rowcolor{Gray}
& \textbf{Cycles} & \textbf{Energy ($\mu$J)} & \textbf{Cycles} & \textbf{Power (mW)} & \textbf{Energy ($\mu$J)} \\
\hline
\multicolumn{6}{|c|}{NewHope-512-CPA-PKE} \\
\hline
\textsf{KeyGen}  & - & - & 18,667 & 7.15 & 1.85 \\
\textsf{Encrypt} & - & - & 53,499 & 7.79 & 5.79 \\
\textsf{Decrypt} & - & - & 29,099 & 6.81 & 2.77 \\
\hline
\multicolumn{6}{|c|}{NewHope-1024-CPA-PKE} \\
\hline
\textsf{KeyGen}  & 1,179,353 & 725.30 & 38,012 & 7.39 & 3.90 \\
\textsf{Encrypt} & 1,663,023 & 1022.76 & 106,611 & 8.10 & 12.00 \\
\textsf{Decrypt} & 194,439 & 119.58 & 56,061 & 9.31 & 7.26 \\
\hline
\multicolumn{6}{|c|}{CRYSTALS-Kyber-512-CPA-PKE} \\
\hline
\textsf{KeyGen}  & 609,923 & 375.10 & 46,187 & 7.61 & 4.90 \\
\textsf{Encrypt} & 721,925 & 443.98 & 66,851 & 8.33 & 7.74 \\
\textsf{Decrypt} & 95,894 & 58.97 & 32,198 & 7.67 & 3.45 \\
\hline
\multicolumn{6}{|c|}{CRYSTALS-Kyber-768-CPA-PKE} \\
\hline
\textsf{KeyGen}  & 1,001,328 & 615.82 & 72,245 & 7.40 & 7.43 \\
\textsf{Encrypt} & 1,116,540 & 686.67 & 94,440 & 7.87 & 10.31 \\
\textsf{Decrypt} & 129,560 & 79.68 & 40,202 & 7.75 & 4.34 \\
\hline
\multicolumn{6}{|c|}{CRYSTALS-Kyber-1024-CPA-PKE} \\
\hline
\textsf{KeyGen}  & 1,610,114 & 990.22 & 100,453 & 7.95 & 11.09 \\
\textsf{Encrypt} & 1,747,687 & 1074.83 & 124,142 & 7.94 & 13.70 \\
\textsf{Decrypt} & 162,204 & 99.76 & 48,205 & 8.42 & 5.65 \\
\hline
\multicolumn{6}{l}{\small{\textsuperscript{$\dagger$} Includes program execution and read/write from/to crypto-processor}} \\
\end{tabular}
\end{table}

\begin{figure}[!t]
\centering
\includegraphics[width=5.4in]{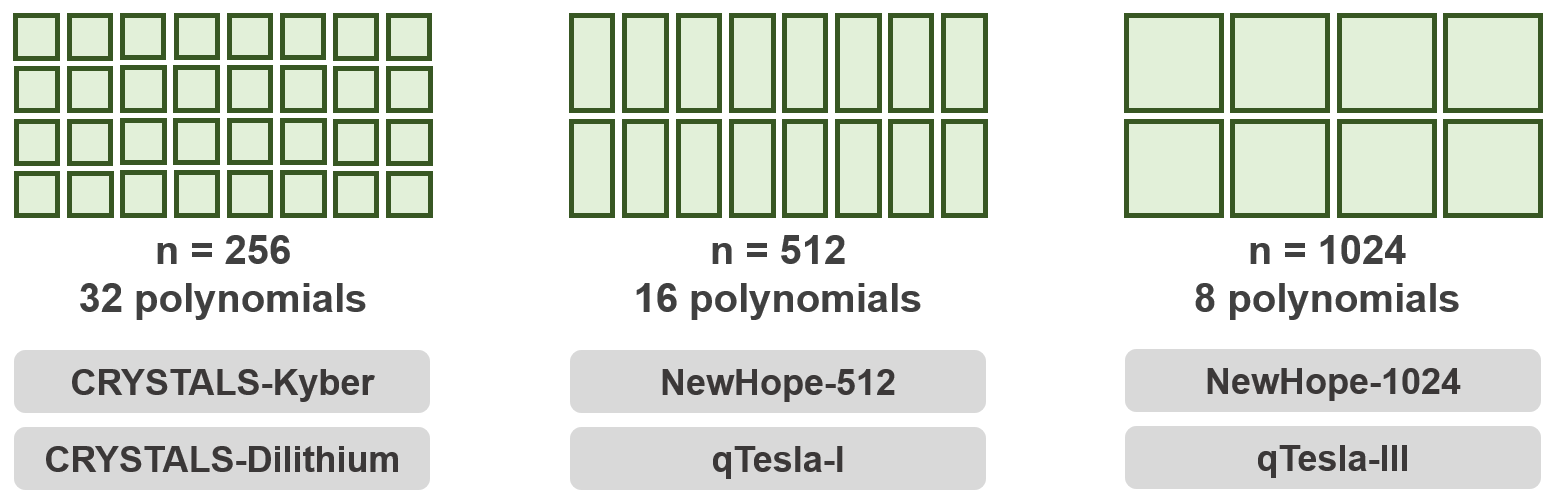}
\caption{Configurations of the Sapphire polynomial cache for Ring-LWE and Module-LWE schemes.}
\label{ring_tiling}
\end{figure}

In Table \ref{table:cpapke_results}, we compare cycle count and energy consumption of our implementations of the Ring-LWE and Module-LWE CPA-PKE schemes with assembly-optimized software on ARM Cortex-M4 micro-processor (from PQM4 \cite{pqm4}), with average cycle counts for 100 executions. The energy consumption of our test chip has been measured at 1.1 V and 72 MHz, while the energy consumption of the Cortex-M4 processor is estimated from cycle counts using average power (61.5 mW or 615 pJ/cycle at 3.0 V and 100 MHz) measured on NUCLEO-F411RE operating at 100 MHz. The cycle count and energy consumption for our implementation include program execution as well as the additional overhead of writing inputs to and reading outputs from the Sapphire cryptographic core. For both NewHope and CRYSTALS-Kyber, we observe up to an order of magnitude improvement in energy-efficiency compared to software, after accounting for voltage scaling. Fig. \ref{ring_tiling} shows how configurability of the Sapphire polynomial cache is utilized to support different ring dimensions.

\begin{figure}[!t]
\centering
\includegraphics[width=6.2in]{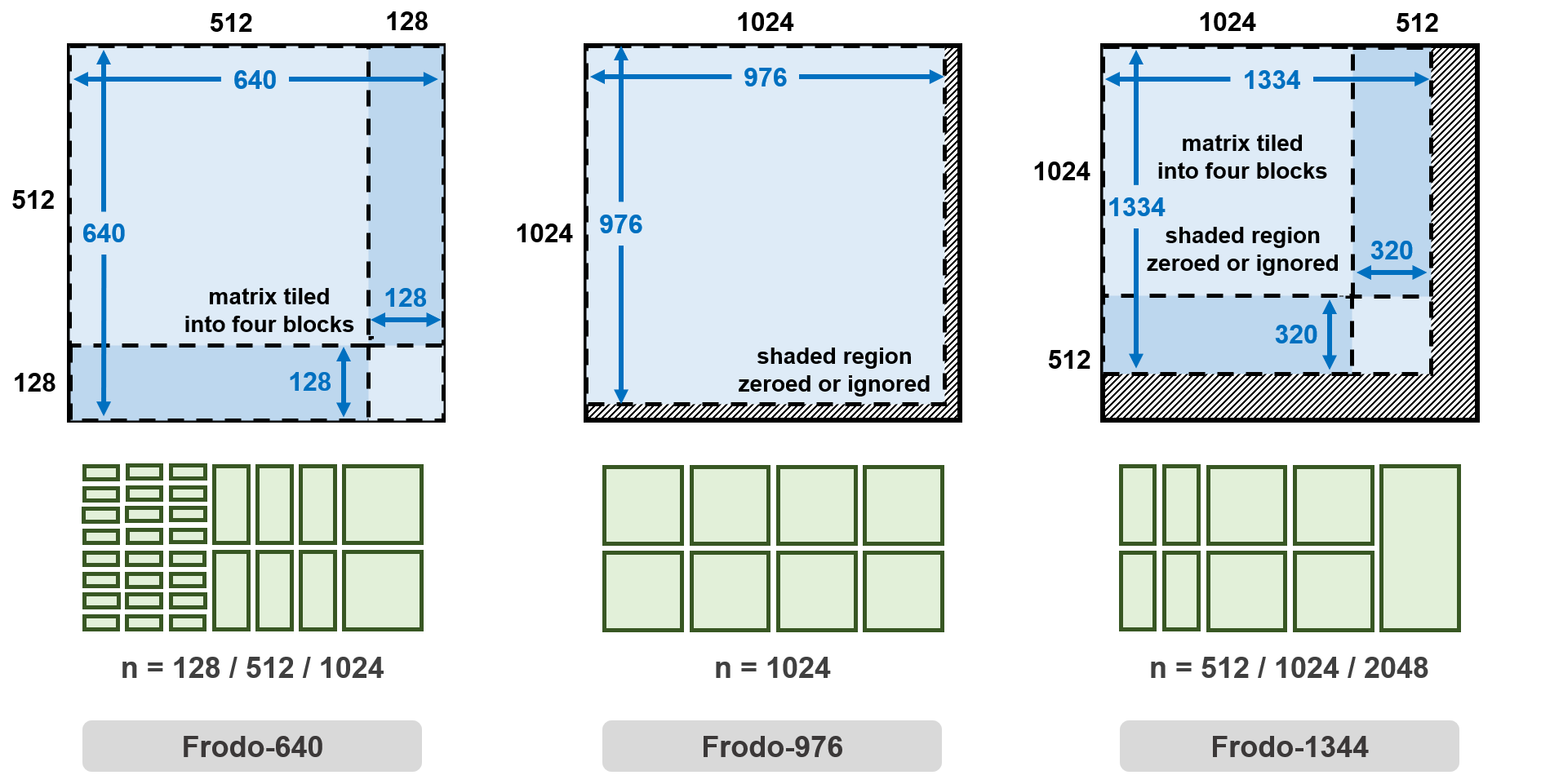}
\caption{Tiling of $n \times n$ square matrices for Frodo-640, Frodo-976 and Frodo-1344.}
\label{frodo_tiling}
\end{figure}

Although our lattice crypto-processor architecture primarily targets Ring-LWE and Module-LWE schemes, we also implement the LWE-based Frodo KEM protocol to demonstrate its flexibility. Since LWE-based algorithms require large matrix multiplications, the arithmetic operations dominate total computation cost unlike Ring-LWE and Module-LWE where sampling is the most expensive operation. Since the matrix dimensions are not powers of two, we tile the rows or columns so that we can use the crypto-processor's array operations effectively, as shown in Fig. \ref{frodo_tiling}. For Frodo-640, we split each 640-element array into two arrays of size 512 and 128. For Frodo-976, we simply use arrays of size 1024 with the last 48 elements zeroed out or ignored, as applicable. For Frodo-1344, we use arrays of size 1536, formed by splitting them into two arrays of size 1024 and 512, with the last 192 elements (of the 512-dimension array) zeroed out or ignored, as applicable. However, this tiling scheme makes our version of Frodo incompatible with the reference software implementation.

Frodo involves three large matrix multiplications: \textbf{AS}, \textbf{S$'$A} and \textbf{S$'$B}, where \textbf{A}, \textbf{S}, \textbf{S$'$} and \textbf{B} have dimensions $n \times n$, $n \times \bar{n}$, $\bar{m} \times n$ and $n \times \bar{n}$ respectively with $n \in \{640, 976, 1344\}$ and $\bar{m} = \bar{n} = 8$. We ensure that \textbf{S$'$} is stored in row-major form and \textbf{B} is stored in column-major form, which simplifies calculating \textbf{S$'$B} using the schoolbook matrix multiplication technique. The \texttt{poly\_op} instruction is used to coefficient-wise multiply a row of the multiplier matrix with a column of the multiplicand matrix, and the \texttt{sum\_elems} instruction computes the sum of its elements to generate one element of the output matrix (please refer to \ref{sec:appendix_b} for details of our custom instructions). For calculating the matrix \textbf{AS}, we generate \textbf{A} in row-major form (using rejection sampling, with zero chance of rejection since $q$ is a power of two) and \textbf{S} in column major form (using CDT-based discrete Gaussian sampling) so that the same techniques still work. For $n \in \{640, 976\}$, the matrix \textbf{S} is generated two columns at a time to reduce the number of outer loop iterations, as illustrated in the pseudo-code below: \\ \\
\textcolor{IFDEF}{\mbox{\texttt{\#if (n == 1344)}}} \\
\mbox{\texttt{for (j = 0; j < nbar; j = j + 1) \{}} \\
\textcolor{IFDEF}{\mbox{\texttt{\#else}}} \\
\mbox{\texttt{for (j = 0; j < nbar/2; j = j + 2) \{}} \\
\textcolor{IFDEF}{\mbox{\texttt{\#endif}}} \\
\mbox{\texttt{~~~    cdt\_sample (prng = SHAKE-256, seed = r1, ..., poly = 0)}} \\
\textcolor{IFDEF}{\mbox{\texttt{~~~    \#if (n != 1344)}}} \\
\mbox{\texttt{~~~    cdt\_sample (prng = SHAKE-256, seed = r1, ..., poly = 1)}} \\
\textcolor{IFDEF}{\mbox{\texttt{~~~    \#endif}}} \\
\mbox{\texttt{~~~    for (i = 0; i < n; i = i + 1) \{}} \\
\mbox{\texttt{~~~~~~~        rej\_sample (prng = SHAKE-128, seed = r0, ..., poly = 4)}} \\
\textcolor{IFDEF}{\mbox{\texttt{~~~~~~~    \#if (n != 1344)}}} \\
\mbox{\texttt{~~~~~~~        poly\_copy (poly\_dst = 5, poly\_src = 4)}} \\
\textcolor{IFDEF}{\mbox{\texttt{~~~~~~~    \#endif}}} \\
\mbox{\texttt{~~~~~~~        poly\_op (op = MUL, poly\_dst = 4, poly\_src = 0)}} \\
\mbox{\texttt{~~~~~~~        AS[i][j] = sum\_elems (poly = 4)}} \\
\textcolor{IFDEF}{\mbox{\texttt{~~~~~~~    \#if (n != 1344)}}} \\
\mbox{\texttt{~~~~~~~        poly\_op (op = MUL, poly\_dst = 5, poly\_src = 1)}} \\
\mbox{\texttt{~~~~~~~        AS[i][j+1] = sum\_elems (poly = 5)}} \\
\textcolor{IFDEF}{\mbox{\texttt{~~~~~~~    \#endif}}} \\
\mbox{\texttt{~~~    \}}} \\
\mbox{\texttt{\}}} \\ \\
Since both matrices \textbf{S$'$} and \textbf{A} are generated on-the-fly in row-major fashion, this makes calculating \textbf{S$'$A} a bit complicated. We multiply each element of the \texttt{i}-th row of \textbf{A} with the \texttt{i}-th element of the \texttt{j}-th row of \textbf{S$'$} to generate a partial sum. These \texttt{i} partial sums are incrementally added together to compute the \texttt{j}-th row of the output matrix \textbf{S$'$A}. Once again, we generate \textbf{S} two columns at a time to reduce the number of outer loop iterations. The corresponding pseudo-code is shown below: \\ \\
\textcolor{IFDEF}{\mbox{\texttt{\#if (n == 1344)}}} \\
\mbox{\texttt{for (j = 0; j < nbar; j = j + 1) \{}} \\
\textcolor{IFDEF}{\mbox{\texttt{\#else}}} \\
\mbox{\texttt{for (j = 0; j < nbar/2; j = j + 2) \{}} \\
\textcolor{IFDEF}{\mbox{\texttt{\#endif}}} \\
\mbox{\texttt{~~~    cdt\_sample (prng = SHAKE-256, seed = r1, ..., poly = 0)}} \\
\mbox{\texttt{~~~    init (poly = 6)}} \\
\textcolor{IFDEF}{\mbox{\texttt{~~~    \#if (n != 1344)}}} \\
\mbox{\texttt{~~~    cdt\_sample (prng = SHAKE-256, seed = r1, ..., poly = 1)}} \\
\mbox{\texttt{~~~    init (poly = 7)}} \\
\textcolor{IFDEF}{\mbox{\texttt{~~~    \#endif}}} \\
\mbox{\texttt{~~~    for (i = 0; i < n; i = i + 1) \{}} \\
\mbox{\texttt{~~~~~~~        rej\_sample (prng = SHAKE-128, seed = r0, ..., poly = 4)}} \\
\mbox{\texttt{~~~~~~~        reg = (poly = 0)[i]}} \\
\mbox{\texttt{~~~~~~~        poly\_op (op = CONST\_MUL, poly\_dst = 2, poly\_src = 4)}} \\
\mbox{\texttt{~~~~~~~        poly\_op (op = ADD, poly\_dst = 6, poly\_src = 2)}} \\
\textcolor{IFDEF}{\mbox{\texttt{~~~~~~~    \#if (n != 1344)}}} \\
\mbox{\texttt{~~~~~~~        reg = (poly = 1)[i]}} \\
\mbox{\texttt{~~~~~~~        poly\_op (op = CONST\_MUL, poly\_dst = 3, poly\_src = 4)}} \\
\mbox{\texttt{~~~~~~~        poly\_op (op = ADD, poly\_dst = 7, poly\_src = 3)}} \\
\textcolor{IFDEF}{\mbox{\texttt{~~~~~~~    \#endif}}} \\
\mbox{\texttt{~~~    \}}} \\
\mbox{\texttt{\}}} \\ \\
where the \texttt{reg = (poly)[i]} instruction is used to save the \texttt{i}-th element of the array in the 24-bit internal register \texttt{reg}, the \texttt{init (poly)} instruction creates an array of zeros and the \texttt{CONST\_MUL} operation multiplies each element of an array with the value stored in \texttt{reg}  (please refer to \ref{sec:appendix_b} for details of our instructions). The \textbf{AS + E} and \textbf{S$'$A + E$'$} computations require 10.9M and 9.9M cycles respectively for Frodo-640, and 25.3M and 23.2M cycles respectively for Frodo-976, and 67.1M and 62.7M cycles respectively for Frodo-1344, which constitute majority of the total cycle count. This is quite different from the Ring-LWE and Module-LWE schemes, where polynomial sampling accounts for 60-70\% of the total computation cost. Please note that memory usage of Frodo-1344-CCA-KEM-\textsf{Decaps} exceeds the 64 KB processor data memory on our test chip; hence it was evaluated only in simulation, with power consumption extrapolated from measured power for Frodo-640 and Frodo-976.

In Tables \ref{table:ccakem_results} and \ref{table:sign_results}, we have compared cycle count and energy consumption of assembly-optimized Cortex-M4 software \cite{pqm4} with our hardware-accelerated implementation on our test chip operating at 1.1 V and 72 MHz, with average cycle counts for 100 executions. Clearly, our design achieves up to an order of magnitude improvement in energy-efficiency and performance compared to state-of-the-art software. We note that Module-LWE schemes, although a bit slower compared to Ring-LWE, offer parameters with better scalability in terms of security and efficiency compared to Ring-LWE schemes. Among the key encapsulation schemes, NewHope and CRYSTALS-Kyber are two orders of magnitude more efficient than Frodo, owing to the inherent structure in ideal and module lattices where the key operation is polynomial multiplication as opposed to matrix multiplication in standard lattices. Among the digital signature schemes evaluated, qTESLA allows faster signature generation and verification compared to CRYSTALS-Kyber. However, our implementation of the key generation step in qTESLA is quite expensive since it uses CDT-based discrete Gaussian sampling with large tables and high precision. This is not a big concern since signature key-pairs are generated infrequently; also, more specialized hardware can be added to our architecture to support such distribution parameters, albeit at the cost of logic area.

\begin{figure}[!hbt]
\centering
\includegraphics[width=6.2in]{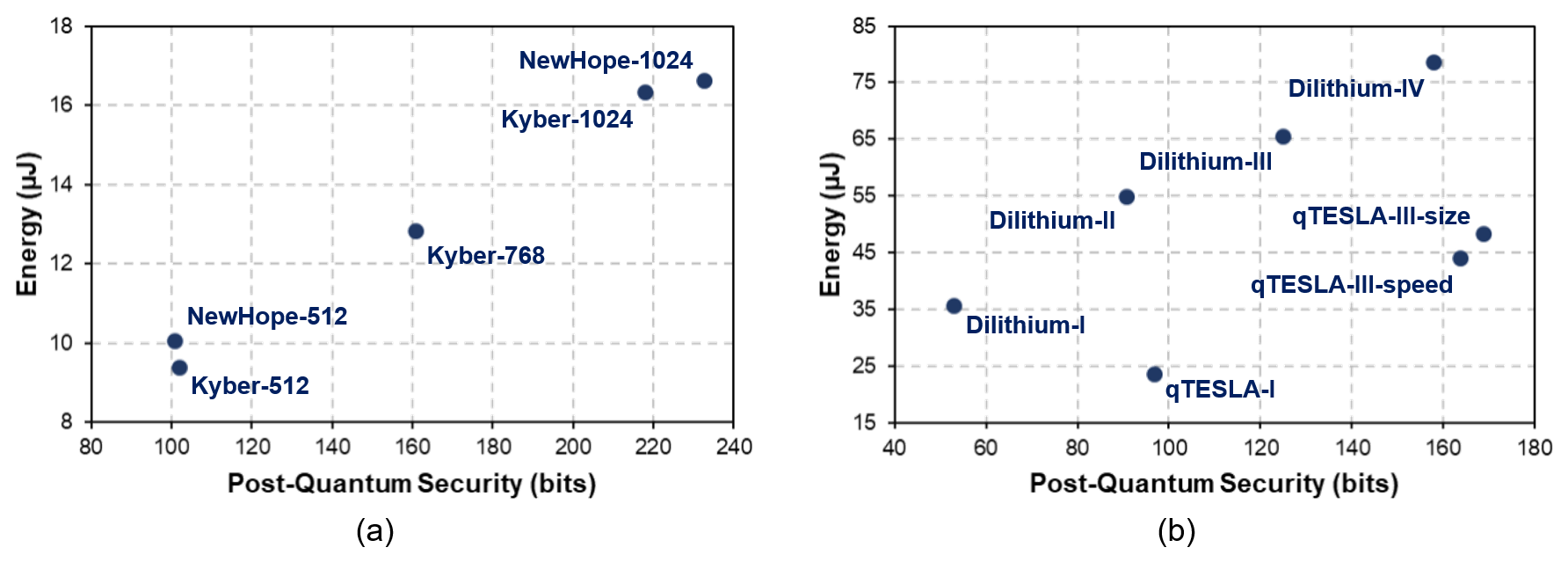}
\caption{Energy consumption of Ring-LWE and Module-LWE-based (a) CCA-KEM-\textsf{Encaps} and (b) \textsf{Sign} algorithms at different post-quantum security levels.}
\label{energy_security_plot}
\end{figure}

In Fig. \ref{energy_security_plot}, we plot the measured energy consumption of the Ring-LWE and Module-LWE-based CCA-KEM-\textsf{Encaps} and \textsf{Sign} algorithms at different post-quantum security levels, as implemented on our test chip operating at at 1.1 V and 72 MHz. Due to the configurability of our lattice crypto-processor, we are able to implement all these different modes and achieve energy scalability through efficiency versus security trade-offs.

\clearpage

\begin{table}[!hbt]
\renewcommand{\arraystretch}{1.45}
\caption{Measured energy and performance of key encapsulation schemes}
\label{table:ccakem_results}
\centering
\begin{tabular}{|l|c|c|c|c|c|}
\hline
\rowcolor{Gray}
\textbf{Protocol} & \multicolumn{2}{c|}{\textbf{Cortex-M4} \cite{pqm4}} & \multicolumn{3}{c|}{\textbf{This work}} \\ \cline{2-6}
\rowcolor{Gray}
& \textbf{Cycles} & \textbf{Energy} & \textbf{Cycles} & \textbf{Power} & \textbf{Energy} \\
\rowcolor{Gray}
& & \textbf{($\mu$J)} & & \textbf{(mW)} & \textbf{($\mu$J)} \\
\hline
\multicolumn{6}{|c|}{NewHope-512-CCA-KEM} \\
\hline
\textsf{KeyGen} & - & - & 52,063 & 6.04 & 4.37 \\
\textsf{Encaps} & - & - & 136,077 & 5.30 & 10.02 \\
\textsf{Decaps} & - & - & 142,295 & 5.80 & 11.46 \\
\hline
\multicolumn{6}{|c|}{NewHope-1024-CCA-KEM} \\
\hline
\textsf{KeyGen} & 1,243,729 & 764.89 & 97,969 & 6.13 & 8.35 \\
\textsf{Encaps} & 1,963,184 & 1207.34 & 236,812 & 5.05 & 16.59 \\
\textsf{Decaps} & 1,978,982 & 1217.07 & 258,872 & 5.89 & 21.17 \\
\hline
\multicolumn{6}{|c|}{CRYSTALS-Kyber-512-CCA-KEM} \\
\hline
\textsf{KeyGen} & 726,921 & 447.06 & 74,519 & 5.77 & 5.97 \\
\textsf{Encaps} & 987,864 & 607.54 & 131,698 & 5.12 & 9.37 \\
\textsf{Decaps} & 1,018,946 & 626.65 & 142,309 & 5.69 & 11.25 \\
\hline
\multicolumn{6}{|c|}{CRYSTALS-Kyber-768-CCA-KEM} \\
\hline
\textsf{KeyGen} & 1,200,291 & 738.18 & 111,525 & 5.28 & 8.19 \\
\textsf{Encaps} & 1,446,284 & 889.46 & 177,540 & 5.19 & 12.80 \\
\textsf{Decaps} & 1,477,365 & 908.58 & 190,579 & 5.86 & 15.52 \\
\hline
\multicolumn{6}{|c|}{CRYSTALS-Kyber-1024-CCA-KEM} \\
\hline
\textsf{KeyGen} & 1,771,729 & 1089.61 & 148,547 & 5.95 & 12.27 \\
\textsf{Encaps} & 2,142,912 & 1317.89 & 223,469 & 5.25 & 16.3 \\
\textsf{Decaps} & 2,188,917 & 1346.18 & 240,977 & 5.91 & 19.76 \\
\hline
\multicolumn{6}{|c|}{Frodo-640-CCA-KEM} \\
\hline
\textsf{KeyGen} & 81,293,476 & 49995.49 & 11,453,942 & 6.65 & 1057.65 \\
\textsf{Encaps} & 86,178,252 & 52999.62 & 11,609,668 & 7.01 & 1129.95 \\
\textsf{Decaps} & 87,170,982 & 53610.15 & 12,035,513 & 6.88 & 1150.83 \\
\hline
\multicolumn{6}{|c|}{Frodo-976-CCA-KEM} \\
\hline
\textsf{KeyGen} & - & - & 26,005,326 & 6.70 & 2420.97 \\
\textsf{Encaps} & - & - & 29,749,417 & 7.05 & 2912.95 \\
\textsf{Decaps} & - & - & 30,421,175 & 6.94 & 2932.13 \\
\hline
\multicolumn{6}{|c|}{Frodo-1344-CCA-KEM} \\
\hline
\textsf{KeyGen} & - & - & 67,994,170 & 6.75 & 6374.45 \\
\textsf{Encaps} & - & - & 71,501,358 & 7.10 & 7050.83 \\
\textsf{Decaps} & - & - & 72,526,695 & 7.00 & 7051.21 \\
\hline
\end{tabular}
\end{table}

\clearpage

\begin{table}[!hbt]
\renewcommand{\arraystretch}{1.45}
\caption{Measured energy and performance of digital signature schemes}
\label{table:sign_results}
\centering
\begin{tabular}{|l|c|c|c|c|c|}
\hline
\rowcolor{Gray}
\textbf{Protocol} & \multicolumn{2}{c|}{\textbf{Cortex-M4} \cite{pqm4}} & \multicolumn{3}{c|}{\textbf{This work}} \\ \cline{2-6}
\rowcolor{Gray}
& \textbf{Cycles} & \textbf{Energy} & \textbf{Cycles} & \textbf{Power} & \textbf{Energy} \\
\rowcolor{Gray}
& & \textbf{($\mu$J)} & & \textbf{(mW)} & \textbf{($\mu$J)} \\
\hline
\multicolumn{6}{|c|}{qTESLA-I} \\
\hline
\textsf{KeyGen} & 17,545,901 & 10790.73 & 4,846,949 & 7.89 & 531.55 \\
\textsf{Sign} & 6,317,445 & 3885.23 & 168,273 & 9.99 & 23.34 \\
\textsf{Verify} & 1,059,370 & 651.51 & 38,922 & 7.99 & 4.32 \\
\hline
\multicolumn{6}{|c|}{qTESLA-III-size} \\
\hline
\textsf{KeyGen} & 58,227,852 & 35810.13 & 11,479,190 & 7.71 & 1229.18 \\
\textsf{Sign} & 19,869,370 & 12219.66 & 348,429 & 9.97 & 48.23 \\
\textsf{Verify} & 2,297,530 & 1412.98 & 69,154 & 7.59 & 7.27 \\
\hline
\multicolumn{6}{|c|}{qTESLA-III-speed} \\
\hline
\textsf{KeyGen} & 30,720,411 & 18893.05 & 11,898,241 & 7.64 & 1262.39 \\
\textsf{Sign} & 11,987,079 & 7372.05 & 317,083 & 9.97 & 43.91 \\
\textsf{Verify} & 2,225,296 & 1368.56 & 67,712 & 7.30 & 6.86 \\
\hline
\multicolumn{6}{|c|}{CRYSTALS-Dilithium-I} \\
\hline
\textsf{KeyGen} & - & - & 95,202 & 6.82 & 9.00 \\
\textsf{Sign} & - & - & 376,392 & 6.77 & 35.41 \\
\textsf{Verify} & - & - & 142,576 & 7.73 & 15.31 \\
\hline
\multicolumn{6}{|c|}{CRYSTALS-Dilithium-II} \\
\hline
\textsf{KeyGen} & - & - & 130,022 & 7.24 & 13.08 \\
\textsf{Sign} & - & - & 514,246 & 7.68 & 54.82 \\
\textsf{Verify} & - & - & 184,933 & 7.49 & 19.23 \\
\hline
\multicolumn{6}{|c|}{CRYSTALS-Dilithium-III} \\
\hline
\textsf{KeyGen} & 2,322,955 & 1428.62 & 167,433 & 7.36 & 17.11 \\
\textsf{Sign} & 9,978,000 & 6136.47 & 634,763 & 7.40 & 65.26 \\
\textsf{Verify} & 2,322,765 & 1428.50 & 229,481 & 7.41 & 23.63 \\
\hline
\multicolumn{6}{|c|}{CRYSTALS-Dilithium-IV} \\
\hline
\textsf{KeyGen} & - & - & 223,272 & 6.89 & 21.38 \\
\textsf{Sign} & - & - & 815,636 & 6.93 & 78.53 \\
\textsf{Verify} & - & - & 276,221 & 7.44 & 28.55 \\
\hline
\end{tabular}
\end{table}

\clearpage

In Table \ref{table:overall_comparison}, we compare our design with existing hardware-accelerated implementations of NIST Round 2 lattice-based protocols. Our crypto-processor is significantly smaller than the multiple designs generated using high-level synthesis in \cite{basu_pqchw_2019}, and is also more flexible and energy-efficient. Our Kyber implementation is faster than \cite{albrecht_rsa_2018} which uses RSA, AES and SHA hardware accelerators on the SLE 78 security controller platform to accelerate lattice cryptography. Efficiency of our design is greater than or comparable to state-of-the-art FPGA implementations of Ring-LWE \cite{guneysu_newhopefpga_2017, sepulveda_pqriscv_2019}. Notably, \cite{sepulveda_pqriscv_2019} also uses a RISC-V processor with NTT and SHA accelerators to implement the NewHope protocol. However, our implementation of Frodo, which re-purposes the Ring/Module-LWE hardware for LWE computations, is not as efficient as the dedicated LWE accelerator in \cite{guneysu_frodo_2018}. Finally, we also compare our design with state-of-the-art pre-quantum elliptic curve cryptography hardware \cite{banerjee_isscc_2018, hutter_nacl_2015}, and we observe our implementation of CCA-secure lattice-based key encapsulation using NewHope-512 to be around $5 \times$ more efficient compared to elliptic curve Diffie-Hellman key exchange using the NIST P-256 curve at comparable pre-quantum security level.


\begin{table}[!t]
\scriptsize
\renewcommand{\arraystretch}{1.45}
\caption{Comparison of our design with state-of-the-art hardware}
\label{table:overall_comparison}
\centering
\begin{tabular}{|l|c|c|c|c|c|c|c|c|c|}
\hline
\rowcolor{Gray}
\textbf{Design} & \textbf{Platform} & \textbf{Tech} & \textbf{VDD} & \textbf{Freq} & \textbf{Protocol} & \textbf{Area} & \textbf{Cycles} & \textbf{Energy} \\
\rowcolor{Gray}
& & \textbf{(nm)} & \textbf{(V)} & \textbf{(MHz)} & & \textbf{(kGE)} & & \textbf{($\mu$J)} \\
\hline
\multirow{7}{1.4cm}{\textbf{This work}} & \multirow{7}{*}{ASIC} & \multirow{7}{*}{40} & \multirow{7}{*}{1.1} & \multirow{7}{*}{72} & NewHope-512-CCA-KEM-\textsf{Encaps} & \multirow{7}{*}{106} & 136,077 & 10.02 \\
 & & & & & NewHope-1024-CPA-PKE-\textsf{Encrypt} & & 106,611 & 12.00 \\
 & & & & & Kyber-512-CCA-KEM-\textsf{Encaps} & & 131,698 & 9.37 \\
 & & & & & Kyber-768-CPA-PKE-\textsf{Encrypt} & & 94,440 & 10.31 \\
 & & & & & Kyber-768-CCA-KEM-\textsf{Encaps} & & 177,540 & 12.80 \\
 & & & & & Frodo-640-CCA-KEM-\textsf{Encaps} & & 11,609,668 & 1129.95 \\
 & & & & & Dilithium-II-\textsf{Sign} & & 514,246 & 54.82 \\
\hline
\multirow{3}{1.4cm}{Basu et al. \cite{basu_pqchw_2019} \textsuperscript{$\dagger$}} & \multirow{3}{*}{ASIC} & \multirow{3}{*}{65} & \multirow{3}{*}{1.2} & 169 & NewHope-512-CCA-KEM-\textsf{Encaps} & 1273 & 307,847 & 69.42 \\
 & & & & 200 & Kyber-512-CCA-KEM-\textsf{Encaps} & 1341 & 31,669 & 6.21 \\
 & & & & 158 & Dilithium-II-\textsf{Sign} & 1603 & 155,166 & 50.42 \\
\hline
\multirow{2}{1.4cm}{Albrecht $\,$ et al. \cite{albrecht_rsa_2018}} & \multirow{2}{*}{SLE 78} & \multirow{2}{*}{-} & \multirow{2}{*}{-} & \multirow{2}{*}{50} & Kyber-768-CPA-PKE-\textsf{Encrypt} & \multirow{2}{*}{-} & 4,747,291 & \multirow{2}{*}{-} \\
 & & & & & Kyber-768-CCA-KEM-\textsf{Encaps} & & 5,117,996 & \\
\hline
\multirow{2}{1.4cm}{Oder et al. \cite{guneysu_newhopefpga_2017}} & \multirow{2}{*}{FPGA} & \multirow{2}{*}{-} & \multirow{2}{*}{-} & \multirow{2}{*}{117} & \multirow{2}{*}{NewHope-1024-Simple-\textsf{Encrypt}} & \multirow{2}{*}{-} & \multirow{2}{*}{179,292} & \multirow{2}{*}{-} \\
 & & & & & & & & \\
\hline
\multirow{2}{1.4cm}{Howe et al. \cite{guneysu_frodo_2018}} & \multirow{2}{*}{FPGA} & \multirow{2}{*}{-} & \multirow{2}{*}{-} & \multirow{2}{*}{167} & \multirow{2}{*}{Frodo-640-CCA-KEM-\textsf{Encaps}} & \multirow{2}{*}{-} & \multirow{2}{*}{3,317,760} & \multirow{2}{*}{-} \\
 & & & & & & & & \\
\hline
\multirow{2}{1.4cm}{Fritzmann et al. \cite{sepulveda_pqriscv_2019}} & \multirow{2}{*}{FPGA} & \multirow{2}{*}{-} & \multirow{2}{*}{-} & \multirow{2}{*}{-} & \multirow{2}{*}{NewHope-1024-CPA-PKE-\textsf{Encrypt}} & \multirow{2}{*}{-} & \multirow{2}{*}{589,285} & \multirow{2}{*}{-} \\
 & & & & & & & & \\
\hline
\multirow{2}{1.4cm}{Hutter $\,\,\,\,$ et al. \cite{hutter_nacl_2015} \textsuperscript{$\dagger$}} & \multirow{2}{*}{ASIC} & \multirow{2}{*}{130} & \multirow{2}{*}{1.2} & \multirow{2}{*}{1} & \multirow{2}{*}{Curve25519-\textsf{ECDHE}} & \multirow{2}{*}{50} & \multirow{2}{*}{1,622,354} & \multirow{2}{*}{113.56} \\
 & & & & & & & & \\
\hline
\multirow{2}{1.4cm}{Banerjee $\,$ et al. \cite{banerjee_isscc_2018}} & \multirow{2}{*}{ASIC} & \multirow{2}{*}{65} & \multirow{2}{*}{1.2} & \multirow{2}{*}{20} & NIST-P256-\textsf{ECDHE} & \multirow{2}{*}{149} & 680,000 & 24.07 \\
 & & & & & NIST-P256-\textsf{ECDSA-Sign} & & 180,000 & 6.48 \\
\hline
\multicolumn{9}{l}{\textsuperscript{$\dagger$} Only post-synthesis area and energy consumption reported} \\
\end{tabular}
\end{table}

\subsection{Side-Channel Analysis}

Side-channel security is an important aspect of all public-key cryptography implementations and lattice-based cryptography is not an exception. In order to prevent information leakage through timing side channels, the most important requirement is to ensure that the timing and memory access patterns of underlying computations are independent of the secret data being computed upon. In our implementation, this is achieved either by making the computations constant-time, e.g., binomial sampling, discrete Gaussian sampling, NTT and polynomial arithmetic, or by using rejection sampling, e.g, sampling numbers from $[0, q)$ or $[-\eta, \eta]$ or probabilistic rejection during signature schemes. Since our cryptographic core and RISC-V processor both have a single-level memory hierarchy, the possibility of cache timing attacks is also eliminated.

\begin{figure}[!t]
\centering
\includegraphics[width=6.0in]{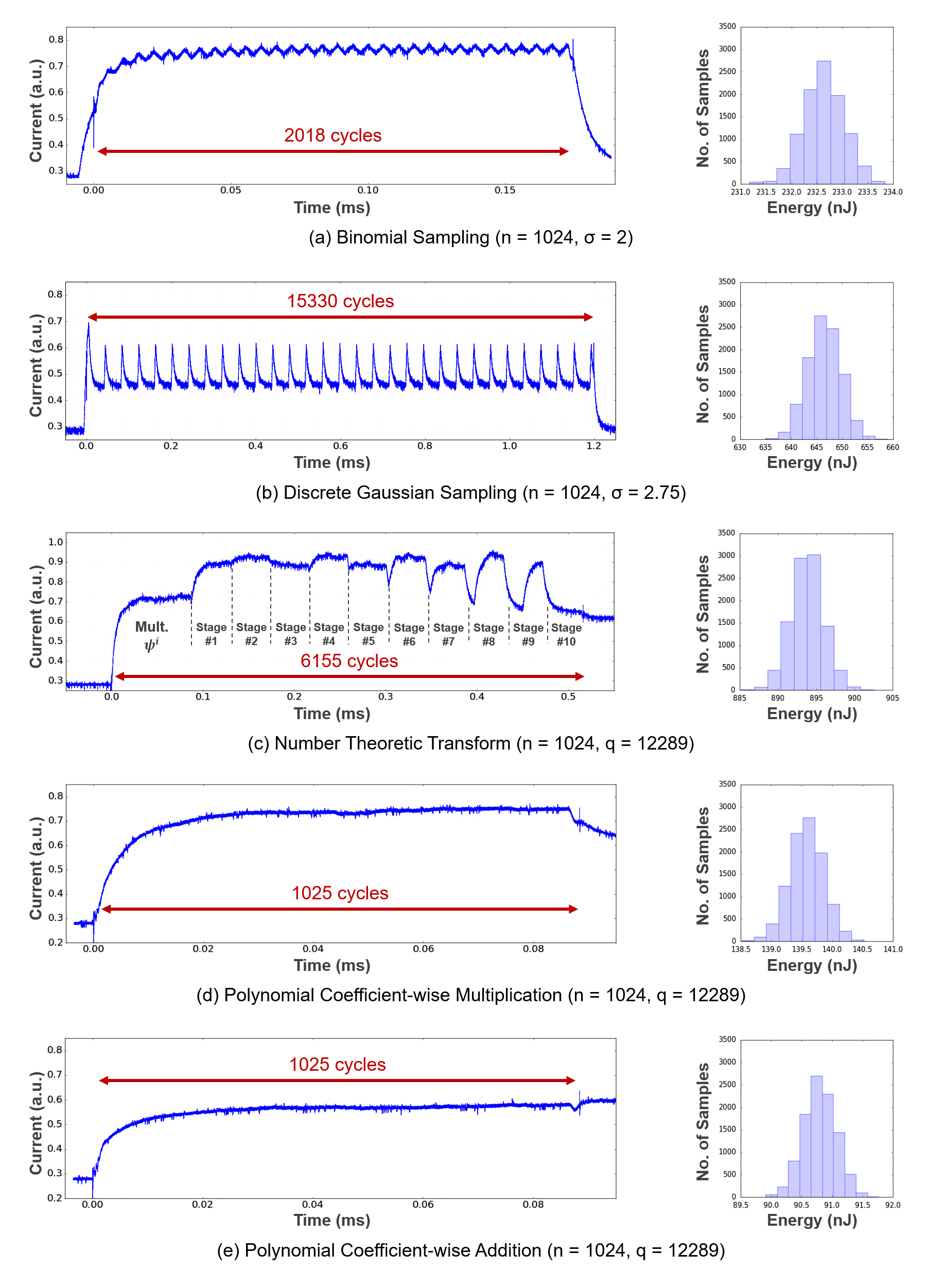}
\caption{Measured power waveforms for different polynomial sampling, transform and arithmetic operations along with histograms of energy consumption for 10,000 measurements for each operation, obtained from our test chip operating at 1.1 V and 12 MHz.}
\label{sca_waveforms}
\end{figure}

\begin{figure}[!t]
\centering
\includegraphics[width=5.0in]{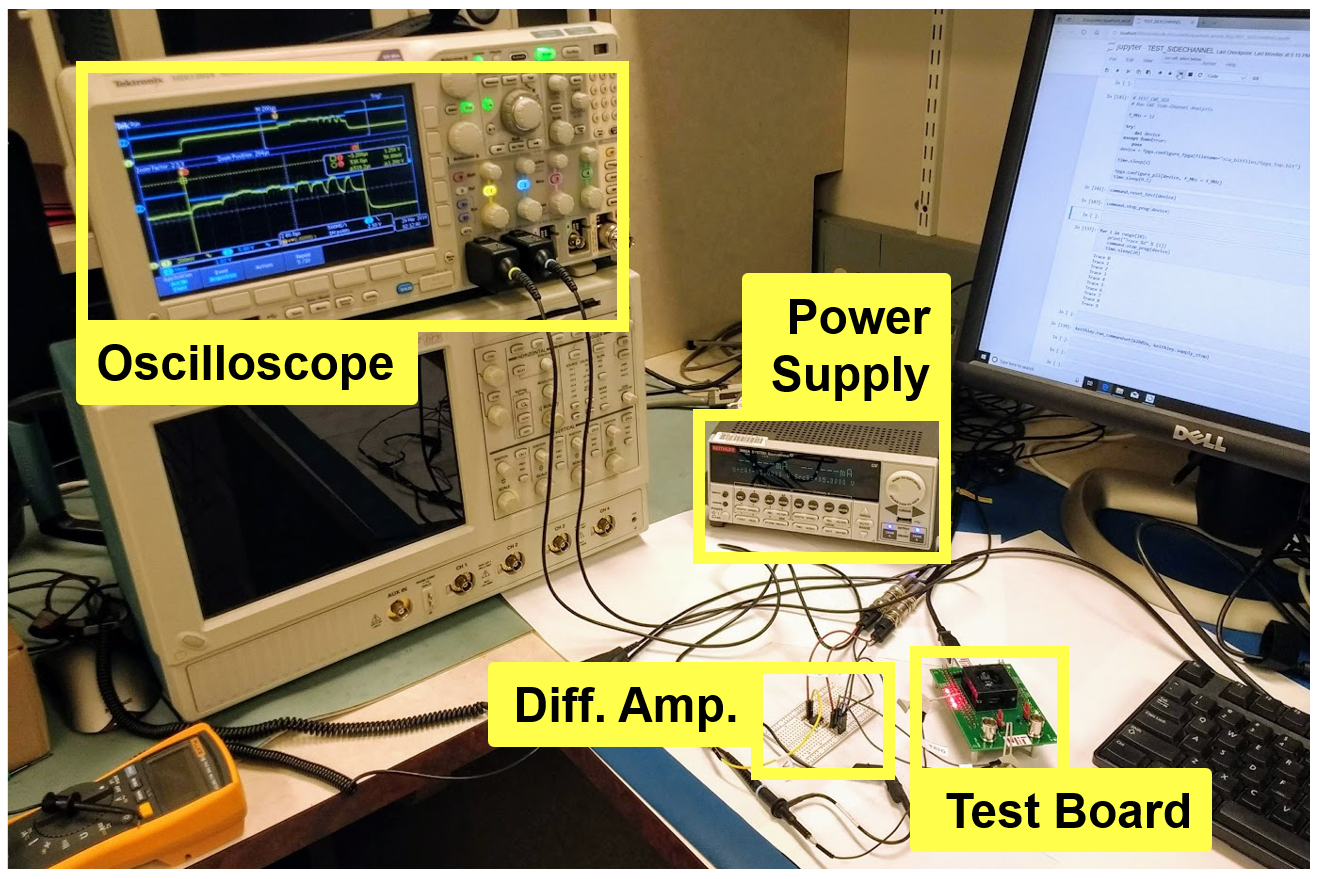}
\caption{Power side-channel measurement setup.}
\label{sca_setup}
\end{figure}

Our power side-channel measurement setup is shown in Fig. \ref{sca_setup}. Our test board has an 18 $\Omega$ resistor connected in series between the power supply and the VDD pin of our test chip. The voltage across this resistor, proportional to the chip's current draw, is magnified using a non-inverting differential amplifier (consists of an AD8001 op-amp chip, with 6~dB flat gain up to 100~MHz, in the non-inverting configuration with resistors of appropriate sizes) and then observed through a 2.5 GS/s Tektronix MDO3024 mixed domain oscilloscope.

The execution times of binomial sampling, discrete Gaussian sampling, NTT, polynomial coefficient-wise multiplication and addition (with $n = 1024$ and $q = 12289$) were measured for 10,000 random executions to verify that these computations are indeed constant-time. The corresponding power waveforms and energy consumption histograms, measured from our test chip operating at 1.1 V and 12 MHz, are shown in Fig. \ref{sca_waveforms}.


Typical simple power analysis (SPA) attacks on lattice cryptography implementations exploit information leakage through conditional branching or data-dependent execution times during the modular arithmetic computations in NTT or polynomial coefficient-wise multiplication \cite{park_spa_2016, primas_sca_2017, aysu_brlwe_2018}. As explained in Fig. \ref{sca_waveforms}, our implementation of polynomial arithmetic is constant-time. To quantitatively evaluate SPA resistance of our design, we perform a difference-of-means test \cite{kocher_dpa_2011, aysu_brlwe_2018, ebrahimi_pqiot_2019} on three polynomial operations -- NTT, coefficient-wise multiplication and coefficient-wise addition -- which are traditionally used as attack points. In this test, we try to differentiate two sets of measurements -- those with a particular coefficient (`0'-th coefficient in our case) in the input polynomial set to 0 (denoted as set `0' or $S_0$) versus the same coefficient set to $q-1$ (denoted as set `1' or $S_1$) -- by comparing their means separately for each point in the mean power trace. The difference-of-means is calculated for increasing number of measurements and plotted as a function of the number of traces $N$. The corresponding 99.99\% confidence interval for having a zero difference of means between these two sets is calculated as $t_c \cdot \sqrt{(\sigma_0^2 + \sigma_1^2)/N}$, where $\sigma_0$ and $\sigma_1$ are the standard deviations of the two sets $S_0$ and $S_1$ respectively and $t_c$ is the critical $t$-statistic for $N-1$ degrees of freedom and cumulative probability $= 1 - (1 - 0.9999)/2 = 0.99995$. As long as the absolute difference-of-means is smaller than the confidence interval, it is a strong indicator that the sets $S_0$ and $S_1$ are indistinguishable.

\begin{figure}[!b]
\centering
\includegraphics[width=6.0in]{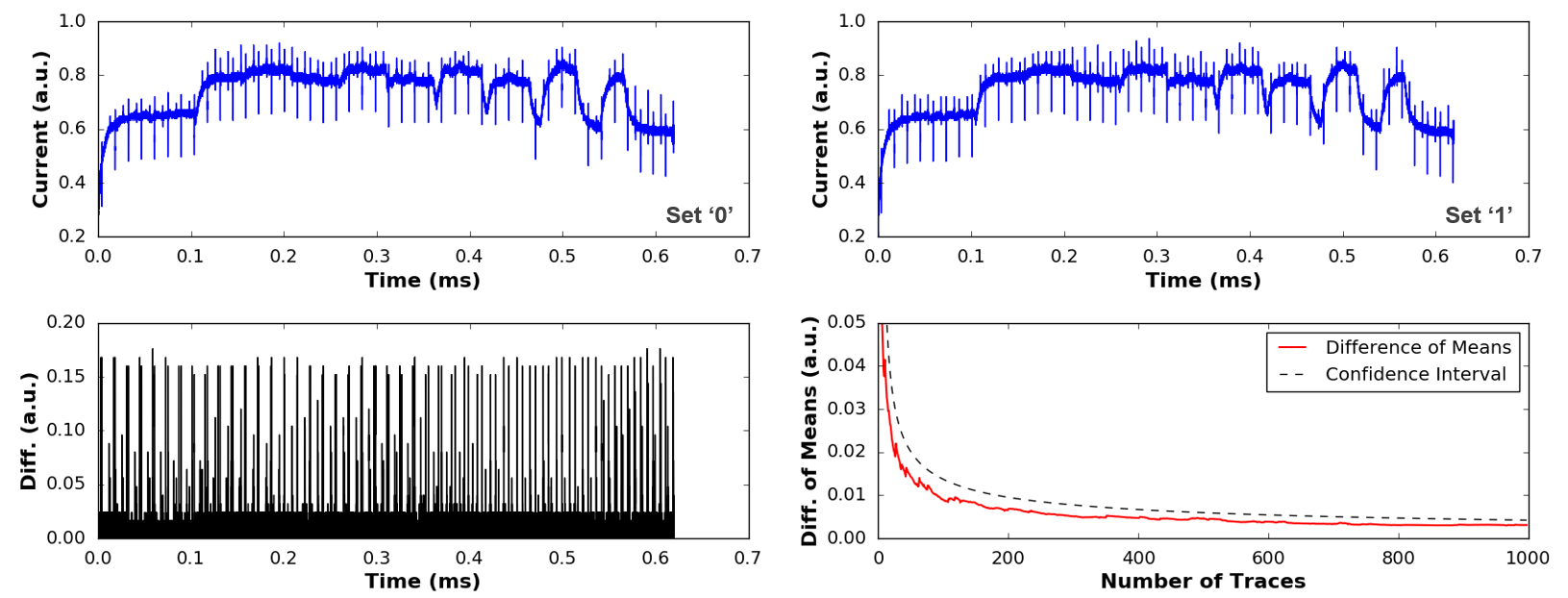}
\caption{Difference-of-means test for polynomial NTT with representative power traces from set $S_0$ (top left) and $S_1$ (top right), difference waveform (bottom left) and difference of means versus number of traces with 99.99\% confidence interval (bottom right).}
\label{spa_ntt}
\end{figure}

\begin{figure}[!t]
\centering
\includegraphics[width=6.0in]{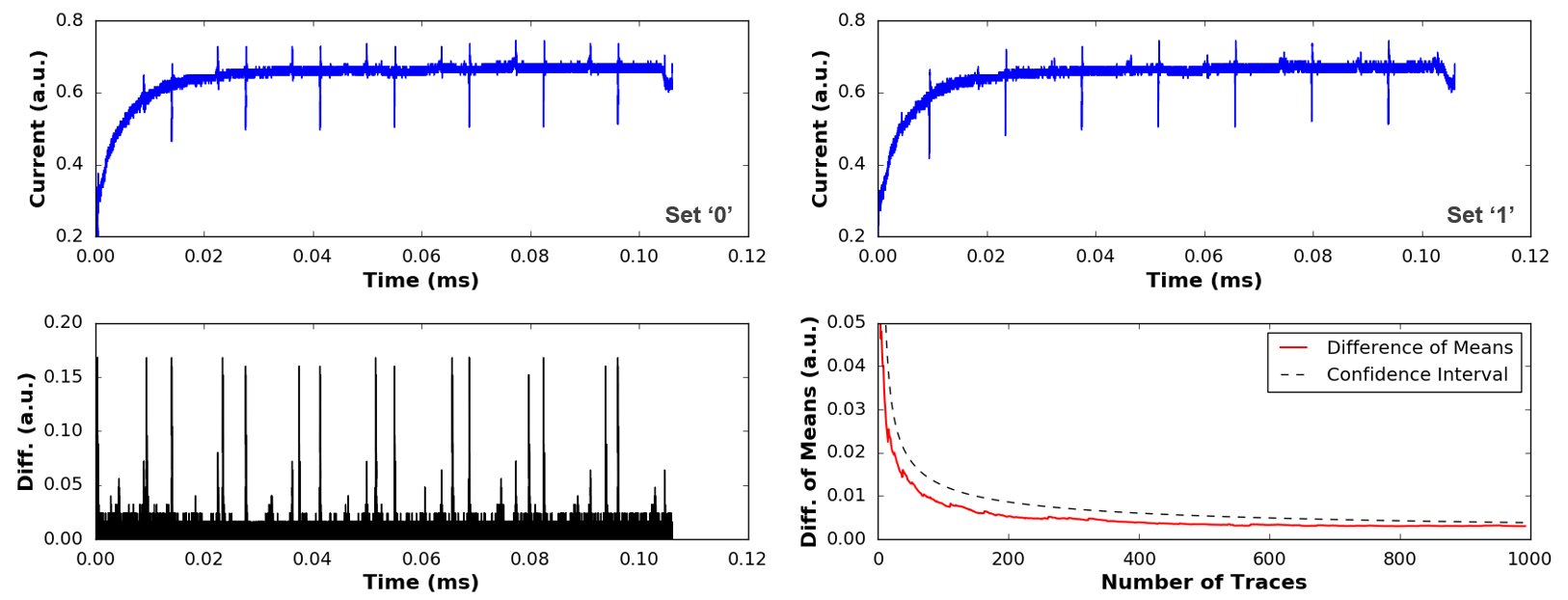}
\caption{Difference-of-means test for polynomial coefficient-wise multiplication with representative power traces from set $S_0$ (top left) and $S_1$ (top right), difference waveform (bottom left) and difference of means versus number of traces with 99.99\% confidence interval (bottom right).}
\label{spa_mul}
\end{figure}

\begin{figure}[!t]
\centering
\includegraphics[width=6.0in]{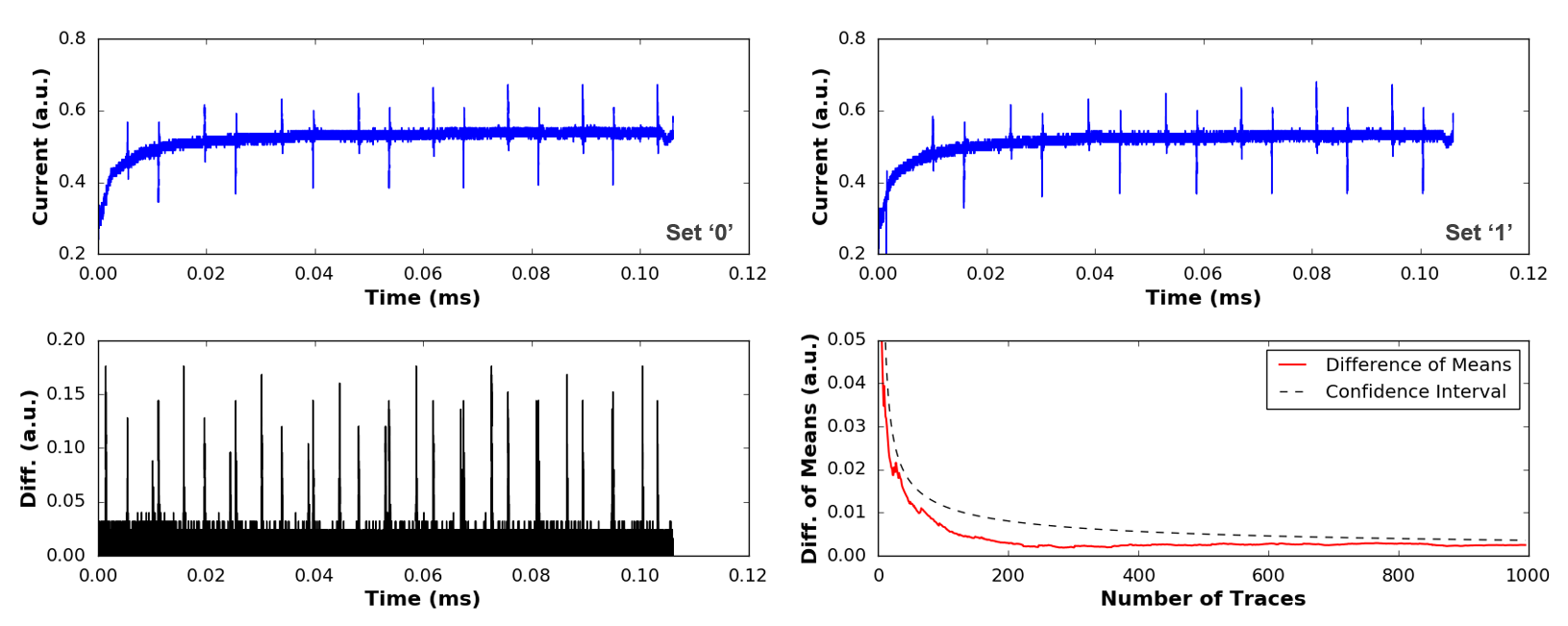}
\caption{Difference-of-means test for polynomial coefficient-wise addition with representative power traces from set $S_0$ (top left) and $S_1$ (top right), difference waveform (bottom left) and difference of means versus number of traces with 99.99\% confidence interval (bottom right).}
\label{spa_add}
\end{figure}

In Figures \ref{spa_ntt}, \ref{spa_mul} and \ref{spa_add}, we provide preliminary difference-of-means test results for three polynomial operations (with $n = 1024$ and $q = 12289$) as measured from our test chip operating at 1.1 V and 10 MHz. Sampling rate of the oscilloscope was set to 500 MS/s for NTT and 2.5 GS/s for coefficient-wise multiplication and addition. The red lines denote measured difference-of-means, and the dashed lines mark the 99.99\% confidence interval for ideal zero difference-of-means. These results validate that our design is secure against SPA side-channel attacks.

\begin{figure}[!b]
\centering
\includegraphics[width=5.5in]{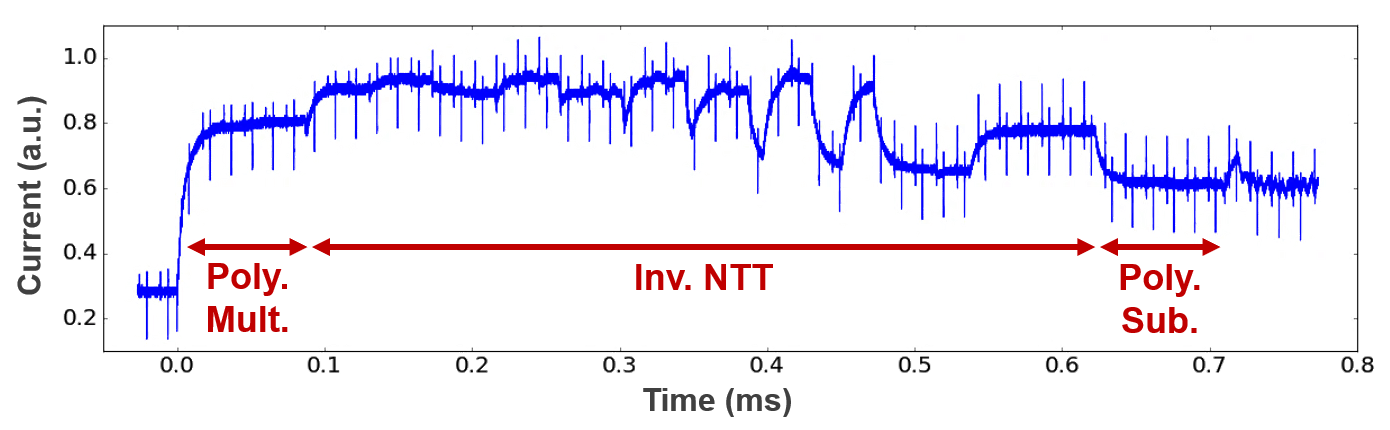}
\caption{Power trace for the NewHope-1024-\textsf{CPA-PKE.Decrypt} algorithm, measured from our test chip operating at 1.1 V and 12 MHz.}
\label{cpapke_waveform}
\end{figure}

The protocol implementations discussed earlier do not have any explicit countermeasures against differential power analysis (DPA) attacks. Although DPA attacks can be mitigated by using ephemeral keys, it is still important to analyze how these protocols can be made DPA-secure. Masking-based countermeasures have been proposed in \cite{ingrid_masked_2015, ingrid_masked_2016, guneysu_masked_2018} for Ring-LWE encryption. Since our crypto-processor is programmable, such masked protocols can be implemented using the right mix of software and hardware acceleration. For example, we consider NewHope-CPA-PKE and discuss how the masked decryption algorithm, inspired by \cite{ingrid_masked_2015, ingrid_masked_2016, guneysu_masked_2018}, can be implemented using our hardware. A simplified version of the CPA-PKE scheme, excluding any key / ciphertext compression / decompression and encoding / decoding and implementation-specific details, is provided below: \\ \\
\textbf{function} NewHope-\textsf{CPA-PKE.KeyGen}($seed$): \\
\hspace*{5mm} \textsf{Sample} $\hat{a}, s, e \in R_q$ \\
\hspace*{5mm} $\hat{b} \leftarrow \hat{a} \odot \hat{s} + \hat{e}$ \\
\hspace*{5mm} \textbf{return} ($pk = (\hat{a}, \hat{b})$, $sk = \hat{s}$) \\
\\
\textbf{function} NewHope-\textsf{CPA-PKE.Encrypt}($pk, coin, \mu \in \{0, \cdots, 255\}^{32}$): \\
\hspace*{5mm} \textsf{Sample} $s', e', e'' \in R_q$ \\
\hspace*{5mm} $\hat{u} \leftarrow \hat{a} \odot \hat{s}' + \hat{e}'$ \\
\hspace*{5mm} $v \leftarrow$ \textsf{Encode}$(\mu) \in R_q$ \\
\hspace*{5mm} $v' \leftarrow b \cdot s' + e'' + v$ \\
\hspace*{5mm} \textbf{return} $c = (\hat{u}, v')$ \\
\\
\textbf{function} NewHope-\textsf{CPA-PKE.Decrypt}($sk, c$): \\
\hspace*{5mm} $v'' \leftarrow v' - u \cdot s$ \\
\hspace*{5mm} $\mu \leftarrow$ \textsf{Decode}$(v'') \in \{0, \cdots, 255\}^{32}$ \\
\hspace*{5mm} \textbf{return} $\mu$ \\ \\
where $\mu$ is the 32-byte message to be encrypted, $\hat{x}$ is the NTT representation of polynomial $x \in R_q$, $\odot$ denotes coefficient-wise multiplication (in the transform domain) and $\cdot$ denotes polynomial multiplication in $R_q$. The \textsf{Encode} function converts message $\mu$ into a polynomial in $R_q$. To allow robustness against errors, each bit of the 256-bit message is encoded into $\lfloor n/256 \rfloor$ coefficients. For example, for $n = 1024$, the $i$-th, $(256+i)$-th, $(512+i)$-th and $(768+i)$-th coefficients are set to 0 or $\lfloor q/2 \rfloor$ depending on whether the $i$-th bit in $\mu$ is 0 or 1 respectively, for $i \in \{0, \cdots, 255\}$. The \textsf{Decode} function maps $\lfloor n/256 \rfloor$ coefficients of a polynomial back to the original message bit. For example, for $n = 1024$, it takes the $i$-th, $(256+i)$-th, $(512+i)$-th and $(768+i)$-th coefficients (each in the range $\{0, \cdots, q-1\}$, subtracts $\lfloor q/2 \rfloor$ from each of them, accumulates their absolute values, and finally sets the $i$-th message bit to 0 if the sum is larger than $q$ or to 1 otherwise, for $i \in \{0, \cdots, 255\}$. Further details about these functions are available in the NewHope specification document \cite{alkim_newhope_2019}. The \textsf{Decrypt} algorithm requires one polynomial coefficient-wise multiplication $\hat{u} \odot \hat{s}$, one inverse NTT (including multiplication with $n^{-1}\psi^{-i}$) to compute $u \cdot s$, and one polynomial coefficient-wise subtraction $v' - u \cdot s$. Figure \ref{cpapke_waveform} shows the corresponding measured power waveform for $n = 1024$.

Similar to the encryption scheme studied in \cite{ingrid_masked_2016}, we note that NewHope-CPA-PKE is also additively homomorphic, that is, if $c_1 = (\hat{u}_1, v'_1)$ and $c_2 = (\hat{u}_2, v'_2)$ are the ciphertexts corresponding to messages $\mu_1$ and $\mu_2$ respectively, under the same key-pair, then $(\hat{u}_1 + \hat{u}_2, v'_1 + v'_2)$ will be the ciphertext corresponding to $\mu_1 \oplus \mu_2$. Following the works of \cite{ingrid_masked_2015, ingrid_masked_2016, guneysu_masked_2018}, this property can be exploited to randomize the decryption algorithm (as a first-order DPA countermeasure) as explained below:
\begin{enumerate}
\itemsep0em
\item Generate a secret random message $\mu_r$
\item Encrypt $\mu_r$ to its corresponding ciphertext $c_r = (\hat{u}_r, v'_r)$
\item Compute $c_m = (\hat{u} + \hat{u}_r, v' + v'_r)$, where $c = (\hat{u}, v')$ is the original ciphertext
\item Decrypt masked ciphertext $c_m$ to obtain $\mu_m = \mu \oplus \mu_r$, where $\mu$ is the original message
\item Recover original message $\mu = \mu_m \oplus \mu_r$
\end{enumerate}
Therefore, the masked decryption now requires generation of a random message along with invocations of both the \textsf{Encrypt} and \textsf{Decrypt} functions. As explained earlier, these functions can be implemented entirely using Sapphire custom programs, so the masking involves minimal software overheads. Referring to the cycle counts and energy consumption of NewHope-1024-CPA-PKE in Table \ref{table:cpapke_results}, we note that the masked decryption is about $3\times$ less efficient compared to the unmasked version, both in terms of energy and performance. Since $\mu_r$ is independent from the original message $\mu$, the ciphertext $c_r$ can be pre-computed offline in order to reduce online computation time and energy consumption. As explained in \cite{ingrid_masked_2016}, this technique does not require any modifications to the \textsf{Decode} function. However, addition of ciphertexts increases the noise in them, thus increasing the decryption failure rate. Each of the two polynomials in the ciphertext contains one noise term whose coefficients are derived from the zero-mean binomial distribution with support $[-k, k]$ and standard deviation $\sigma = \sqrt{k/2}$ ($k = 8$ for NewHope). When two such ciphertexts are added, the resulting noise distribution (still binomial) now has support $[-2k, 2k]$ with standard deviation $\sigma = \sqrt{2k/2} = \sqrt{k}$, that is, the noise variance is doubled. For $k = 16$, which is also used in NewHope-Simple, the decryption failure probability will go up from $2^{-216}$ \cite{alkim_newhope_2019} to $2^{-60}$ \cite{alkim_newhopesimple_2016}. As discussed in \cite{ingrid_masked_2016}, standard deviation of the error distribution can be decreased to allow correct decryptions at the cost of a minor deterioration in security. So, one possibility is to set $k = 4$ in the unmasked scheme (so that $k = 8$ for masked decryption and failure probability remains $2^{-216}$). The corresponding decrease in security level is from 289 bits to 268 bits, as obtained from the LWE hardness estimator \cite{player_hardness_2015} using the following Sage module: \\ \\
\texttt{load("https://bitbucket.org/malb/lwe-estimator/raw/HEAD/estimator.py")} \\
\texttt{n = 1024; q = 12289; stddev = sqrt(4/2); alpha = sqrt(2*pi)*stddev/q} \\
\texttt{\_ = estimate\_lwe(n, alpha, q, reduction\_cost\_model=BKZ.sieve)}

%% file: body/07_conclusion.tex
\section{Conclusion and Future Work}
\label{sec:conclusion}

In this work, we have presented a configurable lattice cryptography processor supporting different parameters for NIST Round 2 lattice-based key encapsulation and digital signature protocols such as NewHope, qTESLA, CRYSTALS-Kyber, CRYSTALS-Dilithium and Frodo. Efficient modular arithmetic, sampling and NTT memory architectures together provide an order of magnitude improvement in performance and energy-efficiency compared to state-of-the-art software and hardware implementations. Our ASIC implementation was fabricated in a 40nm low-power CMOS process and all measurement results are obtained from our test chip operating at 1.1 V and 72 MHz. Our protocol implementations are secure against timing and simple power analysis attacks, and we also discuss how masking countermeasures against differential power analysis can be implemented using the programmability of our crypto-processor.

Since our design supports configurable lattice parameters, it will be interesting to explore other lattice-based protocols such as Saber \cite{vercauteren_saber_2019} and Round5 \cite{saarinen_round5_2019}, which are based on the LWR (learning with rounding) problem \cite{peikert_lwr_2012}. More concrete analysis of DPA-secure masked implementations, for CPA-PKE, CCA-KEM and signature schemes, along with leakage tests and impact on performance and energy-efficiency, will also be performed in the future. Finally, non-lattice-based post-quantum protocols can also be implemented on our platform, using a mix of hardware acceleration and software, since they can still benefit from our efficient implementation of modular arithmetic and SHA-3 computations.

%% file: body/appendix_a.tex
\section{Modular Reduction Parameters}
\label{sec:appendix_a}

As mentioned in Section \ref{sec:ntt}, our modular multiplier with pseudo-configurable prime modulus uses efficient Barrett reduction, with the parameters $m$, $k$ and $q$ coded in digital logic, for a set of chosen primes. These parameters and the corresponding reduction implementations are detailed here. Please note that $m$ and $q$ are written in the form $2^{l_1} \pm 2^{l_2} \pm \cdots \pm 1$ only when the number of such integers $l_1, l_2, \cdots$ is less than 5.

\begin{algorithm}[ht]
\caption{Reduction $\text{mod} \; 7681$}
\begin{algorithmic}[1]
\REQUIRE $q = 2^{13} - 2^{9} + 1, m = 273 = 2^{8} + 2^{4} + 1, k = 21, x \in [0, q^2)$
\ENSURE $z = x \; \text{mod} \; q$
\STATE $t \leftarrow (x \ll 8) + (x \ll 4) + x$
\STATE $t \leftarrow t \gg 21$
\STATE $t \leftarrow (t \ll 13) - (t \ll 9) + t$
\STATE $z \leftarrow x - t$
\IF{$z \ge q$}
\STATE $z \leftarrow z - q$
\ENDIF
\RETURN $z$
\end{algorithmic}
\end{algorithm}

\begin{algorithm}[ht]
\caption{Reduction $\text{mod} \; 12289$}
\begin{algorithmic}[1]
\REQUIRE $q = 2^{13} + 2^{12} + 1, m = 10921, k = 27, x \in [0, q^2)$
\ENSURE $z = x \; \text{mod} \; q$
\STATE $t \leftarrow 10921 \cdot x$
\STATE $t \leftarrow t \gg 27$
\STATE $t \leftarrow (t \ll 13) + (t \ll 12) + t$
\STATE $z \leftarrow x - t$
\IF{$z \ge q$}
\STATE $z \leftarrow z - q$
\ENDIF
\RETURN $z$
\end{algorithmic}
\end{algorithm}

\begin{algorithm}[ht]
\caption{Reduction $\text{mod} \; 40961$}
\begin{algorithmic}[1]
\REQUIRE $q = 2^{15} + 2^{13} + 1, m = 52427, k = 31, x \in [0, q^2)$
\ENSURE $z = x \; \text{mod} \; q$
\STATE $t \leftarrow 52427 \cdot x$
\STATE $t \leftarrow t \gg 31$
\STATE $t \leftarrow (t \ll 15) + (t \ll 13) + t$
\STATE $z \leftarrow x - t$
\IF{$z \ge q$}
\STATE $z \leftarrow z - q$
\ENDIF
\RETURN $z$
\end{algorithmic}
\end{algorithm}

\begin{algorithm}[ht]
\caption{Reduction $\text{mod} \; 120833$}
\begin{algorithmic}[1]
\REQUIRE $q = 2^{17} - 2^{14} + 2^{13} - 2^{11} + 1, m = 71089, k = 33, x \in [0, q^2)$
\ENSURE $z = x \; \text{mod} \; q$
\STATE $t \leftarrow 71089 \cdot x$
\STATE $t \leftarrow t \gg 33$
\STATE $t \leftarrow (t \ll 17) - (t \ll 14) + (t \ll 13) - (t \ll 11) + t$
\STATE $z \leftarrow x - t$
\IF{$z \ge q$}
\STATE $z \leftarrow z - q$
\ENDIF
\RETURN $z$
\end{algorithmic}
\end{algorithm}

\begin{algorithm}[ht]
\caption{Reduction $\text{mod} \; 133121$}
\begin{algorithmic}[1]
\REQUIRE $q = 2^{17} + 2^{11} + 1, m = 64527 = 2^{16} - 2^{10} + 2^{4} - 1, k = 33, x \in [0, q^2)$
\ENSURE $z = x \; \text{mod} \; q$
\STATE $t \leftarrow (x \ll 16) - (x \ll 10) + (x \ll 4) - x$
\STATE $t \leftarrow t \gg 33$
\STATE $t \leftarrow (t \ll 17) + (t \ll 11) + t$
\STATE $z \leftarrow x - t$
\IF{$z \ge q$}
\STATE $z \leftarrow z - q$
\ENDIF
\RETURN $z$
\end{algorithmic}
\end{algorithm}

\begin{algorithm}[ht]
\caption{Reduction $\text{mod} \; 184321$}
\begin{algorithmic}[1]
\REQUIRE $q = 2^{17} + 2^{15} + 2^{14} + 2^{12} + 1, m = 46603, k = 33, x \in [0, q^2)$
\ENSURE $z = x \; \text{mod} \; q$
\STATE $t \leftarrow 46603 \cdot x$
\STATE $t \leftarrow t \gg 33$
\STATE $t \leftarrow (t \ll 17) + (t \ll 15) + (t \ll 14) + (t \ll 12) + t$
\STATE $z \leftarrow x - t$
\IF{$z \ge q$}
\STATE $z \leftarrow z - q$
\ENDIF
\RETURN $z$
\end{algorithmic}
\end{algorithm}

\begin{algorithm}[ht]
\caption{Reduction $\text{mod} \; 8380417$}
\begin{algorithmic}[1]
\REQUIRE $q = 2^{23} - 2^{13} + 1, m = 8396807 = 2^{23} + 2^{13} + 2^{3} - 1, k = 46, x \in [0, q^2)$
\ENSURE $z = x \; \text{mod} \; q$
\STATE $t \leftarrow (x \ll 23) + (x \ll 13) + (x \ll 3) - x$
\STATE $t \leftarrow t \gg 46$
\STATE $t \leftarrow (t \ll 23) - (t \ll 13) + t$
\STATE $z \leftarrow x - t$
\IF{$z \ge q$}
\STATE $z \leftarrow z - q$
\ENDIF
\RETURN $z$
\end{algorithmic}
\end{algorithm}

\begin{algorithm}[ht]
\caption{Reduction $\text{mod} \; 8058881$}
\begin{algorithmic}[1]
\REQUIRE $q = 8058881, m = 8731825, k = 46, x \in [0, q^2)$
\ENSURE $z = x \; \text{mod} \; q$
\STATE $t \leftarrow 8731825 \cdot x$
\STATE $t \leftarrow t \gg 46$
\STATE $t \leftarrow 8058881 \cdot t$
\STATE $z \leftarrow x - t$
\IF{$z \ge q$}
\STATE $z \leftarrow z - q$
\ENDIF
\RETURN $z$
\end{algorithmic}
\end{algorithm}

\begin{algorithm}[ht]
\caption{Reduction $\text{mod} \; 4205569$}
\begin{algorithmic}[1]
\REQUIRE $q = 2^{22} + 2^{13} + 2^{11} + 2^{10} + 1, m = 4183069, k = 44, x \in [0, q^2)$
\ENSURE $z = x \; \text{mod} \; q$
\STATE $t \leftarrow 4183069 \cdot x$
\STATE $t \leftarrow t \gg 44$
\STATE $t \leftarrow (t \ll 22) + (t \ll 13) + (t \ll 11) + (t \ll 10) + t$
\STATE $z \leftarrow x - t$
\IF{$z \ge q$}
\STATE $z \leftarrow z - q$
\ENDIF
\RETURN $z$
\end{algorithmic}
\end{algorithm}

\begin{algorithm}[ht]
\caption{Reduction $\text{mod} \; 4206593$}
\begin{algorithmic}[1]
\REQUIRE $q = 2^{22} + 2^{13} + 2^{12} + 1, m = 2091025 = 2^{21} - 2^{13} + 2^{11} + 2^{4} + 1, k = 43, x \in [0, q^2)$
\ENSURE $z = x \; \text{mod} \; q$
\STATE $t \leftarrow (x \ll 21) - (x \ll 13) + (x \ll 11) + (x \ll 4) + x$
\STATE $t \leftarrow t \gg 43$
\STATE $t \leftarrow (t \ll 22) + (t \ll 13) + (t \ll 12) + t$
\STATE $z \leftarrow x - t$
\IF{$z \ge q$}
\STATE $z \leftarrow z - q$
\ENDIF
\RETURN $z$
\end{algorithmic}
\end{algorithm}

\begin{algorithm}[ht]
\caption{Reduction $\text{mod} \; 8404993$}
\begin{algorithmic}[1]
\REQUIRE $q = 2^{23} + 2^{14} + 1, m = 4186127 = 2^{22} - 2^{13} + 2^{4} - 1, k = 45, x \in [0, q^2)$
\ENSURE $z = x \; \text{mod} \; q$
\STATE $t \leftarrow (x \ll 22) - (x \ll 13) + (x \ll 4) - x$
\STATE $t \leftarrow t \gg 45$
\STATE $t \leftarrow (t \ll 23) + (t \ll 14) + t$
\STATE $z \leftarrow x - t$
\IF{$z \ge q$}
\STATE $z \leftarrow z - q$
\ENDIF
\RETURN $z$
\end{algorithmic}
\end{algorithm}

\clearpage

For the prime $q = 65537 = 2^{16} + 1$, we employ an easier reduction technique owing to the special structure of $q$. Any integer $x \in [0, q^2)$ can be written as $x = x_2 2^{32} + x_1 2^{16} + x_0$ where $x_0$ and $x_1$ are 16-bit numbers and $x_2 \in \{0, 1\}$. Since $2^{16} \equiv -1 \, \text{mod} \, q$, we have $x \equiv x_0 - x_1 + x_2 \, \text{mod} \, q$, which must be followed by a conditional addition to bring back the result to $[0,q)$.

\begin{algorithm}[ht]
\caption{Reduction $\text{mod} \; 65537$}
\begin{algorithmic}[1]
\REQUIRE $q = 2^{16} + 1, x = x_2 2^{32} + x_1 2^{16} + x_0 \in [0, q^2)$
\ENSURE $z = x \; \text{mod} \; q$
\STATE $z \leftarrow x_0 - x_1 + x_2$
\IF{$z < 0$}
\STATE $z \leftarrow z + q$
\ENDIF
\RETURN $z$
\end{algorithmic}
\end{algorithm}

%% file: body/appendix_b.tex
\section{Custom Instruction Set Summary}
\label{sec:appendix_b}

In this section, we briefly describe all the custom instructions supported by our crypto-processor. Apart from the polynomials stored in its memory and the 256-bit seed registers \texttt{r0} and \texttt{r1}, these are the core internal registers that can also be manipulated:
\begin{itemize}
\itemsep0em
\item 24-bit temporary registers \texttt{reg} and \texttt{tmp}
\item 16-bit counter registers \texttt{c0} and \texttt{c1}
\item 2-bit \texttt{flag} register to store comparison results (-1, 0 or +1)
\end{itemize}
Following is the list of instructions along with short descriptions:

\begin{table}[!h]
\renewcommand{\arraystretch}{1.25}
\label{table:skywalker_isa_1}
\centering
\begin{tabular}{|p{13cm}|}
\hline
\textbf{Configuration:} set parameters and clock gates \\
\texttt{config (n, q)} \\
\texttt{clock\_config (keccak, ntt, sampler)} \\
\hline
\textbf{Register Operations:} register assignments and arithmetic \\
\texttt{c0 = \#VAL / c0 + \#VAL / c0 - \#VAL} \\
\texttt{c1 = \#VAL / c1 + \#VAL / c1 - \#VAL} \\
\texttt{reg = \#VAL / tmp} \\
\texttt{tmp = \#VAL / tmp (OP) reg} \\
where \texttt{\#VAL} can be any unsigned integer of appropriate size, and \texttt{(OP)} is one of the following operations: \texttt{\{ADD, SUB, MUL, AND, OR, XOR, RSHIFT, LSHIFT\}} \\
\hline
\textbf{Register-Polynomial Operations:} register and polynomial interactions \\
\texttt{reg = max\_elems (poly)} \\
\texttt{reg = sum\_elems (poly)} \\
\texttt{reg = (poly)[\#VAL] / (poly)[c0] / (poly)[c1]} \\
\texttt{(poly)[\#VAL] / (poly)[c0] / (poly)[c1] = reg} \\
\hline
\textbf{Transforms:} number theoretic transform and related computations \\
\texttt{transform (mode, poly\_dst, poly\_src)} \\
\texttt{mult\_psi (poly) / mult\_psi\_inv (poly)} \\
where \texttt{mode} is one of the following: \texttt{\{DIF\_NTT, DIF\_INTT, DIT\_NTT, DIT\_INTT\}} \\
\hline
\end{tabular}
\end{table}

\clearpage

\begin{table}[!hbt]
\renewcommand{\arraystretch}{1.25}
\label{table:skywalker_isa_2}
\centering
\begin{tabular}{|p{13cm}|}
\hline
\textbf{Sampling:} polynomial sampling from various distributions \\
\texttt{bin\_sample (prng, seed, c0, c1, k, poly)} \\
\texttt{cdt\_sample (prng, seed, c0, c1, r, s, poly)} \\
\texttt{rej\_sample (prng, seed, c0, c1, poly)} \\
\texttt{uni\_sample (prng, seed, c0, c1, eta, bitlen, poly)} \\
\texttt{tri\_sample\_1 (prng, seed, c0, c1, m, poly)} \\
\texttt{tri\_sample\_2 (prng, seed, c0, c1, m0, m1, poly)} \\
\texttt{tri\_sample\_3 (prng, seed, c0, c1, rho, poly)} \\
where \texttt{prng} can be \texttt{SHAKE-128} or \texttt{SHAKE-256}, \texttt{seed} can be \texttt{r0} or \texttt{r1}, and \texttt{k, r, s, eta, bitlen, m, m0, m1, rho} are the distribution parameters \\
\hline
\textbf{Polynomial Computations:} polynomial initialization and other operations \\
\texttt{init (poly)} \\
\texttt{poly\_copy (poly\_dst, poly\_src)} \\
\texttt{poly\_op (op, poly\_dst, poly\_src)} \\
\texttt{shift\_poly (ring, poly\_dst, poly\_src)} \\
where \texttt{op} can be one of the following: \texttt{\{ADD, SUB, MUL, BITREV, CONST\_ADD, CONST\_SUB, CONST\_MUL, CONST\_AND, CONST\_OR, CONST\_XOR, CONST\_RSHIFT, CONST\_LSHIFT\}}, and \texttt{ring} can be either \texttt{x\string^N+1} or \texttt{x\string^N-1} \\
\hline
\textbf{Comparison and Branching:} simple branching operations \\
\texttt{flag = eq\_check (poly, poly)} \\
\texttt{flag = inf\_norm\_check (poly, bound)} \\
\texttt{flag = compare (reg / tmp / c0 / c1, \#VAL)} \\
\texttt{if (flag == / != -1 / 0 / +1) goto <label>} \\
where the \texttt{flag} register stores -1, 0 and +1 for the register comparison result being ``lesser than'', ``equal to'' and ``greater than'' respectively, and it stores 1 or 0 depending on whether the equality check and infinity norm check has passed or failed respectively \\
\hline
\textbf{SHA-3 Computations:} hashing operations \\
\texttt{sha3\_init} \\
\texttt{sha3\_256\_absorb (poly)} \\
\texttt{sha3\_512\_absorb (poly)} \\
\texttt{sha3\_256\_absorb (r0 / r1)} \\
\texttt{sha3\_512\_absorb (r0 / r1)} \\
\texttt{r0 / r1 = sha3\_256\_digest} \\
\texttt{r0 || r1 = sha3\_512\_digest} \\
where the seed registers are used to store the hash outputs -- either \texttt{r0} or \texttt{r1} for SHA-3-256, and both \texttt{r0} and \texttt{r1} together for SHA-3-512 \\
\hline
\end{tabular}
\end{table}

\clearpage

%% file: main.bbl
\begin{thebibliography}{10}

\bibitem{shor_quantum_1997}
P.~W. {Shor}, ``{Polynomial-Time Algorithms for Prime Factorization and
  Discrete Logarithms on a Quantum Computer},'' {\em SIAM Journal of
  Computing}, vol.~26, pp.~1484--1509, Oct. 1997.

\bibitem{nist_pq1_2016}
L.~{Chen}, S.~{Jordan}, Y.~{Liu}, D.~{Moody}, R.~{Peralta}, R.~{Perlner}, and
  D.~{Smith-Tone}, ``{Report on Post-Quantum Cryptography},'' Tech. Rep. 8105,
  National Institute of Standards and Technology, Apr. 2016.

\bibitem{nist_pq2_2019}
G.~{Alagic}, J.~{Alperin-Sheriff}, D.~{Apon}, D.~{Cooper}, Q.~{Dang},
  C.~{Miller}, D.~{Moody}, R.~{Peralta}, R.~{Perlner}, A.~{Robinson},
  D.~{Smith-Tone}, and Y.~{Liu}, ``{Status Report on the First Round of the
  NIST Post-Quantum Cryptography Standardization Process},'' Tech. Rep. 8240,
  National Institute of Standards and Technology, Jan. 2019.

\bibitem{regev_lwe_2005}
O.~{Regev}, ``{On Lattices, Learning with Errors, Random Linear Codes, and
  Cryptography},'' in {\em Proceedings of the Thirty-Seventh Annual ACM
  Symposium on Theory of Computing (STOC)}, pp.~84--93, May 2005.

\bibitem{vadim_ringlwe_2013}
V.~{Lyubashevsky}, C.~{Peikert}, and O.~{Regev}, ``{On Ideal Lattices and
  Learning with Errors over Rings},'' {\em Journal of the ACM}, vol.~60,
  pp.~43:1--43:35, Nov. 2013.

\bibitem{langlois_module_2015}
A.~{Langlois} and D.~{Stehle}, ``{Worst-case to Average-case Reductions for
  Module Lattices},'' {\em Designs, Codes and Cryptography}, vol.~75,
  pp.~565--599, Jun. 2015.

\bibitem{brakerski_hardness_2013}
Z.~{Brakerski}, A.~{Langlois}, C.~{Peikert}, O.~{Regev}, and D.~{Stehle},
  ``{Classical Hardness of Learning with Errors},'' in {\em Proceedings of the
  Forty-fifth Annual ACM Symposium on Theory of Computing (STOC)},
  pp.~575--584, Jun. 2013.

\bibitem{regev_quantum_2004}
O.~{Regev}, ``{Quantum Computation and Lattice Problems},'' {\em SIAM Journal
  of Computing}, vol.~33, pp.~738--760, Mar. 2004.

\bibitem{ingrid_ringlwe_2014}
S.~S. {Roy}, F.~{Vercauteren}, N.~{Mentens}, D.~D. {Chen}, and
  I.~{Verbauwhede}, ``{Compact Ring-LWE Cryptoprocessor},'' in {\em
  Cryptographic Hardware and Embedded Systems -- CHES 2014}, pp.~371--391, Sep.
  2014.

\bibitem{ingrid_software_2015}
R.~{de Clercq}, S.~S. {Roy}, F.~{Vercauteren}, and I.~{Verbauwhede},
  ``{Efficient Software Implementation of Ring-LWE Encryption},'' in {\em 2015
  Design, Automation Test in Europe Conference Exhibition (DATE)},
  pp.~339--344, Mar. 2015.

\bibitem{alkim_newhopearm_2016}
E.~{Alkim}, P.~{Jakubeit}, and P.~{Schwabe}, ``{NewHope on ARM Cortex-M},'' in
  {\em Security, Privacy, and Applied Cryptography Engineering -- SPACE 2016},
  pp.~332--349, Dec. 2016.

\bibitem{kuo_newhopefpga_2017}
P.-C. {Kuo}, W.-D. {Li}, Y.-W. {Chen}, Y.-C. {Hsu}, B.-Y. {Peng}, C.-M.
  {Cheng}, and B.-Y. {Yang}, ``{High Performance Post-Quantum Key Exchange on
  FPGAs}.'' Cryptology ePrint Archive, Report 2017/690, 2017.
\newblock \url{https://eprint.iacr.org/2017/690}.

\bibitem{guneysu_newhopefpga_2017}
T.~{Oder} and T.~{Guneysu}, ``{Implementing the NewHope-Simple Key Exchange on
  low-cost FPGAs},'' in {\em International Conference on Cryptology and
  Information Security in Latin America, -- LATINCRYPT 2017}, pp.~371--391,
  Sep. 2017.

\bibitem{cammarota_ringlwe_2018}
H.~{Nejatollahi}, N.~{Dutt}, I.~{Banerjee}, and R.~{Cammarota},
  ``{Domain-specific Accelerators for Ideal Lattice-based Public Key
  Protocols}.'' Cryptology ePrint Archive, Report 2018/608, 2018.
\newblock \url{https://eprint.iacr.org/2018/608}.

\bibitem{bos_frodom4_2018}
J.~W. {Bos}, S.~{Friedberger}, M.~{Martinoli}, E.~{Oswald}, and M.~{Stam},
  ``{Fly, you fool! Faster Frodo for the ARM Cortex-M4}.'' Cryptology ePrint
  Archive, Report 2018/1116, 2018.
\newblock \url{https://eprint.iacr.org/2018/1116}.

\bibitem{guneysu_frodo_2018}
J.~{Howe}, T.~{Oder}, M.~{Krausz}, and T.~{Guneysu}, ``{Standard Lattice-Based
  Key Encapsulation on Embedded Devices},'' {\em IACR Transactions on
  Cryptographic Hardware and Embedded Systems}, vol.~2018, pp.~372--393, Aug.
  2018.

\bibitem{zhang_leia_2018}
S.~{Song}, W.~{Tang}, T.~{Chen}, and Z.~{Zhang}, ``{LEIA: A 2.05mm$^2$ 140mW
  Lattice Encryption Instruction Accelerator in 40nm CMOS},'' in {\em 2018 IEEE
  Custom Integrated Circuits Conference (CICC)}, pp.~1--4, Apr. 2018.

\bibitem{albrecht_rsa_2018}
M.~{Albrecht}, C.~{Hanser}, A.~{Holler}, T.~{Poppelmann}, F.~{Virdia}, and
  A.~{Wallner}, ``{Implementing RLWE-based Schemes Using an RSA
  Co-Processor},'' {\em IACR Transactions on Cryptographic Hardware and
  Embedded Systems}, vol.~2019, pp.~169--208, Nov. 2018.

\bibitem{liu_rlwe_2019}
D.~{Liu}, C.~{Zhang}, H.~{Lin}, Y.~{Chen}, and M.~{Zhang}, ``{A
  Resource-Efficient and Side-Channel Secure Hardware Implementation of
  Ring-LWE Cryptographic Processor},'' {\em IEEE Transactions on Circuits and
  Systems I: Regular Papers}, vol.~66, pp.~1474--1483, Apr. 2019.

\bibitem{basu_pqchw_2019}
K.~{Basu}, D.~{Soni}, M.~{Nabeel}, and R.~{Karri}, ``{NIST Post-Quantum
  Cryptography - A Hardware Evaluation Study}.'' Cryptology ePrint Archive,
  Report 2019/047, 2019.
\newblock \url{https://eprint.iacr.org/2019/047}.

\bibitem{cammarota_survey_2019}
H.~{Nejatollahi}, N.~{Dutt}, S.~{Ray}, F.~{Regazzoni}, I.~{Banerjee}, and
  R.~{Cammarota}, ``{Post-Quantum Lattice-Based Cryptography Implementations: A
  Survey},'' {\em ACM Computing Surveys}, vol.~51, pp.~129:1--129:41, Jan.
  2019.

\bibitem{guneysu_survey_2016}
T.~{Oder}, T.~{Guneysu}, F.~{Valencia}, A.~{Khalid}, M.~{O'Neill}, and
  F.~{Regazzoni}, ``{Lattice-based Cryptography: From Reconfigurable Hardware
  to ASIC},'' in {\em 2016 International Symposium on Integrated Circuits
  (ISIC)}, pp.~1--4, Dec. 2016.

\bibitem{alkim_frodo_2019}
M.~{Naehrig}, E.~{Alkim}, J.~{Bos}, L.~{Ducas}, K.~{Easterbrook},
  B.~{LaMacchia}, P.~{Longa}, I.~{Mironov}, V.~{Nikolaenko}, C.~{Peikert},
  A.~{Raghunathan}, and D.~{Stebila}, ``{FrodoKEM: Learning With Errors Key
  Encapsulation -- Algorithm Specifications And Supporting Documentation},''
  tech. rep., National Institute of Standards and Technology, 2019.
\newblock
  \url{https://csrc.nist.gov/Projects/Post-Quantum-Cryptography/Round-2-Submissions}.

\bibitem{alkim_newhope_2019}
T.~{Poppelmann}, E.~{Alkim}, R.~{Avanzi}, J.~{Bos}, L.~{Ducas}, A.~{de la
  Piedra}, P.~{Schwabe}, D.~{Stebila}, M.~R. {Albrecht}, E.~{Orsini},
  V.~{Osheter}, K.~G. {Paterson}, G.~{Peer}, and N.~P. {Smart}, ``{NewHope --
  Algorithm Specifications And Supporting Documentation},'' tech. rep.,
  National Institute of Standards and Technology, 2019.
\newblock
  \url{https://csrc.nist.gov/Projects/Post-Quantum-Cryptography/Round-2-Submissions}.

\bibitem{bindel_qtesla_2019}
N.~{Bindel}, S.~{Akleylek}, E.~{Alkim}, P.~S. L.~M. {Barreto}, J.~{Buchmann},
  E.~{Eaton}, G.~{Gutoski}, J.~{Kramer}, P.~{Longa}, H.~{Polat}, J.~E.
  {Ricardini}, and G.~{Zanon}, ``{Lattice-based Digital Signature Scheme qTESLA
  -- Submission to NIST's Post-Quantum Project},'' tech. rep., National
  Institute of Standards and Technology, 2019.
\newblock
  \url{https://csrc.nist.gov/Projects/Post-Quantum-Cryptography/Round-2-Submissions}.

\bibitem{bos_kyber_2019}
P.~{Schwabe}, R.~{Avanzi}, J.~{Bos}, L.~{Ducas}, E.~{Kiltz}, T.~{Lepoint},
  V.~{Lyubashevsky}, J.~M. {Schanck}, G.~{Seiler}, and D.~{Stehle},
  ``{CRYSTALS-Kyber -- Algorithm Specifications And Supporting
  Documentation},'' tech. rep., National Institute of Standards and Technology,
  2019.
\newblock
  \url{https://csrc.nist.gov/Projects/Post-Quantum-Cryptography/Round-2-Submissions}.

\bibitem{vadim_dilithium_2019}
V.~{Lyubashevsky}, L.~{Ducas}, E.~{Kiltz}, T.~{Lepoint}, P.~{Schwabe},
  G.~{Seiler}, and D.~{Stehle}, ``{CRYSTALS-Dilithium -- Algorithm
  Specifications And Supporting Documentation},'' tech. rep., National
  Institute of Standards and Technology, 2019.
\newblock
  \url{https://csrc.nist.gov/Projects/Post-Quantum-Cryptography/Round-2-Submissions}.

\bibitem{bernstein_mult_2008}
D.~J. {Bernstein}, ``{Fast Multiplication and its Applications},'' {\em
  Algorithmic Number Theory}, vol.~44, pp.~325--384, 2008.

\bibitem{cormen_algo_2009}
T.~H. {Cormen}, C.~E. {Leiserson}, R.~L. {Rivest}, and C.~{Stein}, {\em
  {Introduction to Algorithms}}.
\newblock The MIT Press, 3rd~ed., 2009.

\bibitem{bai_polymul_2016}
C.~{Du} and G.~{Bai}, ``{Towards Efficient Polynomial Multiplication for
  Lattice-based Cryptography},'' in {\em 2016 IEEE International Symposium on
  Circuits and Systems (ISCAS)}, pp.~1178--1181, May 2016.

\bibitem{howell_algo_2012}
R.~R. {Howell}, {\em {Algorithms: A Top-Down Approach}}.
\newblock Draft, 2012.
\newblock \url{http://people.cs.ksu.edu/~rhowell/algorithms-text}.

\bibitem{naehrig_ntt_2016}
P.~{Longa} and M.~{Naehrig}, ``{Speeding up the Number Theoretic Transform for
  Faster Ideal Lattice-Based Cryptography}.'' Cryptology ePrint Archive, Report
  2016/504, 2016.
\newblock \url{https://eprint.iacr.org/2016/504}.

\bibitem{barrett_red_1986}
P.~{Barrett}, ``{Implementing the Rivest Shamir and Adleman Public Key
  Encryption Algorithm on a Standard Digital Signal Processor},'' in {\em
  Advances in Cryptology -- CRYPTO 86}, pp.~311--323, Aug. 1986.

\bibitem{seo_emblem_2017}
M.~{Seo}, J.~H. {Park}, D.~H. {Lee}, S.~{Kim}, and S.-J. {Lee}, ``{EMBLEM and
  R.EMBLEM -- Error-blocked Multi-Bit LWE-based Encapsulation Mechanism},''
  tech. rep., National Institute of Standards and Technology, 2017.
\newblock
  \url{https://csrc.nist.gov/Projects/Post-Quantum-Cryptography/Round-1-Submissions}.

\bibitem{zhang_pqntrusign_2017}
C.~{Chen}, J.~{Hoffstein}, W.~{Whyte}, and Z.~{Zhang}, ``{NIST PQ Submission:
  pqNTRUSign -- A Modular Lattice Signature Scheme},'' tech. rep., National
  Institute of Standards and Technology, 2017.
\newblock
  \url{https://csrc.nist.gov/Projects/Post-Quantum-Cryptography/Round-1-Submissions}.

\bibitem{ding_kex_2017}
J.~{Ding}, T.~{Takagi}, X.~{Gao}, and Y.~{Wang}, ``{Ding Key Exchange},'' tech.
  rep., National Institute of Standards and Technology, 2017.
\newblock
  \url{https://csrc.nist.gov/Projects/Post-Quantum-Cryptography/Round-1-Submissions}.

\bibitem{smart_lima_2017}
M.~R. {Albrecht}, Y.~{Lindell}, E.~{Orsini}, V.~{Osheter}, K.~G. {Paterson},
  G.~{Peer}, and N.~P. {Smart}, ``{LIMA –- A PQC Encryption Scheme},'' tech.
  rep., National Institute of Standards and Technology, 2017.
\newblock
  \url{https://csrc.nist.gov/Projects/Post-Quantum-Cryptography/Round-1-Submissions}.

\bibitem{banerjee_isscc_2019}
U.~{Banerjee}, A.~{Pathak}, and A.~P. {Chandrakasan}, ``{An Energy-Efficient
  Configurable Lattice Cryptography Processor for the Quantum-Secure Internet
  of Things},'' in {\em 2019 IEEE International Solid-State Circuits Conference
  (ISSCC)}, pp.~46--48, Feb. 2019.

\bibitem{ingrid_polymul_2015}
D.~D. {Chen}, N.~{Mentens}, F.~{Vercauteren}, S.~S. {Roy}, R.~C.~C. {Cheung},
  D.~{Pao}, and I.~{Verbauwhede}, ``{High-Speed Polynomial Multiplication
  Architecture for Ring-LWE and SHE Cryptosystems},'' {\em IEEE Transactions on
  Circuits and Systems I: Regular Papers}, vol.~62, pp.~157--166, Jan. 2015.

\bibitem{noguchi_dpsram_2008}
H.~{Noguchi}, S.~{Okumura}, Y.~{Iguchi}, H.~{Fujiwara}, Y.~{Morita}, K.~{Nii},
  H.~{Kawaguchi}, and M.~{Yoshimoto}, ``{Which is the Best Dual-Port SRAM in
  45-nm Process Technology? — 8T, 10T Single End, and 10T Differential
  —},'' in {\em 2008 IEEE International Conference on Integrated Circuit
  Design and Technology and Tutorial}, pp.~55--58, Jun. 2008.

\bibitem{pease_fft_1968}
M.~C. {Pease}, ``{An Adaptation of the Fast Fourier Transform for Parallel
  Processing},'' {\em Journal of the ACM}, vol.~15, pp.~252--264, Apr. 1968.

\bibitem{pollard_fft_1971}
J.~M. {Pollard}, ``{The Fast Fourier Transform in a Finite Field},'' {\em
  Mathematics of Computation}, vol.~25, pp.~365--374, May 1971.

\bibitem{sepulveda_ntt_2019}
T.~{Fritzmann} and J.~{Sepúlveda}, ``{Efficient and Flexible Low-Power NTT for
  Lattice-Based Cryptography},'' in {\em 2019 IEEE International Symposium on
  Hardware Oriented Security and Trust (HOST)}, pp.~141--150, May 2019.

\bibitem{pqm4}
M.~J. {Kannwischer}, J.~{Rijneveld}, P.~{Schwabe}, and K.~{Stoffelen},
  ``{PQM4}: Post-quantum crypto library for the {ARM} {Cortex-M4},'' 2018.
\newblock \url{https://github.com/mupq/pqm4}.

\bibitem{nucleo_f411re}
STMicroelectronics, ``{NUCLEO-F411RE Development Board}.''
\newblock \url{https://os.mbed.com/platforms/ST-Nucleo-F411RE}.

\bibitem{guneysu_masked_2018}
T.~{Oder}, T.~{Schneider}, T.~{Poppelmann}, and T.~{Guneysu}, ``{Practical
  CCA2-Secure and Masked Ring-LWE Implementation},'' {\em IACR Transactions on
  Cryptographic Hardware and Embedded Systems}, vol.~2018, pp.~142--174, Feb.
  2018.

\bibitem{nist_sha3_2015}
{NIST}, ``{SHA-3 Standard: Permutation-Based Hash and Extendable-Output
  Functions},'' Tech. Rep. FIPS PUB 202, National Institute of Standards and
  Technology, Aug. 2015.

\bibitem{nist_aes_2001}
{NIST}, ``{Advanced Encryption Standard (AES)},'' Tech. Rep. FIPS PUB 197,
  National Institute of Standards and Technology, Nov. 2001.

\bibitem{bernstein_chacha_2008}
D.~J. {Bernstein}, ``{ChaCha, a variant of Salsa20},'' Jan. 2008.
\newblock \url{https://cr.yp.to/chacha/chacha-20080128.pdf}.

\bibitem{bertoni_keccak_2009}
G.~{Bertoni}, J.~{Daemen}, M.~{Peeters}, and G.~{Van Assche}, ``{Keccak
  Specifications},'' 2009.

\bibitem{gueron_ringlwe_2016}
S.~{Gueron} and F.~{Schlieker}, ``{Speeding up R-LWE Post-Quantum Key
  Exchange}.'' Cryptology ePrint Archive, Report 2016/467, 2016.
\newblock \url{https://eprint.iacr.org/2016/467}.

\bibitem{knuthyao_sample_1976}
D.~E. {Knuth} and A.~C. {Yao}, {\em {Algorithms and Complexity: New Directions
  and Recent Results}}, ch.~{The Complexity of Non-Uniform Random Number
  Generation}.
\newblock Academic Press, 1976.

\bibitem{follath_sampling_2014}
J.~{Follath}, ``{Gaussian Sampling in Lattice Based Cryptography},'' {\em Tatra
  Mountains Mathematical Publications}, vol.~60, pp.~1--23, Sep. 2014.

\bibitem{lee_lizard_2017}
J.~H. {Cheon}, S.~{Park}, J.~{Lee}, D.~{Kim}, Y.~{Song}, S.~{Hong}, D.~{Kim},
  J.~{Kim}, S.-M. {Hong}, A.~{Yun}, J.~{Kim}, H.~{Park}, E.~{Choi}, K.~{Kim},
  J.-S. {Kim}, and J.~{Lee}, ``{Lizard Public Key Encryption},'' tech. rep.,
  National Institute of Standards and Technology, 2017.
\newblock
  \url{https://csrc.nist.gov/Projects/Post-Quantum-Cryptography/Round-1-Submissions}.

\bibitem{ingrid_saberm4_2018}
A.~{Karmakar}, J.~M. {Bermudo Mera}, S.~S. {Roy}, and I.~{Verbauwhede},
  ``{Saber on ARM},'' {\em IACR Transactions on Cryptographic Hardware and
  Embedded Systems}, vol.~2018, pp.~243--266, Aug. 2018.

\bibitem{banerjee_isscc_2018}
U.~{Banerjee}, C.~{Juvekar}, A.~{Wright}, {Arvind}, and A.~P. {Chandrakasan},
  ``{An Energy-Efficient Reconfigurable DTLS Cryptographic Engine for
  End-to-End Security in IoT Applications},'' in {\em 2018 IEEE International
  Solid-State Circuits Conference (ISSCC)}, pp.~42--44, Feb. 2018.

\bibitem{waterman_riscv_2014}
A.~{Waterman}, Y.~{Lee}, D.~A. {Patterson}, and K.~{Asanovic}, ``{The RISC-V
  Instruction Set Manual},'' 2014.

\bibitem{dichtl_trng_2007}
M.~{Dichtl} and J.~D. {Golic}, ``{High-Speed True Random Number Generation with
  Logic Gates Only},'' in {\em Cryptographic Hardware and Embedded Systems -
  CHES 2007}, pp.~45--62, Sep. 2007.

\bibitem{fujisaki_2013}
E.~{Fujisaki} and T.~{Okamoto, Tatsuaki}, ``{Secure Integration of Asymmetric
  and Symmetric Encryption Schemes},'' {\em Journal of Cryptology}, vol.~26,
  pp.~80--101, Jan. 2013.

\bibitem{sepulveda_pqriscv_2019}
T.~{Fritzmann}, U.~{Sharif}, D.~{Müller-Gritschneder}, C.~{Reinbrecht},
  U.~{Schlichtmann}, and J.~{Sepulveda}, ``{Towards Reliable and Secure
  Post-Quantum Co-Processors based on RISC-V},'' in {\em 2019 Design,
  Automation Test in Europe Conference Exhibition (DATE)}, pp.~1148--1153, Mar.
  2019.

\bibitem{hutter_nacl_2015}
M.~{Hutter}, J.~{Schilling}, P.~{Schwabe}, and W.~{Wieser}, ``Nacl's
  crypto{\_}box in hardware,'' in {\em Cryptographic Hardware and Embedded
  Systems -- CHES 2015}, pp.~81--101, Sep. 2015.

\bibitem{park_spa_2016}
A.~{Park} and D.~{Han}, ``{Chosen Ciphertext Simple Power Analysis on Software
  8-bit Implementation of Ring-LWE Encryption},'' in {\em 2016 IEEE Asian
  Hardware-Oriented Security and Trust (AsianHOST)}, pp.~1--6, Dec 2016.

\bibitem{primas_sca_2017}
R.~{Primas}, P.~{Pessl}, and S.~{Mangard}, ``{Single-Trace Side-Channel Attacks
  on Masked Lattice-Based Encryption},'' in {\em Cryptographic Hardware and
  Embedded Systems -- CHES 2017}, pp.~513--533, Sep. 2017.

\bibitem{aysu_brlwe_2018}
A.~{Aysu}, M.~{Orshansky}, and M.~{Tiwari}, ``{Binary Ring-LWE Hardware with
  Power Side-Channel Countermeasures},'' in {\em 2018 Design, Automation Test
  in Europe Conference Exhibition (DATE)}, pp.~1253--1258, Mar. 2018.

\bibitem{kocher_dpa_2011}
P.~{Kocher}, J.~{Jaffe}, B.~{Jun}, and P.~{Rohatgi}, ``{Introduction to
  Differential Power Analysis},'' {\em Journal of Cryptographic Engineering},
  vol.~1, pp.~5--27, Apr. 2011.

\bibitem{ebrahimi_pqiot_2019}
S.~{Ebrahimi}, S.~{Bayat-Sarmadi}, and H.~{Mosanaei-Boorani}, ``{Post-Quantum
  Cryptoprocessors Optimized for Edge and Resource-Constrained Devices in
  IoT},'' {\em IEEE Internet of Things Journal}, vol.~6, pp.~5500--5507, Jun.
  2019.

\bibitem{ingrid_masked_2015}
O.~{Reparaz}, S.~S. {Roy}, F.~{Vercauteren}, and I.~{Verbauwhede}, ``{A Masked
  Ring-LWE Implementation},'' in {\em Cryptographic Hardware and Embedded
  Systems -- CHES 2015}, pp.~683--702, Sep. 2015.

\bibitem{ingrid_masked_2016}
O.~{Reparaz}, R.~d. S.~S. {Roy}, F.~{Vercauteren}, and I.~{Verbauwhede},
  ``Additively homomorphic ring-lwe masking,'' in {\em Post-Quantum
  Cryptography}, pp.~233--244, Feb. 2016.

\bibitem{alkim_newhopesimple_2016}
E.~{Alkim}, L.~{Ducas}, T.~{Poppelmann}, and P.~{Schwabe}, ``{NewHope without
  Reconciliation}.'' Cryptology ePrint Archive, Report 2016/1157, 2016.
\newblock \url{https://eprint.iacr.org/2016/1157}.

\bibitem{player_hardness_2015}
M.~R. {Albrecht}, R.~{Player}, and S.~{Scott}, ``{On the Concrete Hardness of
  Learning with Errors},'' {\em Journal of Mathematical Cryptology}, vol.~9,
  p.~169–203, Oct. 2015.

\bibitem{vercauteren_saber_2019}
J.~{D’Anvers}, A.~{Karmakar}, S.~S. {Roy}, and F.~{Vercauteren}, ``{SABER:
  Mod-LWR based KEM},'' tech. rep., National Institute of Standards and
  Technology, 2019.
\newblock
  \url{https://csrc.nist.gov/Projects/Post-Quantum-Cryptography/Round-2-Submissions}.

\bibitem{saarinen_round5_2019}
O.~{Garcia-Morchon}, Z.~{Zhang}, S.~{Bhattacharya}, R.~{Rietman},
  L.~{Tolhuizen}, J.-L. {Torre-Arce}, H.~{Baan}, M.-J.~O. {Saarinen},
  S.~{Fluhrer}, T.~{Laarhoven}, and R.~{Player}, ``{Round5: KEM and PKE based
  on (Ring) Learning with Rounding},'' tech. rep., National Institute of
  Standards and Technology, 2019.
\newblock
  \url{https://csrc.nist.gov/Projects/Post-Quantum-Cryptography/Round-2-Submissions}.

\bibitem{peikert_lwr_2012}
A.~{Banerjee}, C.~{Peikert}, and A.~{Rosen}, ``{Pseudorandom Functions and
  Lattices},'' in {\em Advances in Cryptology -- EUROCRYPT 2012}, pp.~719--737,
  Apr. 2012.

\end{thebibliography}
